%% file: bv.tex
\newif\iffigs\figstrue

\documentclass[paper, 12pt, letterpaper]{JHEP}

\def\Bbb{\bf}
\def\C{{\Bbb C}}
\def\R{{\Bbb R}}
\def\Z{{\Bbb Z}}

\def\bearray{\begin{eqnarray}}
\def\eearray{\end{eqnarray}}
\def\bearraynn{\begin{eqnarray*}}
\def\eearraynn{\end{eqnarray*}}
\def\bfig{\begin{figure}}
\def\efig{\end{figure}}

\def\opeq#1{\advance\lineskip#1 \advance\baselineskip#1
        \advance\lineskiplimit#1}

\newtheorem{Proposition}{Proposition}[section]

\newtheorem{Theorem}{Theorem}[section]
\newtheorem{Lemma}{Lemma}[section]
\newtheorem{Corrolary}{Corrolary}[section]

\newcommand{\be}{\begin{equation}}
\newcommand{\ee}{\end{equation}}
\newcommand{\bea}{\begin{eqnarray}}
\newcommand{\eea}{\end{eqnarray}}
\newcommand{\we}{\wedge}
\newcommand{\bp}{\begin{Proposition}}
\newcommand{\ep}{\end{Proposition}}
\newcommand{\bt}{\begin{Theorem}}
\newcommand{\et}{\end{Theorem}}
\newcommand{\bl}{\begin{Lemma}}
\newcommand{\el}{\end{Lemma}}
\newcommand{\bc}{\begin{Corrolary}}
\newcommand{\ec}{\end{Corrolary}}
\newcommand{\nn}{\nonumber}

\newcommand{\nonu}{\nonumber}
\def \ov {\over }
\def\bea{\begin{eqnarray}}
\def\eea{\end{eqnarray}}

\def\be{\begin{equation}}
\def\ee{\end{equation}}
\def\ba{\begin{eqnarray}}
\def\ea{\end{eqnarray}}

\def\C{{\cal C}}

\def\1{{{(1)}}}\def\2{{{(2)}}}\def\3{{{(3)}}}

\font\mybb=msbm10 at 12pt

\def\bb#1{\hbox{\mybb#1}}

\def\Z {\bb{Z}}

\def\id{\protect{{1 \kern-.28em {\rm l}}}}

\def\p#1{{\phi{}^{(#1)}}}
\def\hp#1{{{\phi'}{}^{(#1)}}}

\def\boldphi{\mbox{\boldmath $\phi$}}

\def\boldwe{\mbox{\boldmath $\wedge$}}
\def\bwe{{{{\boldwe\hspace{-8.9pt}\boldwe}\hspace{-8.8pt}\boldwe}
\hspace{-8.8pt}\boldwe}}
\def\bweft{{{{\boldwe\hspace{-8.1pt}\boldwe}\hspace{-8pt}\boldwe}
\hspace{-8pt}\boldwe}}

\usepackage{graphics}

\title{Graded Chern-Simons field theory and graded topological 
D-branes}

\author{C.~I.~Lazaroiu, R. Roiban and D. Vaman
\\C.~N.~Yang Institute for Theoretical Physics\\
SUNY at Stony BrookNY11794-3840, U.S.A.\\
calin, roiban, dvaman@insti.physics.sunysb.edu
}

\abstract{We discuss graded D-brane systems of the topological A model
on a Calabi-Yau threefold, 
by means of their string field theory. We give a detailed analysis 
of the extended string field action, showing that it 
satisfies the classical master equation, and construct
the associated BV system. The analysis is entirely general 
and it applies to any collection of D-branes (of distinct grades) 
wrapping the same special Lagrangian cycle, being valid in
arbitrary topology. Our discussion employs a 
$\Z$-graded version of the covariant BV formalism, whose 
formulation involves the concept of 
{\em graded supermanifolds}.
We discuss this formalism in detail 
and explain why $\Z$-graded supermanifolds are necessary for a correct 
geometric understanding of BV systems. 
For the particular case of graded D-brane pairs, 
we also 
give a direct construction of the master action, finding complete agreement
with the abstract formalism. We analyze formation 
of acyclic composites and show that, under certain topological assumptions,
all states resulting from the  condensation process of a pair of branes 
with grades differing by one unit  are BRST trivial and thus the
composite can be viewed as a closed string vacuum. We prove
that there are {\em six} types of pairs which must 
be viewed as generally inequivalent. This contradicts the 
assumption that `brane-antibrane' systems exhaust the nontrivial 
dynamics of topological A-branes with the same geometric support.
}

\preprint{YITP-SB 01-37}

\begin{document}

\tableofcontents

\pagebreak

\section{Introduction}
${}$

The issue of D-brane composite formation plays a central role in the
study of Calabi-Yau compactifications of open superstrings. It has
gradually become clear \cite{Douglas_Kontsevich, com1, com3,
  Aspinwall, Diaconescu, sc, Oz_superconn, Oz_triples} that a proper
understanding of D-brane condensation processes holds the key to
disentangling the relevance of the derived category of coherent
sheaves and its A-model analogue, the derived category of Fukaya's
category \cite{Fukaya, Fukaya2, Kontsevich_recent} (or rather, a
generalization thereof \cite{sc}).  Since D-brane condensation
involves off-shell string dynamics, the most systematic approach to
this problem is through the methods of open string field theory. In
fact, it seems that `mirror symmetry with open strings' is best
formulated as a (quasi)-equivalence of open string field theories with
D-branes, an overarching off-shell version which would subsume all
previous statements.  A fundamental ingredient in this program is the
observation of \cite{Douglas_Kontsevich} that the correct description
of D-branes in Calabi-Yau compactifications involves an extra datum,
an integer-valued quantity called the D-brane's grade.  As pointed out
in \cite{com3}, the procedure of including graded D-branes admits a
general string field theoretic description (the so-called `shift
completion' of a D-brane category). Taking this fact into account
leads to a concrete presentation of the {\em extended} moduli space
of open string vacua, an approach which affords computation of various
physical quantities away from points of the geometric moduli space.
Moreover, the papers \cite{com3, Diaconescu} and \cite{sc} gave an
explicit description of certain subsectors of the relevant string
field theories for the case of Calabi-Yau threefolds.

The purpose of this paper is to continue the analysis (initiated in
\cite{sc}) of a string field-theoretic model relevant for the dynamics
of A-type \cite{Ooguri} graded {\em topological} D-branes of a
Calabi-Yau threefold compactification. In \cite{sc}, a sector of the
string field theory of such objects was described in terms of a
$\Z$-graded version of Chern-Simons field theory living on a special
Lagrangian cycle. The model captures the off-shell dynamics of
topological branes of arbitrary grades wrapped over the cycle. As
sketched in that paper, the classical moduli space of vacua of this
theory allows one to recover a piece of the extended boundary moduli
space, an object which plays a crucial role in the homological mirror
symmetry program \cite{Kontsevich}.  More precisely, points of the
extended moduli space can be described as {\em generalized D-brane
  composites}, obtained by condensing boundary condition changing
states between {\em graded} D-branes.  This suggest that one can
extract physical information about such points by studying the quantum
dynamics of the resulting theory.

Since the theory written down in \cite{sc} involves higher rank forms,
its quantization must deal with the issue of reducible gauge algebras.
Therefore, a correct analysis of this theory requires the full force
of the Batalin-Vilkovisky formalism. The purpose of the present paper
is to carry out the classical part of this analysis, thus providing a
precise starting point for a study of quantum dynamics.  Applying such
methods, we will be able to recover the extended action already
written down in \cite{sc}, which plays the role of (classical) master
action for our theory. This enables us to show that the model
constructed in \cite{sc} is a consistent starting point for a
quantum-mechanical analysis. As an application, we give a detailed
discussion of graded D-brane pairs, thus obtaining a realization of
ideas proposed in \cite{Vafa_cs}, though in a somewhat different
context.

Our investigation reveals that there are six types of D-brane pairs
which are physically inequivalent in general.  
This confirms the $\Z_6$ periodicity
of the D-brane grade suggested in \cite{Douglas_Kontsevich} from
worldsheet considerations and contradicts the assumption that
`brane-antibrane' systems exhaust the nontrivial dynamics of
topological A-branes with the same geometric support.

The paper is organized as follows. We start in Section 2 by
recalling the construction of the classical string field action of
\cite{sc}. In section 3, we discuss the associated extended string
field theory, giving a detailed presentation of the structures
involved.  Though this section is somewhat technical, a clear
description of these structures is crucial for a correct understanding
of latter work.  The central objects are the so-called extended
boundary product and extended bilinear form introduced in \cite{sc},
whose construction we explain in detail.  In Section 4, we proceed to
show that the extended action discussed in Section 3 satisfies the
classical master equation with respect to the antibracket induced by a
certain odd symplectic form (which coincides with the extended
bilinear form up to sign factors and a shift of grading).  This
establishes the fact that the the extended action plays the role of
classical BV action for our systems. The proof, which is completely
general, makes use of the geometric version of the BV formalism, as
discussed, for example, in \cite{Kontsevich_Schwarz}.  Our approach is
in fact a certain modification of usual covariant framework, which
differs from the latter by incorporating the ghost grading. This
modified formalism, which is necessary for a correct description of BV
systems, involves the concept of {\em graded supermanifolds} which was
recently introduced in \cite{Voronov}. We therefore start with a
general exposition of the geometric $\Z$-graded formalism, which is of
independent interest for foundational studies of BV quantization.  We
then apply this framework to the extended string field theory of
Section 3. This allows us to give a complete construction of the
relevant BV system, and a very concise and general proof of the fact
that the extended action satisfies the classical master equation. We
also identify the classical gauge which leads to the unextended string
field theory.  After discussing the form of the BRST operator in this
gauge, we proceed with a discussion of the particular case of graded
D-brane pairs. Section 5 recalls the open string interpretation of the
various data discussed in Section 2, and gives their concrete
 description for such systems.  In Section 6, we consider
composite formation for D-brane pairs and the issue of acyclic
condensates. After discussing the worldsheet BRST cohomology and
explaining how it distinguishes between the various pairs, we explain
the interpretation of string field vacua as points of the extended
moduli space, and give an explicit construction of acyclic composites
for certain graded D-brane pairs whose relative grading equals one.
This gives a concrete realization of suggestions made in
\cite{Vafa_cs}.

While the geometric BV framework of Section 4 is extremely powerful
and general, it does not explain the origin of the various components
of the extended string field. To gain some insight into this issue, we
proceed in Section 7 with a direct construction of the BV action for
the case of graded pairs. We show that the more familiar
component approach leads to an action which can be viewed as the
expression of the extended action of Section 3 in a particular system
of linear coordinates, and show how the various ghosts and antifields
arise in standard manner by performing the BV-BRST resolution of our
(closed, but generally reducible) gauge algebras.  We also give a
discussion of the relation between pairs with arbitrarily high
relative grade.  Section 8 connects these results with the geometric
approach of Section 4, and in particular gives a very concise
formulation of the BV-BRST algorithm.  This synthetic description
should be useful for understanding the structure of the gauge algebra
of systems containing more than two graded branes.  We end in Section
9 by presenting our conclusions and a few directions for further
research.

\section{A string field theory for graded topological D-branes} 

This section describes the string field theory of a collection of
graded topological D-branes wrapping the same special Lagrangian cycle
of a Calabi-Yau threefold. This theory was written down in \cite{sc}
starting from a worldsheet analysis and using the framework developed
in \cite{com1, com3}.

\subsection{Graded D-branes}

We start by recalling some basic concepts introduced in
\cite{Douglas_Kontsevich} (see also \cite{Aspinwall} and \cite{sc}).
We are interested in so-called {\em graded topological D-branes} of
the A-model compactified on a Calabi-Yau threefold $X$.  Recall from
\cite{Witten_CS} that an {\em ungraded} topological D-brane can be
described as a pair $(L,E)$ where $L$ is a (connected) special
Lagrangian cycle of $X$ and $E$ is a flat vector bundle on
$L$\footnote{We remind the reader that this is a vector bundle endowed
  with a flat structure, i.e. the equivalence class of a family of
  local trivializations whose transition functions are constant.
  Specifying a flat structure amounts to giving a gauge-equivalence
  class of flat connections.}.  The generalization to {\em graded}
D-branes \cite{Douglas_Kontsevich, sc} is obtained by replacing $L$
with a graded version \cite{Seidel}, which for practical purposes
amounts to fixing an integer $n$ (the brane's {\em grade}) and a
certain orientation of $L$ which depends on $n$. As discussed in
\cite{Douglas_Kontsevich, Aspinwall, sc}, the worldsheet $U(1)$ charge
in the boundary condition changing sectors between two graded D-branes
wrapping $L$ is then shifted by $\pm n$, while the boundary products
in various sectors contain extra signs. Moreover, the boundary metric
receives grade-dependent signs in boundary condition changing sectors,
due to the fact that the relevant orientation of $L$ depends on the
branes' grades.

\subsection{Graded Chern-Simons field theory as a string field 
  theory for graded D-branes}

As explained in detail in \cite{sc}, it is possible to describe
certain graded D-brane systems through a generalization of
Chern-Simons field theory. To be specific, we consider a collection of
graded topological A-type branes wrapping {\em the same} special
Lagrangian cycle $L$ in $X$.  The main assumption for what follows is
that no two D-branes of this collection have the same grade. This
allows one to label the branes by their grades, which form a finite or
infinite set of integers. If $a_n$ denotes the D-brane of grade $n$,
then we denote by $E_n$ its underlying bundle. This notation includes
the specification of a flat structure (flat connection) on each $E_n$.

With these hypotheses, it was shown in \cite{sc} that the string field
theory of the system is a graded version of Chern-Simons field theory,
which generalizes both the usual Chern-Simons description of
\cite{Witten_CS} and the supergroup Chern-Simons proposal of
\cite{Vafa_cs}.  This describes the dynamics of so-called `degree one
graded connections' \cite{Bismut_Lott} on the graded (super-)bundle
${\bf E}=\oplus_{n}{E_n}$.  To define the theory, one must first
describe the so-called {\em total boundary space} ${\cal H}$ of
\cite{sc}, which consists of the off-shell states of open strings
stretching between our branes.

\subsubsection{The total boundary algebra}

We start by considering the algebra ${\cal E}$ of endomorphisms of
${\bf E}$.  This admits a natural $\Z$-grading induced by the bundle
decomposition: \be
\label{dec}
End({\bf E})=\oplus_{k}{End_k({\bf E})}~~, \ee where: \be End_k({\bf
  E})=\oplus_{n-m=k}{Hom(E_m, E_n)}~~.  \ee If $f$ is a morphism from
$E_m$ to $E_n$, then its degree with respect to this grading is given
by: \be \Delta(f)=n-m~~.  \ee It is easy to see that the composition
of morphisms is homogeneous of degree zero with respect to this
grading: \be \Delta(f\circ g)=\Delta(f)+\Delta(g)~~.  \ee It follows
that ${\cal E}$ (with multiplication given by composition of
morphisms) forms a graded associative algebra with respect to the
grading induced by $\Delta$.

The next step is to consider the tensor product ${\cal
  H}=\Omega^*(L)\otimes {\cal E}$ between the exterior algebra of $L$
and the endomorphism algebra ${\cal E}$.  Since both algebras are
$\Z$-graded, ${\cal H}$ is endowed with gradings $rk$ and $\Delta$
induced from its components, as well as with the total grading
$|~.~|=rk +\Delta$.  On decomposable elements $u=\rho\otimes f$, these
are given by: \be rk u=rk
\rho~~,~~\Delta(u)=\Delta(f)~~,~~|u|=rk\rho+\Delta(f)~~.  \ee An
arbitrary element $u$ of ${\cal H}$ can be viewed as a matrix
$u=(u_{mn})$, where $u_{mn}\in \Omega^*(L)\otimes \Gamma(Hom(E_m,
E_n))$ is a bundle-valued form.  Then $\Delta(u_{mn})=n-m$ and
$|u_{mn}|=rk u_{mn}+n-m$~~. The space ${\cal H}$ is also endowed with
a canonical multiplication $\bullet$ (the `total boundary product' of
\cite{sc}), induced from the multiplicative structure of its tensor
components. On decomposable elements $u=\rho\otimes f$ and
$v=\eta\otimes g$, this is given by: \be
\label{dot}
u\bullet v=(-1)^{\Delta(f)rk \eta}(\rho\wedge \eta)\otimes (f\circ
g)~~.  \ee Up to the sign prefactor, the right hand side is simply the
usual wedge product of $End({\bf E})$ -valued forms, which includes
composition of bundle morphisms on the coefficients: \be u\wedge
v=(\rho\wedge \eta)\otimes (f\circ g)~~.  \ee This allows us to write
(\ref{dot}) in a slightly more familiar form: \be u\bullet
v=(-1)^{\Delta(u)rk v}u\wedge v~~. \ee

The boundary product is homogeneous of degree zero with respect to the
total grading: \be |u\bullet v|=|u|+|v|~~.  \ee Hence ${\cal H}$
becomes a graded associative algebra when endowed with the grading
$|~.~|$ and the product $\bullet$. This well-known construction of a
graded associative structure on the tensor product from similar
structures on components is usually denoted by: \be {\cal
  H}=\Omega^*(L){\hat \otimes} {\cal E}~~, \ee where the hat above the
tensor product indicates that the multiplication and grading on the
resulting space are constructed in the canonical manner discussed
above.

A supplementary datum is provided by the existence of a natural
differential $d$ on the product algebra ${\cal H}$. This is the
exterior differential on $End({\bf E})$-valued forms, twisted with the
direct sum connection $A=\oplus_n{A_n}$. Flatness of $A_n$ assures
that $d^2=0$, and definition (\ref{dot}) implies that $d$ acts as a
graded derivation of the product $\bullet$ (with respect to the total
grading): \be d(u\bullet v)=(du)\bullet v+(-1)^{|u|}u\bullet (dv)~~.
\ee Moreover, one has: \be |du|=|u|+1~~.  \ee

The final element needed in the construction is a `trace' on ${\cal
  H}$.  This is induced by the natural traces on $\Omega^*(L)$ and
${\cal E}$, which are defined as follows. For complex-valued forms
$\rho\in \Omega^*(L)$, we define: \be
Tr_\Omega(\rho)=\int_{L}{\rho}~~, \ee while for $End({\bf E})$-valued
morphisms $f$ we have the supertrace of \cite{Quillen}: \be
\label{str}
str(f)=\sum_{m}{(-1)^{m}tr_{m}(f_{mm})}~~, \ee where $tr_{m}$ is the
fiberwise trace in the bundle $End(E_m)$.  Note that $str(f)$ is a
complex-valued function defined on $L$.  When all components $E_n$ of
${\bf E}$ have grades $n$ of the same parity (which in the language of
\cite{Quillen} amounts to taking the even or odd component of ${\bf
  E}$ to be the zero bundle), the supertrace (\ref{str}) reduces to
$\pm$ the ordinary fiberwise trace on the bundle $End({\bf E})$.
 
Both traces are graded-symmetric with respect to the natural degrees
on their spaces of definition: \be
\label{traces}
Tr_\Omega(\rho\wedge \eta)=(-1)^{rk\rho rk\eta}Tr_{\Omega}(\eta\wedge
\rho) ~~,~~ str(f\circ g)=(-1)^{\Delta(f)\Delta(g)}str(g\circ f)~~.
\ee Since $Tr_\Omega(u\wedge v)$ and $str(f\circ g)$ vanish unless $rk
\rho+rk \eta=3$ (remember that $L$ is a 3-cycle !), respectively
$\Delta(f)+\Delta(g)=0$, the graded symmetry properties are equivalent
with: \be Tr_\Omega(\rho\wedge \eta)=Tr_{\Omega}(\eta\wedge
\rho)~~{\rm~and~}~~ str(f\circ g)=(-1)^{\Delta(f)}str(g\circ f)~~.
\ee Using (\ref{traces}), we define a trace on ${\cal H}$ which on
decomposable elements $u=\rho\otimes f$ is given by: \be Tr_{\cal
  H}(u)=\int_{L}{str(f)\rho}~~.  \ee It is easy to check that this is
graded-symmetric: \be Tr_{\cal H}(u\bullet v)=(-1)^{|u||v|}Tr_{\cal
  H}(v\bullet u)~~.  \ee It immediately follows that the
(nondegenerate) bilinear form on ${\cal H}$, defined through: \be
\langle u , v \rangle:=Tr_{\cal H}(u\bullet v)=\int_{L}{str(u\bullet
  v)}~~, \ee is graded-symmetric as well: \be \langle u , v
\rangle=(-1)^{|u||v|}\langle v, u \rangle~~.  \ee It is clear that
$\langle u, v\rangle$ vanishes on bi-homogeneous elements unless both
of the conditions $rk u+rk v=3$ and $\Delta(u)+\Delta(v)=0$ are
satisfied.  It follows that non-vanishing of $\langle u, v\rangle$
requires $|u|+|v|=3$, for elements homogeneous with respect to the
total degree. Due to this selection rule, the graded symmetry property
is in fact equivalent with: \be \langle u , v \rangle=\langle v, u
\rangle~~.  \ee The last properties we shall need are invariance of
the bilinear form with respect to the total boundary product and
differential: \be \langle du, v\rangle+(-1)^{|u|}\langle u,
dv\rangle=0~~,~~ \langle u\bullet v, w\rangle=\langle u, v\bullet
w\rangle~~.  \ee These follows easily upon using the properties of the
differential and supertrace. We end by noting that the trace and
bilinear form can be expressed in more familiar language as follows:
\be Tr_{\cal H}(u)=\int_{L}{str(u)}~~,~~\langle u, v
\rangle=\int_{L}{(-1)^{\Delta(u)rk v} str(u\wedge v)}~~, \ee if one
extends the supertrace to bundle-valued forms through: \be
str(\rho\otimes f):=str(f)\rho~~.  \ee

\subsubsection{The string field action and its gauge algebra} 

The string field theory of \cite{sc} is described by the action: \be
\label{action}
S(\phi)=\int_{L}{str\left[\frac{1}{2}\phi\bullet d\phi+\frac{1}{3}
    \phi\bullet\phi\bullet\phi\right]}~~.  \ee This is defined on the
component ${\cal H}^1=\{\phi\in {\cal H}||\phi|=1\}$ of the total
boundary space. It can also be written in the form: \be
S(\phi)=\frac{1}{2}\langle \phi, d\phi\rangle +\frac{1}{3}\langle
\phi,\phi\bullet \phi\rangle~~, \ee which is discussed, for example,
in \cite{com1}.  As explained in more detail below, the physical
interpretation of the defining data is as follows. $d$ plays the role
of a `total worldsheet BRST charge' on a certain collection of
(topological) open string sectors. The product $\bullet$ is a total
open string product for that collection and the bilinear form $\langle
.,.\rangle$ is a `total BPZ form' (total topological metric).

The action (\ref{action}) is invariant with respect to infinitesimal
gauge transformations of the form: \be
\label{gauge}
\phi\rightarrow
\phi+\delta_\alpha\phi=\phi-d\alpha-[\phi,\alpha]_\bullet~~, \ee where
$\delta_\alpha\phi=-d\alpha-[\phi,\alpha]_\bullet$, with $\alpha\in
{\cal H}^0=\oplus_{\tiny
\begin{array}{c}m,n\\m\geq n\end{array}}
\Gamma(Hom(E_m,E_n))\otimes\Omega^{m-n}(L)$ a charge zero element
$(|\alpha|=0$) of ${\cal H}$. In these relations, $[.,.]_\bullet$
denotes the graded commutator in the total boundary algebra, which on
arbitrary homogeneous elements is given by: \be
\label{comdot}
[u,v]_\bullet=u\bullet v -(-1)^{|u||v|}v\bullet u~~.  \ee It is easy
to check that: \be
\delta_\alpha\delta_\beta\phi-\delta_\beta\delta_\alpha\phi=\delta_{
  [\alpha,\beta]_\bullet}\phi~~, \ee so that the Lie algebra of
transformations of the form (\ref{gauge}) closes off-shell.  Note that
$[\alpha,\beta]_\bullet=\alpha\bullet\beta-\beta\bullet \alpha$ since $\alpha$
and $\beta$ have vanishing $U(1)$ charge.  In fact, relation
(\ref{comdot}) shows that our gauge transformations give a
representation of the (infinite-dimensional) Lie algebra ${\bf
  g}=({\cal H}^0, [.,.]_\bullet)$, which is a subalgebra of the graded
Lie algebra $({\cal H}, [.,.]_\bullet)$.  The gauge group ${\cal G}$
results by exponentiation of ${\bf g}$. This infinite-dimensional
group is rather exotic, since its generators are higher rank forms on
the cycle $L$. Note that we insist on considering {\em all} elements
of ${\cal H}^0$ as generators of this group, even though we start with
a particular background flat connection on $E=\oplus_{n}E_n$ which
happens to split as a direct sum $\oplus_{m}A_m$. Even for a direct
sum background, one {\em cannot} restrict to the `diagonal' subalgebra
$\oplus_{m} \Gamma(Hom(E_m,E_m))(L)$ of ${\bf g}$.  Inclusion of
non-diagonal generators is necessary for consistency of the string
field interpretation, since such a decomposition of the connection is
accidental, and one can deform away from direct sum backgrounds by
condensing boundary condition changing states \cite{sc}.

\subsection{The open string interpretation}

The precise interpretation of (\ref{action}) in terms of open A-type
strings arises upon applying the general formalism discussed in
\cite{com1, com3}. For this, one considers the decomposition: \be
\label{decomposition}
{\cal H}=\oplus_{m,n}{{\cal H}_{nm}}~~, \ee where ${\cal
  H}_{nm}=\Omega^*(L)\otimes \Gamma(Hom(E_m, E_n))$ and notices that
the product $\bullet$ and differential $d$ are compatible with it in
the sense that $\bullet$ vanishes on ${\cal H}_{kn'}\times {\cal
  H}_{nm}$ if $n\neq n'$ and takes ${\cal H}_{kn}\times {\cal H}_{nm}$
into ${\cal H}_{km}$, while $d$ takes ${\cal H}_{nm}$ into ${\cal
  H}_{nm}$.  This implies that the collection of spaces ${\cal
  H}_{nm}$ can be viewed as the morphism spaces of a differential
graded category built on the objects $a_n$ (which, due to our
assumption, are in bijection with the grades $n$ present in the
system).  The category interpretation results upon defining $Hom(a_m,
a_n):={\cal H}_{nm}$. One can further check that the bilinear form
$\langle . , .\rangle$ is compatible with the decomposition
(\ref{decomposition}), in the sense that it vanishes on ${\cal
  H}_{m'n'}\times {\cal H}_{nm}$ unless $n'=n$ and $m'=m$.  Then the
general discussion of \cite{com1} suggests that ${\cal H}_{nm}$ should
be interpreted as the (off-shell) state space of open topological
strings stretching from $a_m$ to $a_n$. This interpretation is indeed
valid, and can be recovered from the topological A-model as discussed
in \cite{sc}.

For example, the fact that the natural grading on ${\cal H}_{nm}$ is
given by $|~.~|$ implies that the worldsheet $U(1)$ charge of states
for the string stretching from $a_m$ to $a_n$ is given by: \be
\label{ghost_mn}
|u|=rk u+n-m~~, \ee which agrees with the observation of
\cite{Douglas_Kontsevich} that the $U(1)$ charge is shifted in
boundary condition changing sectors connecting two graded topological
D-branes.  A direct construction of the string field theory
(\ref{action}) can be found in the paper \cite{sc}, which takes the
sigma model perspective as a starting point.

\section{The extended string field theory}

The theory (\ref{action}) can be extended in a manner reminiscent of
that discussed in \cite{Gaberdiel}. This extension was already written
down in \cite{sc}, which gave a very short discussion of its
structure.  Here we give a more complete exposition.

Since we start with an action based on a graded super-bundle, the
various objects involved in the extension procedure are somewhat
subtle and we shall give a careful discussion of their construction.
We warn the reader that a cursory reading of the present section may
lead to serious misunderstanding of our sign conventions.

\subsection{The extended boundary data}

In order to formulate the extended theory, we must define an {\em
  extended boundary algebra} $({\cal H}_e,d, *)$, a differential
graded associative algebra which extends 
$({\cal H},d,\bullet)$.  We
also need an {\em extended topological metric} $\langle .,.
\rangle_e$, a graded-symmetric nondegenerate bilinear form on ${\cal
  H}_e$ which extends the BPZ form $\langle .,.\rangle$. We shall
consider the three elements $*$, $d$ and $\langle . , . \rangle_e$ in
turn.

The extended boundary algebra $({\cal H}_e, *)$ is obtained (as in
\cite{Gaberdiel}) by considering a (complex) Grassmann algebra $G$ and
constructing the graded associative algebra ${\cal
  H}_e=\Omega^*(L){\hat \otimes}{\cal E} {\hat \otimes}G$. The
extended boundary product $*$ will be the standard product on this
algebra, to be discussed below.  The Grassmann algebra $G$
\footnote{All of the constructions of this paper can in fact be
  carried out with an {\em arbitrary} commutative superalgebra $G$
  with unit.} (whose elements we denote by $\alpha, \beta \dots$)
comes endowed with the Grassmann degree $g$ and a multiplication which
we write as juxtaposition. We allow $G$ to have any number of odd
generators, which we denote by $\xi^\mu$. An element of $G$ has the
form: \be
\label{Ggens}
\alpha=\alpha_0+\sum_{k\geq 1}
\sum_{\mu_1<\dots<\mu_k}{\alpha_{\mu_1\dots
    \mu_k}\xi^{\mu_1}...\xi^{\mu_k}}~~, \ee where $\alpha_0$ and
$\alpha_{\mu_1..\mu_k}$ are complex numbers. We note the existence of
an evaluation map $ev_G$ from $G$ to $\C$, which projects out the odd
generators: \be
\label{evG}
ev_G(\alpha)=\alpha_0~~.  \ee

Since we shall tensor with $G$, we will encounter various
$\Z_2$-valued degrees, for which we use the following convention. We
let $\Z_2=\Z/2\Z=\{{\hat 0}, {\hat 1}\}$, where ${\hat 1}$ is the
unit. For an element $t\in \Z_2$, we define the power $(-1)^t$ by
picking any representative for $t$ in the covering space $\Z$; this is
clearly well-defined since $(-1)^2=+1$. If $n$ is an integer and $t$
is an element of $\Z_2$, then $nt\in \Z_2$ is the product of $t$ with
the mod 2 reduction of $n$.

Associativity of ${\hat {\otimes}}$ allows us to view ${\cal H}_e$ as
either of the tensor products ${\cal H}{\hat \otimes }G$ or
$\Omega^*(L){\hat \otimes}{\hat {\cal E}}$, where ${\hat {\cal
    E}}:={\cal E}{\hat {\otimes}} G$. It will be useful to discuss the
multiplicative structure of ${\cal H}_e$ from both of these
perspectives.  For this, we first recall the multiplicative structure
on the factor ${\hat {\cal E}}$ of the second presentation.

\subsubsection{The algebra ${\hat {\cal E}}$ of Grassmann-valued sections of 
  $End({\bf E})$}

The space ${\hat {\cal E}}$ is the graded associative algebra of
Grassmann-valued sections of $End({\bf E})$.  If ${\hat f}=f\otimes
\alpha$ and ${\hat g}=g \otimes \beta$ are two decomposable elements
of ${\hat {\cal E}}$ (with $f,g$ elements of ${\cal E}$ and $\alpha,
\beta$ elements of $G$), then their canonical multiplication is given
by: \be
\label{hatfg}
{\hat f}{\hat g}= (-1)^{g(\alpha)\Delta(g)}(f\circ g)\otimes
(\alpha\beta)~~.  \ee

One can extend the composition of morphisms $\circ$ from ${\cal E}$ to
a naive multiplication on ${\hat {\cal E}}$ given by: \be
\label{hatfg_circ}
{\hat f}\circ {\hat g}=(f\circ g)\otimes (\alpha\beta)~~.  \ee In
terms of this naive product, the defining relation (\ref{hatfg})
becomes: \be {\hat f}{\hat g}=(-1)^{g({\hat f})\Delta({\hat g})}{\hat
  f}\circ {\hat g}~~, \ee where $g({\hat f})=g(\alpha)$ and
$\Delta({\hat g})=\Delta(g)$ are the degrees induced on ${\hat {\cal
    E}}$ by the $\Z_2$-grading of $G$ and the $\Z$-grading of ${\cal
  E}$. Note that $g({\hat f})$ is simply the Grassmannality of ${\hat
  f}$, while $\Delta({\hat g})=n-m$ if ${\hat g}$ is a morphism from
$E_m$ to $E_n$.

The algebra ${\hat {\cal E}}$ is endowed with the total $\Z_2$-valued
grading induced by the sum $\sigma=\Delta~(mod~2)~+g$ between the
mod~2 reduction of $\Delta$ and the Grassmann degree $g$ on $G$: \be
\sigma(f\otimes \alpha)=\Delta(f)~(mod~2)~+g(\alpha)~~.  \ee The
product (\ref{hatfg}) is homogeneous of degree zero with respect to
this grading: \be \sigma({\hat f}{\hat g})=\sigma({\hat
  f})+\sigma({\hat g})~~.  \ee
 
The supertrace $str$ on ${\cal E}$ extends to a functional $str_e$ on
${\hat {\cal E}}$, which associates a {\em Grassmann-valued} function
$str_e({\hat f})$ (defined on $L$) to each Grassmann-valued section of
${\hat f}$ of $End({\bf E})$. On decomposable elements ${\hat
  f}=f\otimes \alpha$, this `extended supertrace' is given by: \be
str_e(f\otimes \alpha):=str(f)\otimes \alpha~~.  \ee

It is easy to check that the extended supertrace has the property: \be
str_e({\hat f}{\hat g})=(-1)^{\sigma({\hat f})\sigma({\hat g})}
str_e({\hat g}{\hat f})~~.  \ee

\subsubsection{The multiplicative structure on ${\cal H}_e$}

A decomposable element ${\hat u}$ of ${\cal H}_e$ can be presented as:
\be \label{dec1} {\hat u}=\rho\otimes f\otimes \alpha=u\otimes
\alpha=\rho\otimes {\hat f}~~, \ee where $\rho$ is a (complex-valued)
form on $L$, $f$ is an endomorphism of ${\bf E}$ and $\alpha$ is an
element of the Grassmann algebra $G$.  In this relation, we defined
$u=\rho\otimes f$ (an element of ${\cal H}=\Omega^*(L)\otimes {\cal
  E}$) and ${\hat f}=f\otimes \alpha$ (an element of ${\hat {\cal E}}=
{\cal E}\otimes G$).  If ${\hat v}=\eta\otimes g\otimes \beta=v\otimes
\beta=\eta\otimes {\hat g}$ is another element of ${\cal H}_e$ (with
$v=\eta \otimes g$ and ${\hat g}=g\otimes \beta$), then the canonical
product $*$ in ${\cal H}_e$ is given by \footnote{It might be useful
  to show that the product $*$ is indeed associative.  Consider degree
  one Grassmann valued forms in ${\cal H}_e$: $\hat a*(\hat b*\hat
  c)=\hat a*(-1)^{(1-rk b )rkc+ g(b)\Delta(c)}\hat b \hat
  c=(-1)^{(1-rk b )rk c+ g(b)\Delta(c)+(1-rk a )(rk b+rk
    c)+g(a)(\Delta(b)+\Delta(c))} \hat a\hat b\hat c$ . On the other
  hand $(\hat a*\hat b)*\hat c=(-1)^{(1-rk a )rk b + g(a)\Delta
    b}(\hat a\hat b)*\hat c= (-1)^{(1-rk a)rk b+ g(a)\Delta(b)+(2-rk
    a-rk b)rk c+(g(a)+g(b))\Delta(c)}\hat a\hat b\hat c$.  Since the
  signs are the same in both cases the product $*$ satisfies
  associativity.  }: \bea
\label{star1}
{\hat u}*{\hat v}&=&(-1)^{g(\alpha)|v|}(u\bullet v)\otimes
(\alpha\beta)
\nonu\\
&=&(-1)^{(g(\alpha)+\Delta(f))rk \eta}(\rho\wedge \eta)\otimes
({\hat f}{\hat g})\nonu\\
&=& (-1)^{g(\alpha)|v|+\Delta(f)rk \eta}(\rho \wedge \eta) \otimes
(f\circ g)\otimes (\alpha\beta)~~.  \eea The first two equations
correspond to viewing ${\cal H}_e$ as ${\cal H}{\hat \otimes }G$ and
$\Omega^*(L){\hat \otimes}{\hat {\cal E}}$ respectively. The last
treats ${\cal H}_e$ as the triple tensor product $\Omega^*(L){\hat
  \otimes} {\cal E}{\hat \otimes }G$.

The extended boundary space is equipped with the total ($\Z_2$-valued)
degree $deg$ induced from its components: \be deg({\hat u})= (rk
\rho+\Delta(f))~(mod~2)~+g(\alpha)=|u|~(mod~2)~+g(\alpha)= rk
\rho~(mod~2)~+\sigma({\hat f})~~, \ee on elements ${\hat u}$ of the
form (\ref{dec1}).  The extended boundary product (\ref{star1}) is
homogeneous of degree zero with respect to this grading: \be deg
({\hat u}*{\hat v})=deg ({\hat u})+deg ({\hat v})~~. \ee

\subsubsection{The trace and bilinear form on ${\cal H}_e$}
The extended boundary space is endowed with a trace $Tr_e$ which on
decomposable elements ${\hat u}=\rho\otimes f\otimes \alpha$ takes the
form: \be Tr_e({\hat u})=\int_{L}{\rho~str(f)~\alpha}=
\int_{L}{\rho~str_e({\hat f})}~~= Tr_{\cal H}(u)\otimes \alpha~~.  \ee
This associates an element of $G$ to every element of ${\cal H}_e$. It
is easy to check that the extended trace is graded-symmetric with
respect to the total degree: \be
\label{tre}
Tr_e({\hat u}*{\hat g})=(-1)^{deg {\hat u}~deg {\hat v}} Tr_e({\hat
  v}*{\hat u})~~.  \ee One can also write: \be Tr_e({\hat
  u})=\int_{L}{str_e({\hat u})}~~, \ee upon extending $str_e$ to
${\cal H}_e$ through: \be str_e(\rho\otimes {\hat f})=\rho~str_e({\hat
  f})~~.  \ee

We next introduce a (nondegenerate) bilinear form on ${\cal H}_e$
through: \be \langle {\hat u}, {\hat v}\rangle_e:=Tr_e({\hat u}*{\hat
  v}) =\int_{L}{str_e({\hat u}*{\hat v})}~~.  \ee Property (\ref{tre})
assures graded-symmetry of this form with respect to the total degree
on the extended boundary space: \be \langle {\hat u}, {\hat
  v}\rangle_e=(-1)^{deg {\hat u}~deg {\hat v}} \langle {\hat v}, {\hat
  u}\rangle_e~~. \ee It is also easy to check invariance of the
extended bilinear form with respect to the extended boundary product:
\be \langle {\hat u}*{\hat v},{\hat w}\rangle_e=\langle {\hat u},
{\hat v}*{\hat w}\rangle_e~~.  \ee We finally note the worldsheet
charge selection rule: \be
\label{dc}
\langle {\hat u},{\hat v}\rangle_e=0~~{\rm~unless~}~|{\hat u}| +|{\hat
  v}|=3~~.  \ee Since the supertrace only couples the component
$u_{mn}$ to $v_{nm}$, one has in fact separate selection rules for the
gradings $\Delta$ and $rk$: \be
\label{selrules}
\langle {\hat u},{\hat v}\rangle_e=0~~{\rm~unless~}~rk{\hat u}
+rk{\hat v}=3~ {\rm~and~}~\Delta({\hat u})+\Delta({\hat v})=0~~, \ee
for bi-homogeneous elements $u$ and $v$.

\subsubsection{Expression of the extended product in terms 
  of the wedge product and `twisted wedge product' of Grassmann-valued
  forms with coefficients in $End({\bf E})$}

It is possible to express the extended boundary data discussed above
in somewhat more familiar language as follows.  Upon regarding ${\cal
  H}_e$ as the tensor product $\Omega^*(L){\hat \otimes}{\hat {\cal
    E}}$, one has the usual wedge product: \be
\label{wedge}
{\hat u}\wedge {\hat v}=(\rho\wedge \eta)\otimes ({\hat f}\circ{\hat
  g})~~, \ee which uses the composition (\ref{hatfg_circ}) of
Grassmann-valued bundle morphisms.  One can also define a `twisted
wedge product' by\footnote{The twisted wedge product $\bweft$ was used
  in \cite{sc}, where it was denoted simply by $\wedge$, since it is
  the natural extension of the wedge product of bundle-valued forms to
  the graded case. In the present paper, we reserve the notation
  $\wedge$ (later simply written as juxtaposition) for the usual wedge
  product built with the ordinary composition of morphisms.}: \be
\label{hatuv_wedge}
{\hat u}\bwe{\hat v}=(\rho\wedge \eta)\otimes ({\hat f}{\hat g})~~.
\ee The idea behind this definition is that, since ${\hat u}$ and
${\hat v}$ are forms with Grassmann-valued coefficients in a {\em
  graded} bundle, it is natural to consider a `wedge product' which
includes multiplication with respect to the natural product
(\ref{hatfg}) on the coefficient algebra ${\hat {\cal E}}$.

It is easy to see that: \be {\hat u}\bwe{\hat v}=(-1)^{g({\hat
    u})\Delta({\hat v})}{\hat u}\wedge {\hat v} \ee

With these definitions, equation (\ref{star1}) gives: \be
\label{star}
{\hat u}*{\hat v}=(-1)^{(g({\hat u})+\Delta({\hat u}))rk {\hat v}}
{\hat u}\bwe {\hat v}=(-1)^{g({\hat u})|{\hat v}|+\Delta({\hat u})rk
  {\hat v}} {\hat u}\wedge {\hat v}~~, \ee where we extended the
grades $\Delta, g$ and $rk$ to ${\cal H}_e$ in the obvious manner: $
\Delta(\rho\otimes f\otimes \alpha):=\Delta(f)$, $g(\rho\otimes
f\otimes \alpha):=g(f)$, $rk(\rho\otimes f\otimes \alpha):=rk\rho$.

As in the previous section, the product (\ref{hatuv_wedge}) allows for
a formulation of the extended algebraic structure in perhaps more
familiar language. Upon viewing ${\cal H}_e$ as $\Omega^*(L)\otimes
{\hat {\cal E}}$, one can locally expand its elements in the form: \be
\label{gvf}
{\hat u}= \sum_{k=0}^3{dx^{\alpha_1}\wedge ...  \wedge
  dx^{\alpha_k}{\bf U}^{(k)}_{\alpha_1..\alpha_k}(x)} \ee where the
coefficients ${\bf U}_{\alpha_1..\alpha_k}(x)$ are Grassmann-valued
sections of $End({\bf E})$, i.e. elements of ${\hat {\cal E}}$.  Then
the twisted wedge product (\ref{hatuv_wedge}) reads: \be
\label{doo}
{\hat u}\bwe {\hat v}=\sum_{k,l=0}^3{ dx^{\alpha_1}\wedge ...  \wedge
  dx^{\alpha_k}\wedge dx^{\beta_1}\wedge ...  \wedge dx^{\beta_l}({\bf
    U}^{(k)}_{\alpha_1..\alpha_k}(x) {\bf
    V}^{(l)}_{\beta_1..\beta_l}(x))}~~, \ee {\em where the product
  ${\bf U}^{(k)}_{\alpha_1..\alpha_k}(x) {\bf
    V}^{(l)}_{\beta_1..\beta_l}(x)$ is defined as in (\ref{hatfg})}.

We finally note that one can also define a naive extension of the
product $\bullet$ of (\ref{dot}) to Grassmann-valued forms with
coefficients in $End({\bf E})$: \be \label{bullet_extended} {\hat
  u}\bullet {\hat v}=(u\bullet v)\otimes ({\hat f}{\hat
  g})=(-1)^{\Delta({\hat u}) rk {\hat v}}{\hat u}\bwe {\hat
  v}=(-1)^{\Delta({\hat u}) rk {\hat v}+g({\hat u})\Delta({\hat
    v})}{\hat u}\wedge {\hat v}~~, \ee where $({\hat f}{\hat g})$ is
again defined as in (\ref{hatfg}).  With this definition, one obtains:
\be {\hat u}*{\hat v}=(-1)^{g({\hat u})rk {\hat v}}{\hat u}\bullet
{\hat v}~~.  \ee

\subsubsection{The extended differential}

The differential on $\Omega^*(L, End(E))$ extends to ${\cal H}_e$ in
the obvious manner: \be
\label{de}
d{\hat u}=du\otimes \alpha~~, \ee on decomposable elements ${\hat u}=u
\otimes \alpha$.  The symbol $d$ in the second equality is the
differential on ${\cal H}$.  The differential (\ref{de}) is a graded
derivation of $*$, with respect to the total degree: \be d({\hat
  u}*{\hat v})=(d{\hat u})*{\hat v}+(-1)^{deg {\hat u}} {\hat
  u}*d{\hat v}~~.  \ee It is also easy easy to check that: \be \langle
d{\hat u},{\hat v}\rangle_e=-(-1)^{deg {\hat u}}\langle {\hat u},
d{\hat v} \rangle_e~~.  \ee

\subsection{The extended action and restricted odd symplectic form}

The data of the previous subsection allows one to write an extended
action on the subspace ${\cal H}_e^1=\{ {\hat \phi}\in {\cal H}_e |
deg {\hat \phi}={\hat 1}\}$ of ${\cal H}_e$: \be
\label{extended_action}
S_e({\hat \phi})= \int_L{ str_e\left[\frac{1}{2}{\hat \phi} *d{\hat
      \phi}+ \frac{1}{3}{\hat \phi}*{\hat \phi}*{\hat \phi}
  \right]}~~.  \ee This action was written down in \cite{sc} by
analogy with the general extension procedure discussed in
\cite{Gaberdiel, Zwiebach_open} \footnote{As the latter translates in
  our conventions and with certain modifications.}. It is one of the
purposes of this paper to show that $S_e$ plays the role of BV action
for the string field theory (\ref{action}).

On ${\cal H}_e^1$, the product (\ref{star}) agrees with the
multiplication $\bullet$ of (\ref{dot}) precisely when the worldsheet
$U(1)$ charge of the first factor is odd, in which case its Grassmann
degree is even (see equation (\ref{star1})).  If we extend the
evaluation map (\ref{evG}) to a map from ${\cal H}_e$ to ${\cal H}$
defined through: \be ev_G(u\otimes \alpha):=ev_G(\alpha)u=\alpha_0u~~,
\ee we find that: \be {\cal H}^1=ev_G(M_0)~~, \ee where $M_0$ is the
subspace of ${\cal H}_e^1$ given by: \be
\label{M0}
M_0=\{{\hat \phi}\in {\cal H}_e^1||{\hat \phi}|=1\}~~.  \ee This
implies that the restriction of $S_e$ to $M_0$ is related to the
unextended action of (\ref{action}) through: \be
\label{restriction} 
ev_G(S_e({\hat \phi}))=S(ev_G({\hat \phi}))~{\rm~for~}
{\hat \phi}\in M_0~~.  \ee

\noindent For later reference, we note that the restriction: 
\be
\label{omega0}
\omega_0:=\langle . , .\rangle_e|_{{\cal H}_e^1\times {\cal H}_e^1}
\ee of the extended bilinear form to the subspace ${\cal H}_e^1$ is an
{\em antisymmetric} nondegenerate bilinear form whose values are
Grassmann-odd numbers. This follows from the properties of $\langle .
, . \rangle_e$ discussed in the previous section.  Let us express
$\omega_0$ in terms of the wedge products defined in (\ref{wedge}) and
(\ref{hatuv_wedge}).  Upon using relation (\ref{star}) and the fact
that $deg {\hat u} =deg {\hat v} ={\hat 1}$, one obtains: \be
\label{omega0_rescaling}
\omega_0({\hat u}, {\hat v})=(-1)^{(1-rk {\hat u})rk {\hat
    v}}\int_{L}{ str_e({\hat u} \bwe {\hat v})}~~, \ee which in view
of the selection rules (\ref{selrules}) also reads: \be
\label{omega0_wedge}
\omega_0({\hat u}, {\hat v})=(-1)^{rk {\hat v}} \int_{L}{str_e({\hat
    u}\bwe {\hat v})}=(-1)^{rk {\hat v}+g({\hat u}) \Delta({\hat v})}
\int_{L}{str_e({\hat u}\wedge {\hat v})}~~.  \ee Moreover, we notice
that $\omega_0({\hat u}, {\hat v})$ vanishes unless $g({\hat
  u})+g({\hat v})$ is odd (this follows from the constraints $deg
{\hat u}=deg {\hat v}=odd$ and the selection rule (\ref{dc})); this
establishes that $\omega_0$ takes Grassmann-odd values. These
observations will be useful in Section 7.

\subsection{Superspace formulation of the extended action}

The extended action (\ref{extended_action}) can be formulated in
superspace language as follows.  Consider the supermanifold ${\cal
  U}:=\Pi TL$ obtained by applying parity reversal on the fibers of
the tangent bundle of $L$. Superfunctions defined on ${\cal L}$ are
(Grassmann-valued) superfields on $L$, with odd superspace coordinates
$\theta^j$ associated with the tangent vectors
$\partial_j=\frac{\partial}{\partial x^j}$ defined by a coordinate
system $\{x^j\}_{j=1..3}$ on $L$.

Grassmann-valued forms with coefficients in $End({\bf E})$ can be put
into correspondence with superfields valued in $End({\bf E})$ upon
identifying ${\hat u}$ of equation (\ref{gvf}) with: \be
\label{superfield}
{\bf U}(x,\theta)=\sum_{k=0}^3{\theta^{\alpha_1}...\theta^{\alpha_k}
  {\bf U}^{(k)}_{\alpha_1...\alpha_k}(x)}~~.  \ee In order to
translate the action (\ref{extended_action}) in superspace language,
one must use the somewhat unusual convention that the components ${\bf
  U}^{(k)}_{\alpha_1...\alpha_k}(x)\in {\hat {\cal E}}$ of superfields
of the form (\ref{superfield}) are multiplied with the product
(\ref{hatfg}) and that the sign obtained when commuting $\theta^j$
with such a component \ is $(-1)^{\sigma({\bf U})}$.  With these
conventions, one can check \cite{sc} that multiplication of
superfields reproduces the product (\ref{star}) for the associated
forms, which allows one to write the extended action as: \be
\label{extended_action_superspace}
S_e({\bf \Phi})= \int_L d^3x\int{d^3\theta str_e\left[\frac{1}{2}{\bf
      \Phi} {\bf D}{\bf \Phi}+ \frac{1}{3}{\bf \Phi}{\bf \Phi}{\bf
      \Phi} \right]}~~, \ee where ${\bf \Phi}$ is the superfield
associated with ${\hat \phi}$ and ${\bf D}=\theta^j{\partial_j}$.  In
this paper, we shall only use the differential form language of
(\ref{extended_action}).

\subsection{The underlying superbundle and the physical role of $\Z$-grading}

It is clear from our construction that the extended boundary product
$*$ (and thus the extended action (\ref{extended_action})) depend only
on the mod two reduction of the relative D-brane grade $\Delta$. To
formalize this, let us consider the reduction $E=E_{even}\oplus
E_{odd}$ of the $\Z$-grading of ${\bf E}$, where: \bea
E_{even}=\oplus_{n=even}{E_n}~~,~~E_{odd}=\oplus_{n=odd}{E_n}~~.  \eea
This allows us to view ${\bf E}$ as a superbundle \cite{Quillen},
while forgetting the finer data associated with the $\Z$-grading.  It
is clear that the boundary product, extended boundary product and
extended action depend only on this superbundle structure. In
particular, one has only two classes of extended actions. The first
corresponds to the case $E_{even}=0$ or $E_{odd}=0$ and can be
recognized as the extended Chern-Simons action coupled to the bundle
${\bf E}=E_{odd}$ or ${\bf E}=E_{even}$.  The second corresponds to
$E_{even}\neq 0$ and $E_{odd}\neq 0$ and can be viewed as an extended
version of the `supergroup Chern-Simons action' \cite{supergroupCS}
coupled to ${\bf E}=E_{odd}\oplus E_{even}$.  This agrees with ideas
proposed in \cite{Vafa_cs}.

The D-brane grade plays the role of specifying the finer $\Z$-grading
given by the worldsheet $U(1)$ charge.  It is this piece of data which
defines the subspace $\{{\hat \phi}\in {\cal H}^1_e||{\hat \phi}|=1\}$
on which the extended action (\ref{extended_action}) reduces to the
unextended functional (\ref{action}). As we shall see in the next
section, the extended theory can be viewed as a classical BV system,
with $S_e$ playing the role of tree-level master action.  From the BV
perspective, the choice of D-brane grading is what specifies both the
BV ghost number and the so-called {\em classical gauge}. In
particular, two theories which have the same underlying superbundle
but distinct choices of $\Z$-grading have the same tree-level BV
actions but correspond to different choices for these two pieces of
data.  Since the ghost grading is physically relevant (in particular,
as we recall at the end of Section 4.1.3, it specifies the algebra of
classical on-shell gauge-invariant observables, given the other
tree-level BV data), different choices of D-brane grading lead to {\em
  different} physical theories, in spite of the fact that the
difference may not be manifest in the BV action itself.  This is the
crucial conceptual distinction between our approach and the proposals
of \cite{Vafa_cs}.

\section{The extended action as a classical master action}

In this section we show that the extended action satisfies the
classical master equation with respect to a BV bracket induced by an
odd symplectic form associated to the extended bilinear form. We also
show that $S_e$ reduces to the unextended action $S$ in a certain
`classical gauge'.  These extremely general results are valid for an
arbitrary collection of graded branes (of distinct grades) wrapping
the cycle $L$, and hold for any topology of the cycle and background
connection.

Our approach uses a certain variant of the geometric BV framework
developed in \cite{Witten_antibracket, Henneaux_geom, Khudaverdian}
and \cite{Kontsevich_Schwarz,Schwarz_geom, Schwarz_semiclassical,
  Schwarz_symms, Schwarz_superanalogues}.  This formalism has the
advantage that it is computationally compact and well-adapted to
topologically-nontrivial situations. In fact, it turns out that the
current version of the geometric BV formalism is incomplete, and we
shall have to extend it in an appropriate manner. The problem is that
the geometric description presented in the references just cited does
not keep track of the BV ghost number. Indeed, the geometric formalism
is usually discussed in terms of a P-manifold, i.e. a supermanifold
endowed with an odd symplectic form. While this correctly considers
the Grassmann parity of various BV fields, it fails to account for the
ghost grading.  This $\Z$-grading on the space of superfunctions plays
a crucial role in the bottom-up (or homological) approach to BV
quantization \cite{FHST, FH, Stasheff_bv, Henneaux_lectures, Gomis}
and in many questions of direct physical significance. For example, it
is a central result of the BV formalism that the BRST cohomology in
{\em ghost} degree zero computes the space of on-shell gauge-invariant
observables of the system. Since the current geometric formulation
does not consider the ghost grading, it does not allow for a
description of this (and other) fundamental results.  In particular,
two {\em distinct} BV systems can have the same supermanifold
interpretation and the same BV action, so they cannot be distinguished
by the current geometric approach. This can lead to confusion when
applied to our models. To avoid such problems, one must refine the
geometric formulation by explicitly including the ghost grading. This
can be done with the help of $\Z$-{\em graded supermanifolds}, which
were recently discussed in \cite{Voronov}. We start with a brief
account of graded supermanifolds and continue by presenting a
$\Z$-graded version of the geometric BV formalism. We then apply it to
our theories in order to obtain a complete description of the
associated BV systems.  Most of this section is formulated in an
entirely general manner, and may be of independent interest for
foundational studies of BV quantization.

\subsection{Covariant description of classical BV systems in the graded 
  supermanifold approach}

\subsubsection{Supermanifold conventions}
 
We remind the reader that there are two major proposals for a rigorous
definition of supermanifolds, the so-called Berezin-Konstant
\cite{Berezin} and DeWitt-Rogers \cite{DeWitt, Rogers} theories.  The
major difference between the two is that the definition of a
DeWitt-Rogers supermanifold requires the choice of an auxiliary
Grassmann algebra $G$, the `algebra of constants'. Berezin's approach
is based on an `intrinsic' sheaf of superalgebras, which leads to a
formulation in terms of ringed spaces (`superschemes'). In this
theory, the manifold has only even points, while the odd coordinates
appear as a form of `algebraic fuzz'. By contrast, the DeWitt-Rogers
theory constructs supermanifolds which possess both even and odd
points, thus leading to a geometrization of the odd directions; this
geometric description of odd coordinates depends on the algebra of
constants $G$.  It is a basic result that the set of $G$-valued points
(defined in a manner similar to that employed in scheme theory) of a
Berezin supermanifold defines a DeWitt-Rogers supermanifold
\footnote{More precisely, it defines a so-called
  $H^\infty$-supermanifold \cite{Rogers}.}  (though not every
DeWitt-Rogers supermanifold can be obtained in this manner
\cite{Rogers}).  Since the extended theory (\ref{extended_action})
incorporates the auxiliary Grassmann algebra $G$, we shall employ the
formalism due to DeWitt and Rogers.  Thus all of our supermanifolds
are understood in the DeWitt-Rogers sense \footnote{The formalism we
  use can in fact be applied to so-called $G^\infty$-supermanifolds,
  which are a generalization of $H^\infty$-supermanifolds
  \cite{Rogers}.}.

Given a (complex) DeWitt-Rogers supermanifold $M$ (modeled over the
algebra of constants $G$), its tangent space at a point $p$ is a
super-bimodule $T_pM$ over $G$ (see Appendix A), whose left and right
module structures are compatible: \be \alpha
X_p=(-1)^{\epsilon_\alpha\epsilon(X_p)}X_p\alpha~~, \ee for $\alpha$
an element of $G$ and $X_p$ an element of $T_pM$.  We make the
convention that $\epsilon$ denotes the $\Z_2$-degree of an element in
the space to which it belongs.  Thus $\epsilon_\alpha$ is the
Grassmann parity of $\alpha$, while $\epsilon(X_p)$ is the parity of
$X_p$ with respect to the $\Z_2$-grading on $T_pM$. The disjoint union
of the tangent spaces gives the tangent bundle $TM$.

Globally defined $G$-valued functions on $M$ form a commutative
$\Z_2$-graded ring ${\cal F}(M,G)$ with respect to pointwise
multiplication, with: \be \epsilon_F={\hat 0}~{\rm~if~}F(p)\in
G_0~{\rm~for~all~}p\in M~~,~~ \epsilon_F={\hat 1}~{\rm~if~}F(p)\in
G_1~{\rm~for~all~}p\in M~~.  \ee This is also a left- and right-
$\Z_2$-graded algebra over the ring of constants $G$.  Left and right
derivations of this algebra give so-called left and right vector
fields on $M$ (this is discussed in more detail in Appendix B).  The
spaces of left/right vector fields are graded in the obvious manner,
with even and odd derivations corresponding to even and odd vector
fields.  It is customary to identify left and right derivations, and
we shall do so in the following (see Appendix B for details of this
construction). This allows us to speak simply about vector fields.
With this convention, a vector field $X$ can act both to the left (as
a left derivation) and to the right (as a right derivation), with the
two actions related by applying the sign rule. We shall indicate the
left and right actions by superscript arrows pointing respectively to
the right and left. For every function $F$ we thus have: \be
\label{done}
F\stackrel{\leftarrow}{X}=
(-1)^{\epsilon_F\epsilon_X}\stackrel{\rightarrow}{X}F:=dF(X)~~, \ee
where $dF$ is by definition the differential of $F$. This is a
complex-linear function defined on the space of vector fields, which
is also $G$-linear in the obvious sense: \be
\label{lr_linear}
dF(X\alpha)=dF(X)\alpha~~,~~dF(\alpha~X)=(-1)^{\epsilon_\alpha\epsilon_F}
\alpha~dF(X)~~.  \ee It induces $G$-linear functionals $d_pF$ on each
of the tangent spaces $T_pM$.

The space of vector fields is endowed with a Lie bracket, which in
terms of the action on functions is given by: \be
F\stackrel{\longleftarrow}{[X,Y]}:=-
F(\stackrel{\leftarrow}{X}\stackrel{\leftarrow}{Y}-
(-1)^{\epsilon_X\epsilon_Y}\stackrel{\leftarrow}{Y}
\stackrel{\leftarrow}{X}) \Longleftrightarrow
\stackrel{\longrightarrow}{[X,Y]}F=
(\stackrel{\rightarrow}{X}\stackrel{\rightarrow}{Y}-
(-1)^{\epsilon_X\epsilon_Y}
\stackrel{\rightarrow}{Y}\stackrel{\rightarrow}{X})F~~.  \ee This
operation satisfies $\epsilon_{[X,Y]}=\epsilon_X+\epsilon_Y$ and is
graded-symmetric and $G$-bilinear: \be
\label{Lie}
\left[X,Y\right]~~=-(-1)^{\epsilon_X\epsilon_Y}\left[Y,X\right]~~,~~
\left[\alpha X,Y\beta\right] =~~~ \alpha \left[X,Y\right]\beta~~.  \ee
It also satisfies the graded Jacobi identity: \be
[[X,Y],Z]+(-1)^{\epsilon_X(\epsilon_Y+\epsilon_Z)}[[Y,Z],X]+
(-1)^{\epsilon_Z(\epsilon_X+\epsilon_Y)}[[Z,X],Y]=0~~.  \ee Endowed
with this commutator, the space of vector fields becomes a
$\Z_2$-graded Lie algebra (Lie superalgebra).

The space of functionals $\eta(X)$ obeying $G$-linearity constraints
of the type (\ref{lr_linear}) forms a $\Z_2$-graded $G$-bimodule in
the obvious manner. This is the space $\Omega^1(M)$ of one-forms on
$M$. One defines higher rank forms with the help of the wedge product:
\be
\label{fwedge}
\rho\wedge \eta=\rho\otimes
\eta-(-1)^{\epsilon_\rho\epsilon_\eta}\eta\otimes \rho~~, \ee which
has the graded symmetry property: \be \rho\wedge
\eta=(-1)^{\epsilon_\rho\epsilon_\eta+1}\eta\wedge \rho~~.  \ee Upon
taking iterated wedge products one obtains forms of arbitrary ranks
and (\ref{fwedge}) extends to such forms in the obvious fashion.  One
also has an exterior differential, obtained by extending (\ref{done}).
In particular, a two-form $\omega(X,Y)$ on $X$ is a $G$-valued
complex-bilinear functional on vector fields which has the properties:
\bea
\label{forms}
\epsilon_{\omega(X,Y)}~~~
&=&~~~~~\epsilon_X+\epsilon_Y+\epsilon_\omega~~\nn\\
\omega(\alpha X,Y\beta)&=&(-1)^{\epsilon_\alpha \epsilon_\omega}\alpha
\omega(X,Y)\beta~~\\
\omega(X,Y)~~&=&(-1)^{\epsilon_X\epsilon_Y+\epsilon_\omega}\omega(Y,X)~~.\nn
\eea The quantity $\epsilon_\omega\in \{{\hat 0},{\hat 1}\}$ defines
its parity: $\omega$ is even if $\epsilon_\omega={\hat 0}$ and odd if
$\epsilon_\omega={\hat 1}$.  A two-form $\omega$ is called {\em
  symplectic} if it is nondegenerate and closed $(d\omega=0$).

Local coordinates give independent Grassmann-valued functions $z^a$
defined on an open subset of $M$, whose parities we denote by
$\epsilon_a:=\epsilon(z_a)$. Given such coordinates, one has
locally-defined vector fields
$\partial^l_a=(-1)^{\epsilon_a}\partial^r_a$ (of parity $\epsilon_a$),
which are uniquely determined by: \be
\stackrel{\rightarrow}{\partial^l_a}z^b=
z^b\stackrel{\leftarrow}{\partial^r_a}=\delta^b_a~~.  \ee Their action
on a function $F$ defines its left and right derivatives: \be
\stackrel{\rightarrow}{\partial^l_a}F= \frac{\partial_lF}{\partial
  z^a}=(-1)^{\epsilon_F\epsilon_a}dF(\partial^l_a)~~,~~
F\stackrel{\leftarrow}{\partial^r_a}=\frac{\partial_rF}{\partial z^a}=
dF(\partial^r_a)~~.  \ee For the coordinate functions one obtains: \be
dz^a(\partial^r_b)=(-1)^{\epsilon_a}dz^a(\partial^l_b)=\delta^a_b~~.
\ee This allows us to write $dF$ in the form: \be
\label{dF}
dF=\frac{\partial_rF}{\partial
  z^a}dz^a=dz^a\frac{\partial_lF}{\partial z^a}
=(-1)^{\epsilon_a(\epsilon_F+1)} \frac{\partial_lF}{\partial
  z^a}dz^a~~.  \ee Given a vector field $X$, one can expand it locally
as: \be X=X^a_l\partial^l_a=\partial^r_aX^a_r~~, \ee which defines its
left and right coefficients $X^a_l$ and $X^a_r$ (=locally defined
Grassmann-valued functions).  Equation (\ref{dF}) then gives: \be
dF(X)=\frac{\partial_rF}{\partial z^a}X^a_r=
(-1)^{\epsilon_X\epsilon_F}X^l_a\frac{\partial_lF}{\partial z^a}~~.
\ee Let us next consider the local expression of an {\em odd}
symplectic form $\omega$. If one defines its coefficients through: \be
\label{omega_components}
\omega_{ab}:=\omega(\partial^r_a, \partial^r_b)=
(-1)^{\epsilon_a+\epsilon_b}\omega(\partial^l_a, \partial^l_b)
=(-1)^{\epsilon_a}\omega(\partial^l_a, \partial^r_b)
=(-1)^{\epsilon_b}\omega(\partial^r_a, \partial^l_b)~~, \ee then it is
easy to check that: \be \omega=-\frac{1}{2}\omega_{ab}dz^b\wedge
dz^a~~ \ee (note the reversed order in the wedge product). Its value
on an arbitrary pair of vector fields then follows from the
bi-linearity property listed in (\ref{forms}): \be
\omega(X,Y)=(-1)^{(\epsilon_X+\epsilon_a)\epsilon_b}\omega_{ab}X^a_rY^b_r~~.
\ee Definition (\ref{omega_components}) and relations (\ref{forms})
imply the properties: \be
\epsilon(\omega_{ab})=\epsilon_a+\epsilon_b+1~~,~~
\omega_{ab}=-(-1)^{\epsilon_a\epsilon_b}\omega_{ba}~~.  \ee

\subsubsection{$\Z$-graded supermanifolds}

The collection ${\cal F}=({\cal F}(U,G))$ of $G$-valued functions
defined on open subsets $U$ of $G$ forms a sheaf of superalgebras with
respect to the $\Z_2$-grading given by $\epsilon$.  A {\em
  $\Z$}-graded supermanifold \cite{Voronov} is a supermanifold endowed
with a $\Z$-grading $s$ on this sheaf.  This $\Z$-grading is required
to be compatible with pointwise multiplication $s(FG)=s(F)+s(G)$ and
with restriction from an open set to its open subsets. We also require
$s(\alpha F)=s(F\alpha)=s(F)$ for $\alpha\in G$.  The $\Z_2$-grading
$\epsilon$ need not be the mod~2 reduction of $s$; in fact, this is
almost never the case if one works with DeWitt-Rogers
supermanifolds\footnote{The reason is that $\epsilon$ must satisfy
  $\epsilon(F\alpha)=\epsilon_F+\epsilon_\alpha$, while $s$ satisfies
  $s(F\alpha)=s(F)$. Hence $\epsilon =s~(mod~2)$ would require
  $\epsilon_\alpha=0$ for all $\alpha\in G$, which is only possible if
  $G$ has no odd generators.}.

A $\Z$-grading on ${\cal F}$ can be specified by giving an atlas
$\{(U, z^a_U)\}$ of local coordinates and picking integer grades
$s_a^U$ for $z^U_a$ such that the change of coordinates from $z^U_a$
to $z^V_a$ (when $U$ intersects $V$) is compatible with these degrees.
For simplicity, let us restricts the elements of ${\cal F}(U,G)$ to be
functions which are polynomial in coordinates\footnote{One can also
  consider formal power series, which gives a formal $\Z$-graded
  supermanifold. If one wishes to extend this beyond formal power
  series, one has to deal with issues of convergence, which we wish to
  avoid.}.  Such a function has the form: \be
\label{function}
F(p)=\sum_{k=1}^N\sum_{a_1..a_k}{\alpha_{a_1...a_k}
  z^{a_1}_U(p)...z^{a_k}_U(p)}\Leftrightarrow
F=\sum_{k=1}^N\sum_{a_1..a_k}{\alpha_{a_1...a_k}
  z^{a_1}_U...z^{a_k}_U}~~, \ee where $\alpha_{a_1..a_k}$ are elements
of $G$ and the sum runs over monomials of degree smaller than some
positive integer $N$.  We extend $s_a$ to a $\Z$-grading on ${\cal
  F}(U,G)$ by declaring that
$s(z^{a_1}...z^{a_k})=s_{a_1}+...+s_{a_k}$. A function
(\ref{function}) is $s$-homogeneous of degree $\sigma$ if all of the
monomials appearing in its expansion satisfy
$s(z^{a_1}...z^{a_k})=\sigma$. It is clear that this grading is
compatible with pointwise multiplication and restriction to open
subsets of $U$, and satisfies $s(\alpha F)=s(F\alpha)=s(F)$.

If $V$ is another coordinate neighborhood in the distinguished atlas
(such that $U$ intersects $V$), then on the intersection $U\cap V$ one
can express $z^a_V$ as: \be z^a_V=z^a_V(z_U)~~, \ee where we assume
that the transition functions are polynomial. The compatibility
condition requires that $s_a$ coincide with the degree of the function
$z^a_V(z_U)$, defined with respect to the coordinates $z^U$. This
assures us that the degree of a function in ${\cal F}(U\cap V,G)$ does
not depend on the coordinates in which it is computed, and thus we
have a well-defined $\Z$-grading on the sheaf ${\cal F}$\footnote{To
  globalize this argument one needs to assume the existence of an
  appropriate partition of unity etc.}.

Notice that the Grassmann coefficients $\alpha$ play no role the
grading $s$, i.e.  one can formally write $s(\alpha_{a_1..a_k})=0$.
As mentioned above, the Grassmann grading $\epsilon_F$ of a function
$F$ need not coincide with the mod~2 reduction of its $\Z$-grading.
For example, if $F=\alpha z^{a_1}...z^{a_k}$, then
$\epsilon_F=\epsilon_\alpha+\epsilon_{a_1}+...+\epsilon_{a_k}$, but
$s_F=s_{a_1}+..+s_{a_k}$, so that $s_F (mod~2)$ may differ from
$\epsilon_F$ even if one chooses $s_a$ such that $s_a
(mod~2)=\epsilon_a$. This mismatch between $\Z_2$-grading and
$\Z$-grading is due to the presence of the Grassmann algebra of
constants $G$, and thus is an inescapable feature of working with
DeWitt-Rogers supermanifolds.  The $\Z$ and $\Z_2$-gradings $s$ and
$\epsilon$ must be viewed as independent pieces of data.

The integer grading on Grassmann-valued functions allows us to
introduce $\Z$-gradings on the spaces of vector fields and
differential forms.  Since vector fields $X$ are complex-linear maps
from ${\cal F}$ to itself, we shall say that $X$ is $s$-homogeneous of
degree $\sigma$ if: \be
\label{sX}
s(F\stackrel{\leftarrow}{X})=s(F)+s_X~~ \ee for some integer $s_X$
which defines the $s$-degree of $X$.  It is clear that
$s([X,Y])=s_X+s_Y$.

We shall say that a local coordinate system is {\em $s$-homogeneous}
if the associated coordinate functions $z^a$ are $s$-homogeneous
elements of ${\cal F}$. In this case, we denote $s(z^a)$ by $s_a$.
Given an $s$-homogeneous coordinate system, it is clear that
$s(F\stackrel{\leftarrow}{\partial^r_a})=
s(\stackrel{\rightarrow}{\partial^l_a}F) =s(F)-s_a$, which implies:
\be s(\partial^r_a)=s(\partial^l_a)=-s_a~~.  \ee This allows us to
introduce a $\Z$-grading on the tangent spaces $T_pM$ by using
$X_p=(\partial^r_a)_pX^r_a(p)$ and the rule $s(X^r_a(p))=0$ (since
$X^a_r(p)$ is a Grassmann constant, i.e.  an element of $G$).  It is
clear that this grading is independent of the choice of
$s$-homogeneous coordinates; it can be defined more invariantly by
considering localization of vector fields.  This $\Z$-grading, as well
as the $\Z$-grading (\ref{sX}) on vector fields, have no direct
relation to the $\Z_2$-grading $\epsilon$.

If $\eta$ is a linear functional on vector fields, then we define its
$s$-degree through: \be s(\eta(X))=s(X)+s_\eta~~.  \ee In
$s$-homogeneous coordinates, the relation
$dz^a(\partial^r_b)=\delta^a_b$ implies: \be s(dz^a)=s(z^a)=s_a~~.
\ee Moreover, we obtain: \be
s(dF)=s_F~~,{\rm~i.e.~~~}s(dF(X))=s_F+s_X~~, \ee and
$s(\frac{\partial_lF}{\partial z^a})=s(\frac{\partial_rF}{\partial
  z^a})= s_F-s_a$.

This grading extends to multilinear forms in the obvious manner.  For
a two-form, we have: \be s(\omega(X,Y))=s(X)+s(Y)+s_\omega~~ \ee
(notice that $\omega(X,Y)$ is an element of ${\cal F}(M,G)$).  In
$s$-homogeneous local coordinates, this gives: \be
s(\omega_{ab})=s_\omega -s_a-s_b~~, \ee where $\omega_{ab}$ is the
function $z\rightarrow \omega_{ab}(z)$.

\subsubsection{Basics of the geometric framework}

We now give a brief outline of a $\Z$-graded version of the geometric
BV formalism. This is a supergeometric version of the symplectic
formalism of Hamiltonian mechanics, endowed with the supplementary
data of an integer-valued grading (the ghost grading).  Its starting
point is a {\em graded P-manifold}, i.e. a graded DeWitt-Rogers
supermanifold $M$ (modeled\footnote{In our case $n$ will be infinite,
  as always in field theory. We shall neglect the well-known problems
  with infinite-dimensional supermanifolds (see, for example,
  \cite{Schmitt}). In fact, we shall later apply this formalism to
  linear supermanifolds only, for which the treatment can be made
  rigorous in terms of Banach supermanifolds.  The condition `modeled
  on $\R^{n,n}$' means that one has an equal number of even and odd
  coordinates; in our application, this can be formulated in terms of
  countable coordinate frames.} on $\R^{n,n}$), endowed with an {\em
  odd} symplectic form $\omega$ which is $s$-homogeneous of degree
$s_\omega=-1$

Given a P-manifold, the odd symplectic form allows one to define a
(right) Hamiltonian vector field $Q_F$ associated with an arbitrary
(Grassmann-valued) function $F$ on $M$: \be dF(X)=\omega(Q_F,X)~~.
\ee This equation is sensible since both the left and right hand sides
are linear with respect to the right $G$-module structure on vector
fields; non-degeneracy of $\omega$ assures the existence of a unique
solution.

It is clear from this definition that: \be Q_{\alpha
  F}=(-1)^{\epsilon_\alpha}\alpha Q_F~~,~~Q_{F\alpha}=Q_F\alpha~~, \ee
for any Grassmann constant $\alpha$.

Since the symplectic form satisfies $\epsilon_\omega={\hat 1}$ and
$s_\omega=-1$, the vector field $Q_F$ has parity
$\epsilon_{Q_F}=\epsilon_F+{\hat 1}$ and ghost number $s_{Q_F}=s_F+1$.
In local coordinates $z^a$, one has the expansions
$Q_F=Q^a_{F,l}\partial^l_a=\partial^r_aQ^a_{F,r}$, with the
components: \be
Q^a_{F,l}=(-1)^{\epsilon_F+\epsilon_a+1}\frac{\partial_lF}{\partial
  z^b} \omega^{ba}~~,~~
Q^a_{F,r}=(-1)^{\epsilon_a+1}\omega^{ab}\frac{\partial_rF}{\partial
  z^b}~~, \ee where $\omega^{ab}$ is the inverse of the matrix
$\omega_{ab}$: \be
\omega^{ab}\omega_{bc}=\omega_{cb}\omega^{ba}=\delta^a_c~~.  \ee Note
the properties: \be \epsilon(\omega_{ab})=\epsilon_a+\epsilon_b+1~~,
~~s(\omega^{ab})=s_a+s_b+1~~,~~
\omega^{ab}=-(-1)^{(\epsilon_a+1)(\epsilon_b+1)}\omega_{ba}~~.  \ee

Given two functions $F,G$ on $M$\footnote{We sometimes use the symbol
  $G$ to denote a Grassmann-valued function on $M$.  This should not
  be confused with the underlying Grassmann algebra, which is denoted
  by the same letter.}, we define their {\em odd Poisson bracket}
(antibracket) through: \be
\label{antibracket}
\{F,G\}=-\omega(Q_F, Q_G)=-dF(Q_G)=(-1)^{(\epsilon_F+1)(\epsilon_G+1)}
dG(Q_F)= \frac{\partial_rF}{\partial z^a}\omega^{ab}
\frac{\partial_lG}{\partial z^b}~~.  \ee One has
$\epsilon_{\{F,G\}}=\epsilon_F+\epsilon_G+1$ and
$s(\{F,G\})=s_F+s_G+1$.  It is easy to check the properties: \bea
\{FG,H\}=F\{G,H\}+(-1)^{\epsilon_F\epsilon_G}G\{F,H\}~~&,&~~
\{F,GH\}=\{F,G\}H+(-1)^{\epsilon_G\epsilon_H}\{F,H\}G~~\nn\\
\{\alpha F,G\beta\}=\alpha\{F,G\}\beta~~~&,&~~
\{F,G\}=-(-1)^{(\epsilon_F+1)(\epsilon_G+1)}\{G,F\}~~.  \eea as well
as the odd graded Jacobi identity: \be
\{\{F,G\},H\}+(-1)^{(\epsilon_F+1)(\epsilon_G+\epsilon_H)}\{\{G,H\},F\}+
(-1)^{(\epsilon_H+1)(\epsilon_F+\epsilon_G)}\{\{H,F\},G\}~~.  \ee

In particular, the space of $G$-valued functions on $M$ forms an odd
Lie superalgebra \footnote{An odd Lie superalgebra \cite{Manin} is
  simply the parity change of a Lie superalgebra.  This is obtained by
  reversing the parity of all elements, while leaving the Lie bracket
  unchanged. Together with pointwise multiplication of functions, the
  antibracket endows the space ${\cal F}(M,G)$ with the structure of a
  so-called {\em odd Poisson algebra} or {\em Gerstenhaber algebra}.}
with respect to the antibracket. Equation (\ref{antibracket}) shows
that: \be
\label{act}
F\stackrel{\leftarrow}{Q_G}=-\{F,G\}\Longleftrightarrow
~\stackrel{\rightarrow}{Q_G}F=-(-1)^{\epsilon_F(\epsilon_G+1)}\{F,G\}~~.
\ee Together with the Jacobi identity, this implies: \be
\label{morphism}
Q_{\{F,G\}}=[Q_F,Q_G]~~.  \ee Thus the map $F\rightarrow Q_F$ acts as
an `odd morphism' (the composition of a morphism of $\Z_2$-graded Lie
algebras with parity change).  This translates the odd Lie
superalgebra language appropriate for functions into the $\Z_2$-graded
Lie algebra language relevant for vector fields.

In the context of BV quantization, the antibracket is interpreted as
the BV bracket.  For any function $F$, one has\footnote{Note that
  (\ref{antibracket}) implies $\omega(Q_F,Q_F)=-dF(Q_F)=0$ if $F$ is
  odd.}: \be \{F,F\}=-\omega(Q_F, Q_F)=-dF(Q_F)~~,~~
Q_{\{F,F\}}=[Q_F,Q_F]~~.  \ee Hence given an action (even function of
ghost degree zero) $S_{BV}$ on our supermanifold, the classical master
equation can be written in the equivalent forms: \be
\label{master}
\{S_{BV}, S_{BV}\}=0\Leftrightarrow QS_{BV}=0\Leftrightarrow
\omega(Q,Q)=0\Leftrightarrow [Q,Q]=0\Leftrightarrow
Q^2(F)=0~{~\rm~for~all~}F~~, \ee where we defined $Q:=Q_{S_{BV}}$. It
is clear that $\epsilon(Q_F)={\hat 1}$ and $s(Q_F)=+1$.  In
particular, a classical BV system defines a so-called {\em
  QP-manifold} \cite{Kontsevich_Schwarz}, i.e. a P-manifold endowed
with an odd nilpotent vector $Q$ field which preserves the odd
symplectic form.

We remind the reader that an odd vector field $Q$ on a supermanifold
is called {\em nilpotent} (or {\em homological}) if it satisfies
$[Q,Q]=0\Leftrightarrow Q^2F=0{\rm~for~all~}F$. Given such a vector
field, the space ${\cal F}(M,G)$ of globally defined Grassmann-valued
functions (viewed as a complex vector space) becomes a complex with
respect to the differential $Q$. If the underlying supermanifold is
$\Z$-graded, and if $Q$ has $\Z$-degree equal to $+1$, then $({\cal
  F}(M,G),Q)$ is a $\Z$-graded cochain complex. The main result of the
bottom-up approach to BV quantization is that, for $Q=Q_{BV}$, the
cohomology of this complex in integer degree zero computes the space
of gauge-invariant functionals on the shell of the associated
classical action (the relation between the BV action and the classical
action is described in geometric terms in the next section). Since the
space of such observables has direct physical meaning, it is clear
that two BV systems which have different ghost gradings (but the same
underlying manifold and odd symplectic form) must be considered as
distinct. Otherwise, there would be no clear way of recovering the
classical data from the geometric formalism -- in particular, one
would reach the paradox that two classical systems with very different
algebras of on-shell gauge-invariant observables are equivalent,
provided that they differ `only' by the choice of ghost grading, a
statement which is clearly incorrect.  This observation justifies the
need for a $\Z$-graded geometric formalism, and is crucial for a
correct understanding of graded D-brane systems. We believe that a
correct geometric description of BV systems must systematically
consider the ghost grading.  Below, we limit ourselves to
re-formulating some basic results of the geometric framework (which
will be needed in our application) in the graded manifold language.

\subsubsection{Gauges and BRST transformations}

In the geometric formalism, a {\em gauge} corresponds to the choice of
a Lagrangian sub-supermanifold of $M$, i.e. a sub-supermanifold ${\cal
  L}$ whose total dimension is half of the total dimension of $M$
\footnote{If a supermanifold is modeled on $\R^{p|q}$, then its total
  dimension is $p+q$.} and with the property that $\omega$ restricts
to zero on ${\cal L}$.  To make contact with the bottom-up approach,
one must also assume that ${\cal L}$ is $s$-{\em homogeneous}, i.e.
its tangent bundle $T{\cal L}$ decomposes as a direct sum of
$s$-homogeneous subbundles of $TM|_{\cal L}$: \be T{\cal
  L}=\oplus_{s\geq 0}{TM|_{\cal L}(s)}~~, \ee where $TM|_{\cal L}(s)$
is the subbundle of $TM|_{\cal L}$ consisting of elements of ghost
degree equal to $s$ \footnote{It is easy to see that $TM=T_{+}M\oplus
  T_{-}M$ (where $T_{\pm}M=\oplus_{\pm s\geq 0}{T(s)}$) gives a
  Lagrangian decomposition of $TM$; this follows from the condition
  $s_\omega=-1$. An $s$-homogeneous Lagrangian submanifold is an
  integral submanifold for the Frobenius distribution $T_{+}M$.  This
  distribution is clearly integrable, since $s([X,Y])=s(X)+s(Y)\geq 0$
  for all vector fields $X,Y$ satisfying $s(X), s(Y)\geq 0$.}.
 
The path integral in this gauge is given by integrating
$e^{-\frac{i}{\hbar }S_{BV}}|_{\cal L}=e^{-\frac{i}{\hbar }S_{\cal
    L}}$ along ${\cal L}$, where $S_{\cal L}:=S_{BV}|_{\cal L}$ is the
restriction of $S_{BV}$ to ${\cal L}$. This is a global version of the
usual description of gauges in terms of fields and antifields and
gauge-fixing fermions. We shall follow common practice and omit the
word `super' when talking about submanifolds of a supermanifold.  We
remind the reader that an odd vector field on $M$ can be viewed as an
odd section of $TM$ or an even section of the parity changed bundle
$\Pi TM$. It is sometimes convenient to work with even sections only,
in which case the parity of a vector field is made clear by the
presence or absence of parity change on the underlying bundle. We
shall sometimes use this convention in what follows. Therefore, a
section of a bundle will always mean an even section unless explicitly
stated otherwise.

Let us recall from \cite{Kontsevich_Schwarz} how the BRST
transformations of the gauge-fixed action are realized in the
geometric formalism.  Choosing a gauge ${\cal L}$, one constructs a
symmetry of $S_{\cal L}$ as follows.  Upon restricting $Q$ to ${\cal
  L}$, one obtains a section of the bundle $\Pi TM|_{\cal L}$. In
order to produce an (odd) vector field on ${\cal L}$, one considers
the decomposition $TM|_{\cal L}=T{\cal L}\oplus N{\cal L}$, where: \be
N{\cal L}=\oplus_{s<0}{TM|_{\cal L}(s)}~~.  \ee It is clear from the
condition $s_\omega=-1$ that this is a Lagrangian splitting of the
restricted tangent bundle $TM|_{\cal L}$, i.e. $\omega$ vanishes on
$T{\cal L}\times T{\cal L}$ and $N{\cal L}\times N{\cal L}$ and is
non-degenerate on $T{\cal L}\times N{\cal L}$ and on $N{\cal L}\times
T{\cal L}$.  As explained in more detail below, this decomposition of
$TM|_{\cal L}$ is related to the field-antifield split of the local
formalism.

One has a similar decomposition $\Pi TM|_{\cal L}=\Pi T{\cal L}\oplus
\Pi N{\cal L}$ of the parity changed bundle.  If $T, R$ are the
associated projectors of $\Pi TM|_{\cal L}$ onto $\Pi T{\cal L}$ and
$\Pi N{\cal L}$, then the relations: \be
\label{qgen}
q:=TQ|_{\cal L}~~,~~q^*:=RQ|_{\cal L} \ee define sections of $\Pi
T{\cal L}$ and $\Pi N {\cal L}$ which give a decomposition of $Q$ on ${\cal
  L}$: \be
\label{Qdec}
Q|_{\cal L}=q+q^*~~.  \ee It is clear that the operators $T$ and $R$
are $s$-homogeneous of degree zero, so that the vector fields $q$ and
$q^*$ are $s$-homogeneous of degree $+1$.  More generally, vectors
$u\in TM|_{\cal L}$ decompose as $u={\bf u}+{\bf u}^*$, with ${\bf
  u}\in T{\cal L}$ and ${\bf u}^*\in N{\cal L}$.  Upon using this in
the defining relation for $Q$, one obtains: \be dS_{BV}(u)=\omega(u,
Q)=\omega({\bf u}, q^*)+ \omega({\bf u}^*, q)~~.  \ee which combines
with $dS_{BV}(u)=dS_{BV}({\bf u})+dS_{BV}({\bf u}^*)= dS_{\cal L}({\bf
  u})+dS_{BV}({\bf u}^*)$ to give: \be
\label{first}
dS_{\cal L}({\bf u})=\omega({\bf u}, q^*)~~,~~ dS_{BV}({\bf
  u}^*)=\omega({\bf u}^*, q)~~.  \ee The second equation shows that
the first order term in the Taylor expansion of $S_{BV}$ in antifields
is proportional with $q$. The first implies that the value of $q^*$ at
a point $p$ in ${\cal L}$ vanishes precisely when $p$ is critical for
$S_{\cal L}$.  Hence the critical set of $S_{\cal L}$ is the locus
where $Q$ is tangent to ${\cal L}$.  Combining this with equation
(\ref{Qdec}) and using the nilpotence of $Q$ shows that $q$ squares to
zero `on the shell of $S_{\cal L}$': \be [q,
q]=0~~{\rm~on~}~Crit(S_{\cal L})~~.  \ee

We finally note from (\ref{qgen}) and (\ref{master}) that $q$
generates a symmetry of the gauge fixed action: \be
\stackrel{\rightarrow}{q}S_{\cal L}=0~~.  \ee It is clear that $q$ is
the BRST generator in the gauge $\cal L$.

\subsubsection{The coordinate description}

We now sketch how the local description arises in the geometric
formalism. Given a gauge ${\cal L}$, one can locally identify $M$ and
the total space of the bundle $N{\cal L}$.  One chooses
$s$-homogeneous coordinates $z^\alpha$ and $z^*_\alpha$ along ${\cal
  L}$ and the fiber of $N{\cal L}$ such that $(z^*_\alpha, z^\alpha)$
are Darboux coordinates for $\omega$, i.e. : \be
\epsilon(z^*_\alpha)+\epsilon(z^\alpha)={\hat 1}~~{\rm~and~}~~
\omega_{\alpha^*\beta}=-\omega_{\beta\alpha^*}=\delta_{\alpha,\beta}~~.
\ee and such that $s(z^*_\alpha)+s(z^\alpha)=-1$\footnote{Note that
  one need not assume $\epsilon(z^\alpha)=s(z^\alpha) (mod~2)$ and
  $\epsilon(z^*_\alpha)=s(z^*_\alpha) (mod~2)$. In fact, this is
  impossible to arrange if the classical action $S$ contains
  Grassmann-odd variables.}.

In this case, the odd symplectic form reduces to: \be
\label{Darboux}
\omega=dz^*_\alpha\wedge dz^\alpha~~.  \ee One can identify $z^\alpha$
and $z^*_\alpha$ with the fields and antifields of the traditional
formalism.  In such coordinates, the BV bracket has the familiar form:
\be
\label{Darboux_bracket}
\{F,G\}= \frac{\partial_r F}{\partial z^\alpha} \frac{\partial_l
  G}{\partial z^*_\alpha}- \frac{\partial_r F}{\partial z^*_\alpha}
\frac{\partial_l G}{\partial z^\alpha} ~~, \ee and the Lagrangian
manifold ${\cal L}$ is locally described by the equations
$z^*_\alpha=0$.

\subsubsection{The geometric meaning of BV `quantization'}

The procedure of BV `quantization' translates as follows.  One starts
with a so-called `classical gauge' ${\cal L}$, and with a {\em
  classical action} $S_{\cal L}=S_{BV}|_{\cal L}$ defined on ${\cal
  L}$.  This gauge is typically not convenient for the purpose of
quantization, in that $S_{\cal L}$ is degenerate (has degenerate
Hessian) on ${\cal L}$, which in field-theoretic applications leads to
`infinite factors' in the path integral and the impossibility of
defining propagators. A degenerate Hessian signals the existence of
flat directions for $S$, which in conveniently chosen local
coordinates $z=(z^a)$ means that $S$ depends only on a subset
$x=(z^j)$ of $(z^a)$, which in practice is the subset of coordinates
with ghost degree $s(z^j)=0$.  To avoid such problems, one picks
another gauge ${\cal L}'$ such that the Hessian of $S_{\cal L'}$ is
nondegenerate.  This allows for the definition of propagators in the
new gauge, which provides a starting point for perturbative
renormalization of the associated path integral.

The BV procedure can be seen as a systematic approach to performing
the change of gauge from ${\cal L}$ to ${\cal L}'$. This is done by
first extending the classical action $S_{\cal L}$ to the BV action
$S_{BV}$, and then restricting the latter to ${\cal L}'$ to obtain a
candidate for a meaningful definition of the path integral (this
process can be described locally in the traditional language of
gauge-fixing fermions).  In its most general form \cite{Schwarz_geom},
the central result is that two gauges ${\cal L}_1$ and ${\cal L}_2$
are physically equivalent provided that their bodies (even parts) are
homologous in the body of $M$ and that the BV action satisfies the
quantum generalization of the classical master equation. It is
important to realize, however, that this statement need not be valid
in more general situations.  There is no reason to expect that
`topologically inequivalent' gauges ${\cal L}_1$ and ${\cal L}_2$ lead
to equivalent path integrals.

\subsubsection{The BV algorithm}

The full BV data is rarely known in practical applications. In a
typical situation, one is only given the action $S:=S_{\cal L}$ in a
classical gauge. Since the classical gauge is degenerate, one has no
apriori knowledge of any of the data $M$, $\omega$ or $S_{BV}$. In
this case, one recovers a BV system $(M,\omega, S_{BV}, {\cal L})$
(such that $S_{BV}|_{\cal L}=S$) in the following constructive manner.
First, one has to decide on a gauge algebra, i.e. an algebra of
symmetries of $S$. It should be stressed that the choice of gauge
algebra is not uniquely determined by $S$, since one can insist to
choose a strict subalgebra of the maximal algebra of gauge symmetries
of the classical action\footnote{We shall encounter this phenomenon in
  Section 7, when discussing the BV action for D-brane pairs with
  relative grading greater than one.}; this choice is dictated by the
physical interpretation of the model.  Then one constructs a BV system
through the following two-step procedure:

\ 

(1) Perform the BRST extension of $S(x)$, by applying the BRST
procedure for the given gauge algebra (the precise choice of algebra
influences the result of this step).  This enlarges the classical
system from the set of classical fields $x$ to the set of BV fields
$z^\alpha=(x, c_1, c_2,..)$, where $c_k$ are ghosts at generation $k$.
It also produces a nilpotent odd vector field $q$ (the BRST generator)
acting on the enlarged collection of fields.  The number of ghost
generations is dictated by the degree of reducibility of the gauge
algebra. After extracting the complete set of ghosts $c_k$, introduce
antifields $z^*_\alpha=(x^*, c_1^*,c_2^*,..)$ for the classical fields
and ghosts.  The BV fields and antifields $(z^\alpha, z^*_\alpha)$ are
identified with Darboux coordinates of $M$.  This allows one to
locally recover both the supermanifold $M$ and the odd symplectic form
$\omega$ upon using relation (\ref{Darboux}). The Lagrangian
submanifold ${\cal L}$ which defines the classical gauge is recovered
through the equation $z^*_\alpha=0$. The fibers of the Lagrangian
complement $N$ of $T{\cal L}$ are locally defined by the directions
$z^\alpha=ct$.

Knowledge of the correct collection of BV fields and antifields allows
one to enlarge the degenerate classical action $S(x)$ by adding the
so-called first order action $S_1(z^*,z)=z^*_\alpha q^\alpha(z)$.  $S$
and $S_1$ are the first two terms in the expansion of the BV action in
antifields.

\ 

(2) The odd symplectic form $\omega$ recovered in the first step
defines the BV bracket $\{.,.\}$. To recover the full BV action, one
must solve the master equation $\{S_{BV}, S_{BV}\}=0$ with the ansatz
$S_{BV}=\sum_{k\geq 0}{S_k}$, where $S_k$ is the $k^{th}$ order term
in the Taylor expansion in antifields. The first two terms $S_0=S$ and
$S_1$ are known from Step (1).

\subsection{Realization of the geometric data and check of the master equation}

\subsubsection{The graded P-manifold}

\paragraph{The supermanifold}

Remember that the unextended total boundary space ${\cal H}$ is a
$\Z$-graded vector space with respect to the worldsheet $U(1)$ charge
$|~.~|$. The mod $2$ reduction of $|~.~|$ makes ${\cal H}$ into a
vector superspace ($\Z_2$-graded vector space): \be {\cal H}={\cal
  H}_{even}\oplus {\cal H}_{odd}~~, \ee where: \bea {\cal
  H}_{even}&=&\{u\in {\cal H}||u|=even\}= \oplus_{k+n-m=even~}
{\Omega^k(L)\otimes \Gamma(Hom(E_m,E_n))}~~,\nn\\
{\cal H}_{odd}~&=&\{u\in {\cal H}||u|=odd~~\}=\oplus_{k+n-m=odd~~}
{\Omega^k(L)\otimes \Gamma(Hom(E_m,E_n))}~~.  \eea

To construct the supermanifold relevant for the geometric BV
formalism, we consider a new $\Z$-grading $s$ on ${\cal H}$ which is
related to $|~.~|$ by: \be s(u)=1-|u|~~.  \ee The vector space ${\cal
  H}$ endowed with this grading will be denoted by ${\tilde {\cal
    H}}$. The mod~2 reduction of $s$ makes ${\tilde {\cal H}}$ into a
vector superspace: \be {\tilde {\cal H}}={\tilde {\cal
    H}}_{even}\oplus {\tilde {\cal H}}_{odd}~~, \ee where ${\tilde
  {\cal H}}_{even}={\cal H}_{odd}$ and ${\tilde {\cal H}}_{odd~}
={\cal H}_{even}$.

It is clear that the superspaces ${\cal H}$ and ${\tilde {\cal H}}$
differ by parity change.  They define infinite-dimensional complex
linear supermanifolds (in the sense of Berezin), which we denote by
$L({\tilde {\cal H}})$ and $L({\cal H})$. Their $G$-valued points
define DeWitt-Rogers supermanifolds: \be M:=L({\tilde {\cal
    H}})(G)=({\tilde {\cal H}}\otimes G)^0= {\tilde {\cal
    H}}_e^0={\cal H}_e^1~~,~~ P=L({\cal H})(G)=({\cal H}\otimes
G)^0={\cal H}_e^0~~, \ee where we defined ${\tilde {\cal
    H}}_e:={\tilde {\cal H}}\otimes G$.  In these equations, ${\cal
  H}_e$ and ${\tilde {\cal H}}_e$ are viewed as vector superspaces,
and we used the obvious relation: \be {\tilde {\cal H}}_e=\Pi {\cal
  H}_e~~.  \ee

\paragraph{The graded manifold structure}

We will mainly be interested in the linear supermanifold $M$, on which
we now introduce a structure of $\Z$-graded manifold.  For this,
consider a homogeneous basis $e_a$ of ${\cal H}$, with
$|e_a|:=|a|\Rightarrow s(e_a)=s_a:=1-|a|$.  Then every element ${\hat
  \phi}$ of ${\cal H}^1_e$ has the expansion: \be {\hat
  \phi}=\sum_{a}{e_a\otimes {\hat \phi}^a}~~, \ee with ${\hat
  \phi}^a\in G$ and $g({\hat \phi}^a)=deg {\hat \phi}-|e_a| (mod~2)=
1-|e_a| (mod~2)=s_a (mod~2)$.  This allows us to define maps
$z^a:{\cal H}_e\rightarrow G$ through $z^a({\hat \phi})={\hat
  \phi}^a$, which have parities $\epsilon_a=g_a=s_a (mod~2)$ as
$G$-valued functions.  They give (global) coordinates on the
supermanifold $M$. Coordinates for $M$ obtained in this manner will be
called {\em homogeneous linear coordinates}.

The collection of homogeneous linear coordinates associated to all
homogeneous bases $(e_a)$ of ${\cal H}$ forms a distinguished atlas
for our supermanifold. Such coordinates are endowed not only with a
$\Z_2$-degree $\epsilon_a=\epsilon(z^a)$, but also with a $\Z$-valued
degree $s(z^a):=s_a=s(e_a)$, such that $\epsilon_a=s_a (mod~2)$. As
explained in subsection 4.2., this can be used to define a
$\Z$-grading $s$ on the sheaf ${\cal F}$ of $G$-valued functions, if
one restricts the latter to consist of functions which are polynomial
in coordinates. This grading on ${\cal F}$ will play the role of ghost
grading in the BV formalism.

\paragraph{Vector fields as nonlinear operators}
Since $M$ is a linear supermanifold, vector fields on $M$ can be
viewed as maps of the form $X:{\hat \phi}\rightarrow ({\hat \phi},
{\bf X}({\hat \phi}))\in M\times {\tilde {\cal H}}_e$, where ${\bf X}$
is a (generally nonlinear) operator from $M$ to ${\tilde {\cal H}}_e$.
Even ($\epsilon_X={\hat 0}$) and odd ($\epsilon_X={\hat 1}$) vector
fields correspond to operators ${\bf X}$ from $M$ to ${\cal H}_e^1$
and $M$ to ${\cal H}_e^0$ respectively.

In homogeneous linear coordinates $z^a$, the vector fields
$\partial^r_a$ correspond to the constant operators ${\hat
  \phi}\rightarrow e_a\otimes 1_G$. One has $s(\partial^r_a)=-s_a$ and
$\epsilon(\partial^r_a)=\epsilon_a=s_a (mod~2)$.  For an arbitrary
vector field $X=\partial^r_aX^a_r$, one has ${\bf X}({\hat
  \phi})=e_a\otimes X^a_r({\hat \phi})$ (with $X^a_r({\hat \phi})\in
G$). Since $z^a({\bf X}({\hat \phi}))=X^a_r({\hat \phi})$, the vector
field $X$ can be recovered from the operator ${\bf X}$ through the
relation: \be (z^a\stackrel{\leftarrow}{X})({\hat \phi})=X^a({\hat
  \phi})= z^a({\bf X}({\hat \phi}))~~.  \ee Note that one can expand:
\be {\bf X}({\hat \phi})={\bf X}_0+{\bf X}_1({\hat \phi})+ {\bf
  X}_2({\hat \phi}, {\hat \phi})+{\bf X}_3({\hat \phi}, {\hat \phi},
{\hat \phi})+...~~, \ee where ${\bf X}_k: ({\tilde {\cal H}}\otimes
G)^{\otimes k}\rightarrow {\tilde {\cal H}}\otimes G$ are
$G$-multilinear operators. It is not hard to check that the vector
field $X$ is $s$-homogeneous of degree $\sigma$ if and only if ${\bf
  X}_k$ are $s$-homogeneous of degree $-\sigma$, i.e.: \be s({\bf
  X}_k({\hat \phi}_1, ..., {\hat \phi}_k))= s({\hat
  \phi}_1)+...+s({\hat \phi}_k)-\sigma~~.  \ee

By localizing the description of vector fields to a point ${\hat
  \phi}$ of $M$, one obtains the identification: \be T_{\hat
  \phi}M={\tilde {\cal H}}\otimes G={\tilde {\cal H}}_e~~, \ee which
holds both as an isomorphism of right G-supermodules and as an
isomorphism of $\Z$-graded vector spaces (the $\Z$-grading on both
sides being given by the ghost degree $s$).  It follows that the total
space of the tangent bundle to $M$ can be identified with: \be
\label{tb}
TM=M\times {\tilde {\cal H}}_e~~.  \ee The left G-supermodule
structure on $T_{\hat \phi}M$ is defined through\footnote{Note that
  the exponent in (\ref{lrmod}) involves $\epsilon(X_{\hat \phi})= deg
  X_{\hat \phi}+{\hat 1}$ and {\em not} $deg X_{\hat \phi}$.  This is
  due to the presence of parity reversal in the isomorphism of
  superspaces $T_{\hat \phi}M={\tilde {\cal H}}_e=\Pi{\cal H}_e$.}:
\be
\label{lrmod}
\alpha X_{\hat \phi}= (-1)^{\epsilon_\alpha\epsilon(X_{\hat
    \phi})}X_{\hat \phi}\alpha~~, \ee where $\alpha\in G$ and
$\epsilon_\alpha=g(\alpha)$.

\paragraph{The odd symplectic form}

The extended bilinear form $\langle . , . \rangle_e$ on ${\cal H}_e$
allows us to define the following two-form on $M$: \be
\label{omega}
\omega_{\hat \phi}(X_{\hat \phi}, Y_{\hat \phi}):=
(-1)^{\epsilon(X_{\hat \phi})}\langle X_{\hat \phi}, Y_{\hat
  \phi}\rangle_e~~, \ee where ${\hat \phi}$ is a point in $M$ and
$X_{\hat \phi}, Y_{\hat \phi}\in T_{\hat \phi}M={\tilde {\cal
    H}}\otimes G$ are tangent vectors to $M$ at ${\hat \phi}$.  It is
easy to check that $\omega$ is an {\em odd symplectic form} on $M$,
and that $s_\omega=-1$.  The last statement follows upon choosing a
homogeneous basis $s_a$ of ${\cal H}$ and considering the coefficients
of $\omega$ in this basis: \be \omega_{ab}=(-1)^{\epsilon_a}\langle
e_a, e_b\rangle=ct~~, \ee where we used the identification
$\partial^r_a=e_a\otimes 1_G$ and the fact that $\langle e_a\otimes
1_G, e_b\otimes 1_G\rangle_e= \langle e_a, e_b\rangle$. The selection
rules (\ref{selrules}) allow us to restrict to the case
$|e_a|+|e_b|=3\Leftrightarrow s_a+s_b=-1$. In this case, one has
$s(\omega_{ab})=-s_a-s_b-1$, where we used the fact that $\omega_{ab}$
is a constant and thus $s(\omega_{ab})=0$.  This implies that
$s_\omega=-1$.  We conclude that $(M, \omega)$ is a (DeWitt-Rogers)
graded P-manifold.  Note that the restriction of $\omega$ to the even
component $TM^0\approx {\cal H}_e^1$ of the tangent bundle coincides
with the form $\omega_0=\langle ., .\rangle_e|_{M\times M}$ of
equation (\ref{omega0}).  In fact, $\omega$ is completely determined
by this restriction and by the requirement that it must be an odd
form. This observation will be useful in Section 7.

The extended action (\ref{extended_action}) is an (even) function
defined on $M$. To check that its ghost degree equals zero, we first
notice that the operator $d$ and extended boundary product $*$ satisfy
$s(d{\hat \phi})=s({\hat \phi})-1$ and $s({\hat \phi}_1*{\hat \phi}_2)
=s({\hat \phi}_1)+s({\hat \phi}_2)-1$. Since $d$ and $*$ are right
$G$-linear and bilinear respectively, they define a nonlinear operator
${\bf W}({\hat \phi})=\frac{1}{2}d{\hat \phi}+ \frac{1}{3}{\hat
  \phi}*{\hat \phi}$ which is $s$-homogeneous of degree $-1$.  This in
turn defines a vector field $W$ of ghost degree $+1$ on $M$.  On the
other hand, the identity operator ${\bf I}:{\hat \phi}\rightarrow
{\hat \phi}$ defines a vector field $I$ of ghost degree zero.  It is
then clear that $S_e$ can be written in the purely geometric form: \be
S_e=\omega(I, W)~~, \ee which obviously has ghost degree zero (since
$\omega$ has ghost degree $-1$).  We are now ready to apply the
geometric formalism to the system $(M, S_e, \omega)$ in order to show
that the extended action satisfies the classical master equation.

\subsubsection{The odd vector field $Q$ }

As discussed above, the odd Hamiltonian vector field $Q:=Q_{S_e}$ can
be viewed as a (non-linear) map ${\bf Q}$ from $M$ to $P$.  To obtain
an explicit formula for ${\bf Q}$, we compute the variation of
$S_e({\hat \phi})$ under an infinitesimal change of ${\hat \phi}$: \be
\delta S_e({\hat \phi})=\langle d{\hat \phi}+\frac{1}{2}[{\hat
  \phi},{\hat \phi}]_*, \delta {\hat \phi} \rangle_e= -\omega_{\hat
  \phi}( d{\hat \phi}+\frac{1}{2}[{\hat \phi},{\hat \phi}]_*, \delta
{\hat \phi})~~, \ee which means that the differential of $S_e$ has the
form: \be dS_e(X)=\omega(Q, X)~~, \ee with: \be
\label{bfQ}
{\bf Q}({\hat \phi})=-(d{\hat \phi}+\frac{1}{2}[{\hat \phi},{\hat
  \phi}]_*)~~.  \ee

\subsubsection{Check of the master equation}

Relation (\ref{bfQ}) implies: \be \omega(Q,Q)({\hat \phi})=
\omega_{\hat \phi}({\bf Q}({\hat \phi}), {\bf Q}({\hat \phi}))
=-\langle d{\hat \phi}, d{\hat \phi}\rangle_e- 2\langle d{\hat \phi},
{\hat\phi}^{*2}\rangle_e- \langle {\hat \phi}^{*2}, {\hat
  \phi}^{*2}\rangle_e~~, \ee where we use the notation $\phi^{*n}$ to
indicate the $n^{th}$ power of $\phi$ computed with the product $*$.
It is easy to check that all three terms vanish upon using the
properties of $\omega$ and the condition $deg {\hat \phi}={\hat 1}$.
Indeed: \be \langle d{\hat \phi}, d{\hat \phi}\rangle_e= \langle {\hat
  \phi}, d^2{\hat \phi}\rangle_e=0~~, \ee \bea \langle d{\hat \phi},
{\hat \phi}^{*2}\rangle_e= \langle{\hat \phi}, (d{\hat \phi})*{\hat
  \phi}\rangle_e &-&\langle{\hat \phi}, {\hat \phi}*d{\hat
  \phi}\rangle_e= -\langle d{\hat \phi})*{\hat \phi}, {\hat
  \phi}\rangle_e
-\langle {\hat \phi}^{*2}, d{\hat \phi}\rangle_e=~~~~~~\nn\\
&=&-2\langle d{\hat \phi}, {\hat \phi}^{*2}\rangle_e\Longrightarrow
\langle d{\hat \phi}, {\hat \phi}^{*2}\rangle_e=0~~, \eea and finally:
\bea \langle{\hat \phi}^{*2}, {\hat \phi}^{*2}\rangle_e
&=&\langle{\hat \phi}, {\hat \phi}^{*3}\rangle_e=- \langle{\hat
  \phi}^{*3}, {\hat \phi}\rangle_e= -\langle{\hat \phi}^{*2}, {\hat
  \phi}^{*2}\rangle_e~~ \Longrightarrow \langle{\hat \phi}^{*2}, {\hat
  \phi}^{*2}\rangle_e=0~~.\nn \eea We conclude that $\omega({\bf Q},
{\bf Q})=0$, which in view of equations (\ref{master}) implies that
$S_e$ satisfies the classical master equation $\{S_e,S_e\}=0$, with
respect to the BV bracket induced by $\omega$.

\subsubsection{The classical gauge}

Let us consider the decompositions: \bea
M=\oplus_{s}{M_s}~~,~~P=\oplus_{s}{P_s}~~, \eea where $M_s={\tilde
  {\cal H}}^s\otimes G_{s~(mod~2)}= {\cal H}^{1-s}\otimes
G_{s~(mod~2)}$, $P_s={\tilde {\cal H}}^s\otimes G_{(1-s)~(mod~2)}=
{\cal H}^{1-s}\otimes G_{(1-s)~(mod~2)}$ denote the the subspaces of
$M$ and $P$ spanned by elements ${\hat \phi}$ of ghost number $s$.
The space $M_0={\tilde {\cal H}}^0\otimes G_0= {\cal H}^1\otimes G_0$
was considered in equation (\ref{M0}) of Section 3.2.  As mentioned
there, the unextended string field $\phi$ can be related to the
component ${\hat \phi}_0$ of ${\hat \phi}$ along this subspace: \be
{\hat \phi}_0=\phi\otimes 1_G+{\hat \phi}_0^\mu\xi_\mu~~, \ee where
$\xi^\mu$ are the odd generators of $G$ (see relation \ref{Ggens}).
Since it is easy to include the evaluation map $ev_G$ in relations
such as (\ref{restriction}), we shall denote ${\hat \phi}_0$ by $\phi$
in order to simplify notation. This allows us to view the unextended
string field as the component of ${\hat \phi}$ along $M_0$.

The selection rule (\ref{dc}) for the symplectic form implies that the
subspaces: \bea
\label{lagr_dec}
{\cal L}&=&\{{\hat \phi}\in M|s({\hat \phi})\geq 0\}= \oplus_{s\geq
  0}{M_s}
~~\nn\\
{\cal N}&=&\{{\hat \phi}\in M|s({\hat \phi})<0\}=\oplus_{s<0}{M_s}
\eea give a Lagrangian decomposition of $(M, \omega)$ (viewed as an
odd symplectic vector space).  In particular, ${\cal L}$ is a
Lagrangian submanifold of $M$, and we can identify $T{\cal L}={\cal
  L}\times {\cal L}$ and $N{\cal L}={\cal L}\times {\cal N}$ (then
$TM|_{\cal L}(s)={\cal L}\times M_s$).

The same selection rule shows that the restriction of $S_e$ to ${\cal
  L}$ coincides with $S$ (up to application of $ev_G$). It follows
that $S_e$ plays the role of BV action for the classical action $S$,
while ${\cal L}$ describes the associated `classical gauge'. Since
$S_e|_{\cal L}=S$ depends only on the component of ${\hat \phi}$ along
$M_0$, the classical action $S_{\cal L}$ has degenerate Hessian on
${\cal L}$. Note that the classical gauge ${\cal L}$ is entirely
determined by the ghost grading $s$.

Following standard BV procedure, we define BV {\em fields}
$\boldphi\in {\cal L}$ and {\em antifields} $\boldphi^*\in {\cal N}$
as the components of ${\hat \phi}$ along ${\cal L}$ and ${\cal N}$.
This gives the decomposition ${\hat \phi}=\boldphi+\boldphi^*$.  The
component $\phi\in M_0$ of $\boldphi$ is the unextended string field,
while the higher components $c_{s}\in M_{s}$ ($s\geq 1$) play the role
of ghosts.  The highest component $\phi^*\in M_{-1}$ of $\boldphi^*$
is the antifield of $\phi$, while the lower components $c^*_{s}\in
M_{-1-s}$ $(s\geq 1$) are the antifields of $c_s$. Hence one has the
decomposition: \be
\label{fa_expansion}
{\hat \phi}=... +c^*_2+ c^*_1+\phi^*+\phi+c_1+c_2+...~~, \ee which we
also write in the form: \be {\hat \phi}=\oplus_{s}{\phi_s}~~, \ee
where $\phi_0:=\phi$, $\phi_{-1}=\phi^*$, $\phi_s=c_s$ and
$\phi_{-1-s}=c_s^*$ for $s\geq 1$. We have $\boldphi=\oplus_{s\geq
  0}{\phi_s}$ and $\boldphi^*=\oplus_{s<0}{\phi_s}= \oplus_{s\geq
  0}{\phi^*_s}$.  Note the relations: \be
\phi_{-1-s}=\phi^*_s~{\rm~for~}s\geq 0~,~ rk \phi^*_s + rk
\phi_s=3~,~\Delta(\phi^*_s)+\Delta(\phi_s)=0~,~ g(\phi^*_s) +
g(\phi_s)={\hat 1}~~, \ee which are due to the selection rules for
$\omega$.

\subsubsection{BRST transformations in the classical gauge}

Applying parity change to the bundle decomposition $TM|_{\cal
  L}=T{\cal L}\oplus T{\cal N}$ gives $\Pi TM|_{\cal L}=\Pi T{\cal
  L}\oplus \Pi N{\cal L}$, with $\Pi T{\cal L}={\cal L}\times \Pi
{\cal L}$ and $\Pi N{\cal L}={\cal L}\times \Pi {\cal N}$, with $\Pi
{\cal L}$ and $\Pi {\cal N}$ given by the following complementary
subspaces of ${\cal H}_e$: \be
\label{pN}
\Pi {\cal L}=\oplus_{s\geq 0}{P_s}~~,~~ \Pi {\cal
  N}=\oplus_{s<0}{P_s}~~, \ee where used the obvious identity $\Pi
M_s=P_s$.  Since ${\cal L}$ is a vector space and the bundle $\Pi
T{\cal L}={\cal L}\times \Pi {\cal L}$ is trivial, the odd vector
field $q$ of (\ref{qgen}) can be identified with an even nonlinear
operator ${\bf q}$ from ${\cal L}$ to $\Pi {\cal L}$.  Relation
(\ref{qgen}) then translates as: \be
\label{BRST}
{\bf q}(\boldphi)= T({\bf Q}(\boldphi))=
-T\left(d\boldphi+\frac{1}{2}[\boldphi,\boldphi]_*\right)~~, \ee where
$T$ is the projector of $P$ onto $\Pi {\cal L}$, taken parallel with
the subspace $\Pi {\cal N}$ (the projector on parity changed BV
fields).

\subsubsection{Expansion of $S_e$ in antifields}

We end this section by writing the extended action in a form which
will be useful later. It is easy to check that: \be
\label{Sexp}
S_e({\hat \phi})=S(\phi)-\langle \boldphi^*, {\bf Q}
(\boldphi)\rangle_e+\langle \boldphi,\boldphi^*~*\boldphi^*\rangle_e=
S(\phi)-\langle \boldphi^*, {\bf q} (\boldphi)\rangle_e+\langle
\boldphi,\boldphi^*~*\boldphi^*\rangle_e~~.  \ee This expression
corresponds to the expansion of $S_e$ in antifields around the
classical gauge ${\cal L}$.

\section{Graded D-brane pairs \label{sec:gradedpairs}}

\subsection{`Component' description of graded pairs} 

If one restricts to the case of graded D-brane pairs, our string field
theory can be described as follows (figure 1).  Using the labels $a,b$
to denote the associated graded branes, with underlying flat bundles
$E_a$ and $E_b$ living on $L$, one can take the grade of $a$ to be
zero without loss of generality.  If $n$ denotes the grade of $b$,
then the relative grade $grade(b)-grade(a)$ is $n$. The theory has
four boundary sectors, which we denote by $Hom(a,a)$, $Hom(b,b)$ (the
diagonal sectors) and $Hom(a,b)$, $Hom(b,a)$ (the off-diagonal, or
boundary condition changing sectors). They are the off-shell spaces of
states for strings stretching from $a$ to $a$, $b$ to $b$, $a$ to $b$
and $b$ to $a$ respectively. The localization arguments of
\cite{Witten_CS}, combined with the shift of $U(1)$ charge discussed
in \cite{Douglas_Kontsevich}, lead to the identifications: \bea
\label{homs}
Hom^k(a,a)=\Omega^k(End(E_a)) &,& Hom^k(b,b)=
\Omega^k(End(E_b)) \nn\\
Hom^k(a,b)=\Omega^{k-n}(Hom(E_a,E_b)) &,& Hom^k(b,a)=\Omega^{k+n}
(Hom(E_b,E_a))~, \eea where $k$ is the worldsheet $U(1)$ charge. The
relation between the charge $|~.~|$ and form rank is given by
(\ref{ghost_mn}): \be |u_{AB}|=rk u_{AB}+grade(B)-grade(A)~~{\rm~for~}
u_{AB}\in Hom(A, B)~, \ee for all $A,B\in \{a,b\}$.  In equations
(\ref{homs}), we let $k$ take any integer value and define the space
of forms of negative ranks (or ranks greater than three) on $L$ to be
zero. Thus non-vanishing states in the sectors $Hom(a,b)$ and
$Hom(b,a)$ have charges $k=n,n+1,n+2$ or $n+3$ and
$k=-n,-n+1,-n+2,-n+3$ respectively.  It follows that the total
boundary space ${\cal H}=Hom(a,a)\oplus Hom(b, a)\oplus Hom(a,b)\oplus
Hom(b,b)$ has non-vanishing components of charges $k=\{-n,\dots
,\,-n+3\} \cup \{n,\dots ,\, n+3\}\cup \{0,\dots,\,3\}$.

\hskip 1.0 in
\begin{center} 
  \scalebox{0.6}{\input{pair.pstex_t}}
\end{center}
\begin{center} 
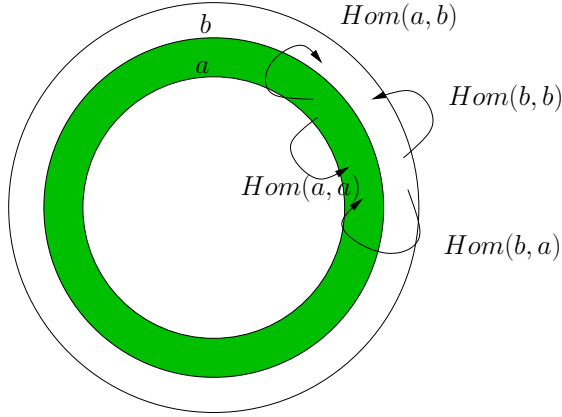
  Figure 1. {\footnotesize Boundary sectors for a pair of graded
    D-branes wrapping the same special Lagrangian cycle. The two
    D-branes $a$ and $b$ are thickened out for clarity, though their
    (classical) thickness is zero.}
\end{center}

The various boundary sectors carry BRST operators $d_{aa}, d_{ab},
d_{ba}$ and $d_{bb}$ given by the covariant differentials twisted with
the flat connections $A_a$ and $A_b$.  One also has a modification of
the boundary product: \be
\label{cdot}
u_{BC}\bullet u_{AB}=(-1)^{[grade(B)-grade(C)]\,rk u_{AB}}
\,u_{BC}\wedge u_{AB}~~, \ee for all $A,B,C\in \{a,b\}$.

Upon using this data, one can write the explicit form of the string
field action (\ref{action}) for such a system. In this case, the
graded bundle is ${\bf E}=E_a\oplus E_b$, and the string field $\phi$
is a (worldsheet charge one) element of the total boundary space
${\cal H}=\Omega^*(L)\otimes \Gamma(End({\bf E}))$.  It can be
represented as a matrix of bundle-valued forms: \be
\phi=\left[\begin{array}{cc} \phi^{(1)}_{aa}~~~&~~~\phi^{(1+n)}_{ba}\\
    \phi^{(1-n)}_{ab}&\phi^{(1)}_{bb}\end{array}\right]~~,
\label{eq:morphisms}
\ee with: \bea \phi^{(1)}_{aa}\in
Hom^1(a,a)=\Omega^1(End(E_a))&,&\phi^{(1)}_{bb}\in Hom^1(b,b)=
\Omega^1(End(E_b))\nn\\
\phi^{(1-n)}_{ab}\in Hom^1(a,b)=\Omega^{1-n}(Hom(E_a,E_b)) &,&
\phi^{(1+n)}_{ba}\in Hom^1(b,a)=\Omega^{1+n}(Hom(E_b,E_a))\nn~,~~~~~
\eea where the superscripts in round brackets indicate form rank.

Given two states $u,v\in {\cal H}$, represented by the matrices
$u=\left[\begin{array}{cc} u_{aa}&u_{ba}\\
    u_{ab} & u_{bb}\end{array}\right]$ and
$v=\left[\begin{array}{cc} v_{aa}&v_{ba}\\
    v_{ab} & v_{bb}\end{array}\right]$, their boundary product
(\ref{dot}) is given by: \be u\bullet v:=\left[\begin{array}{ccc}
    u_{aa}\bullet v_{aa}+u_{ba}\bullet v_{ab}
    &{~}& u_{aa}\bullet v_{ba} +u_{ba}\bullet v_{bb}\\
    u_{ab}\bullet v_{aa}+ u_{bb}\bullet v_{ab} &{~}& u_{bb}\bullet
    v_{bb} + u_{ab}\bullet v_{ba}
\end{array}\right]~~.
\label{cdot'}
\ee We also have the fiberwise supertrace on $End({\bf E})$: \be
str\left[\begin{array}{cc} u_{aa}&u_{ba}\\
    u_{ab} &
    u_{bb}\end{array}\right]=tr_{a}(u_{aa})+(-1)^ntr_{b}(u_{bb})~~,
\ee and the total worldsheet BRST operator: \be
d\left[\begin{array}{cc} u_{aa}&u_{ba}\\
    u_{ab} & u_{bb}\end{array}\right]=
\left[\begin{array}{cc} d_{aa}u_{aa}&d_{ba}u_{ba}\\
    d_{ab}u_{ab} & d_{bb}u_{bb}\end{array}\right]~~, \ee as well as
the boundary bilinear form: \be \langle u, v
\rangle=\int_{L}{str(u\bullet v)}=\int_{L}{\left[~tr_a(u_{aa}\bullet
    v_{aa}+ u_{ba}\bullet v_{ab})+(-1)^n tr_b(u_{ab}\bullet
    v_{ba}+u_{bb}\bullet v_{bb})~\right]}~~, \ee where $tr_a$ and
$tr_b$ denote the fiberwise trace on the bundles $End(E_a)$ and
$End(E_b)$.

With these notations, the string field action (\ref{action}) expands
as: \bea
\label{action_pair}
S(\phi)&=&\int_{L}{str\left[\,{1\ov 2}\phi \bullet d
    \phi +\frac{1}{3}\phi\bullet \phi\bullet \phi \right]} \\
=&\frac{1}{2}&\int_{L}{\left[tr_a(\phi_{aa}\bullet d\phi_{aa}+
    \phi_{ba}\bullet d\phi_{ab})+
    (-1)^n~tr_b(\phi_{bb}\bullet d\phi_{bb}+\phi_{ab}\bullet d\phi_{ba})\right]}+\nn\\
&\frac{1}{3}&\int_{L}{\left[tr_a(\phi_{aa}\bullet \phi_{aa}\bullet
    \phi_{aa}+ \phi_{aa}\bullet \phi_{ba}\bullet
    \phi_{ab}+\phi_{ba}\bullet \phi_{bb}\bullet \phi_{ab}
    +\phi_{ba}\bullet \phi_{ab}\bullet \phi_{aa})\right]}+\nn\\
&\frac{1}{3}&\int_{L}{(-1)^n\left[tr_b(\phi_{bb}\bullet
    \phi_{bb}\bullet \phi_{bb}+ \phi_{bb}\bullet \phi_{ab}\bullet
    \phi_{ba}+\phi_{ab}\bullet \phi_{aa}\bullet \phi_{ba}
    +\phi_{ab}\bullet \phi_{ba}\bullet \phi_{bb})\right]} ~~.\nonumber
\eea Since the rank of a form on the three-cycle $L$ lies between $0$
and $3$, one can distinguish the cases $n=-2,-1,0,1,2$ as well as the
`diagonal' case $|n|\geq 3$.
Notice that one can easily translate from negative to positive $n$ by
reversing the roles of $a$ and $b$, so we can further restrict to
$n=0,1$ or $2$.  The case $n=0$ gives the usual Chern-Simons theory on
the direct sum bundle $E_a\oplus E_b$. The case $n\geq 3$ gives a
graded sum of two Chern-Simons theories.  Hence the interesting cases
are $n=1$ and $n=2$.

\subsection{Comparison of worldsheet BRST cohomologies}

Let us compare the physical (charge one) worldsheet BRST cohomology
for the cases $n=0,1,2$. The cohomology of the BRST operator $d$ has
the direct sum decomposition: \be H^1_d({\cal H})=H^1(End(E_a))\oplus
H^1(End(E_b))\oplus H^{1+n}(Hom(E_b,E_a)) \oplus
H^{1-n}(Hom(E_a,E_b))~~.  \ee Hence it suffices to compare the
`off-diagonal components' $H^1_{od}=H^{1+n}(Hom(E_b,E_a))\oplus
H^{1-n}(Hom(E_a,E_b))$.  Since $H^{1+n}(Hom(E_b,E_a))\approx
H^{2-n}(Hom(E_a, E_b))$ by Poincar\'e duality, one obtains:

\ 

(1) ($n=0$) $H^1_{od}\approx H^1(Hom(E_a,E_b))\oplus
H^2(Hom(E_a,E_b))$.

(2) ($n=1$) $H^1_{od}\approx H^0(Hom(E_a,E_b))\oplus
H^1(Hom(E_a,E_b))$.

(3) ($n=2$) $H^1_{od}\approx H^0(Hom(E_a,E_b))$.

(4) ($n\geq 3$) $H^1_{od}\approx 0$.
 
\ 

It is clear from this that $H^1_d({\cal H})$ will generally be
different for various $n$. For the simple case when $E_a\approx
E_b={\cal O}_L$ (the trivial line bundle on $L$), endowed with the
trivial flat structures, one obtains:
 
\ 

(1) ($n=0$) $dim H^1_{od}=2b_1(L)$.

(2) ($n=1$) $dim H^1_{od}=1+b_1(L)$.

(3) ($n=2$) $dim H^1_{od}=1$.

(4) ($n\geq 3$) $dim H^1_{od}=0$~~,
 
\ 

\noindent where $b_1(L)$ is the first Betti number of $L$ 
(we assume that $L$ is connected, so that $dim H^0(L)=1$). Since $L$
is compact, the dimension $dim H^1_d({\cal H})=2b_1(L)+dim H^1_{od}$
counts the number of independent physical degrees of freedom. It
follows that our theories are generally inequivalent.  The case
$b_1(L)\geq 2$ (for example a special Lagrangian 3-torus, with
$b_1(L)=3$) allows us to distinguish between all four classes based on
this simple argument.  Including the `conjugate' cases $n<0$, we
conclude that there are in general {\em six} distinct types of D-brane
pairs, which shows that the $\Z$-valued D-brane grade has physical
consequences.  This is true even though D-brane pairs whose relative
gradings coincide modulo two (such as the pairs with $n=-1$, $n=+1$,
or the pairs $n=-2$, $n=0$ and $n=2$) have the same classical master
action. As mentioned above, the difference between such theories can
be understood at the BV level as resulting from inequivalent choices
for the ghost grading and classical gauge.

\section{Composite formation and acyclicity} 

\subsection{Vacuum shifts and D-brane composites}

As discussed in some detail in \cite{sc} (upon following the general
framework of \cite{com1, com3}), D-brane composite formation can be
described by the simple device of shifting the string vacuum.  This
results from the observation \cite{com1} that a vacuum shift will
generally break the decomposition of the total boundary space ${\cal
  H}$ into boundary sectors, thereby forcing a change in our D-brane
interpretation\footnote{This process admits an elegant mathematical
  description in the language of differential graded categories
  \cite{com1}, but no familiarity with category theory is required for
  reading the present paper.}. In the case of graded D-brane pairs,
this process can be described as follows.  Suppose that we are given a
(worldsheet charge one) solution $\phi$ to the string field equations
of motion: \be
\label{eom}
\frac{\delta S}{\delta \phi}=0\Longleftrightarrow d\phi+\phi\bullet
\phi=0~~, \ee and with the property that at least one of the
components $\phi^{(1-n)}_{ab}$ and $\phi^{(1+n)}_{ba}$ is nonzero. If
one shifts the string vacuum through $\phi$, then the BRST operator
around the new vacuum is given by: \be
\label{d'}
d'_\phi~u=du+[\phi,u]_\bullet=du+\phi\bullet u -(-1)^{|u|}u\bullet
\phi ~~, \ee where $[u,v]_\bullet= u\bullet v-(-1)^{|u||v|}v\bullet u$
stands for the {\em graded} commutator with respect to the boundary
product $\bullet$ and $U(1)$ charge $|~.~|$, and $u$ is an element of
the total boundary space ${\cal H}$. The string field equations of
motion (\ref{eom}) are equivalent with the condition that $d'_\phi$
squares to zero. The important observation is that $d'_\phi$ does not
preserve the original boundary sectors $Hom(a,a)$, $Hom(a,b)$ etc, due
to non-vanishing of either $\phi^{(1-n)}_{ab}$ or $\phi^{(1+n)}_{ba}$.
Hence one cannot interpret the new background as a collection of two
D-branes. In fact, there is no decomposition of ${\cal H}$ into new
boundary sectors which would satisfy the basic conditions for such an
interpretation (these conditions can be formulated \cite{com1} as the
existence of a category structure compatibile with $d'_\phi$ and with
the boundary product and bilinear form).  Hence the new vacuum must be
interpreted as a single object, namely as a composite of the original
branes.  In this case, ${\cal H}$ is viewed as the boundary sector of
strings `stretching from this composite to itself', though whether
such an interpretation can be implemented in some sort of sigma model
requires a case by case analysis. The issue of some sigma model
representation for such composites is secondary for our purpose, since
we are interested in {\em off-shell} string dynamics, for which it is
natural to take the string field theory approach as fundamental. From
our perspective, {\em all} D-brane composites constructed in this
manner are equally `fundamental', and their inclusion in the theory is
a dynamical requirement. Whether the entirety of this dynamics admits
some open sigma model interpretation is irrelevant for our
considerations.

\subsection{Graded D-brane composites as a  
  representation of the extended moduli space}

Let us consider an ungraded topological D-brane $a$ wrapping $L$, i.e.
a flat bundle $E$ on $L$, with underlying connection $A$. The string
field theory for strings whose endpoints lie on $a$ is described
\cite{Witten_CS} by the Chern-Simons field theory on $E$.  The
associated moduli space ${\cal M}$ of string vacua is the moduli space
of flat connections on $E$, which can be described through rank one
solutions $\phi\in \Omega^1(End(E))$ to the Maurer-Cartan equation for
the commutator algebra of the graded associative algebra
$(\Omega^*(L),\wedge)$, which is the boundary algebra of topological
A-type strings stretching from $a$ to $a$: \be
\label{mc0}
d\phi+\frac{1}{2}[\phi,\phi]_\wedge=0 \Longleftrightarrow
d\phi+\phi\wedge \phi=0~~, \ee divided through the gauge group
generated by transformations of the form: \be
\label{gauge_pair0}
\phi\rightarrow \phi-d\alpha-[\phi,\alpha]_\wedge~~.  \ee In these
equations, $[u,v]_\wedge=u\wedge v -(-1)^{rk u~rk v} v \wedge u$ is
the graded commutator built from the usual wedge product of
$End(E)$-valued forms and taken with respect to the grading given by
form rank. Since $rk\alpha=rk\phi -1$ it follows that in equation
(\ref{gauge_pair0}) the graded commutator is a genuine commutator.
The tangent space to this moduli space is given by $H^1(End(E))$, as
can be seen by considering the linearized form of (\ref{mc}) and
(\ref{gauge_pair0}).

An important problem in open topological string theory and mirror
symmetry is to understand the significance of the so-called {\em
  extended moduli space} ${\cal M}_e$, obtained by considering
deformations along directions in the other cohomology groups, namely
$H^0(End(E))$, $H^2(End(E))$ and $H^3(End(E))$. The extended moduli
space can be defined formally through the so-called `extended
deformation theory' of \cite{Kontsevich_Felder}, which gives a precise
technique for constructing a (formal) supermanifold ${\cal M}_e$ whose
tangent space at the origin is given by the full cohomology group
$H^*(End(E))$. Since $H^1(End(E))$ is a subspace of $H^*(End(E))$, the
unextended moduli space ${\cal M}$ can be identified with a
submanifold of ${\cal M}_e$.  Points in the complement ${\cal
  M}_e-{\cal M}$ are understood as `generalized' vacua of the open
string theory in the given boundary sector, i.e. topological string
backgrounds which do not have an immediate geometric interpretation.
 
The formal approach of extended deformation theory tells us nothing
about the physical significance of the generalized backgrounds
described by points lying in ${\cal M}_e-{\cal M}$. This problem has
obstructed efforts to fulfill the program outlined in
\cite{Witten_nlsm, Witten_mirror} of gaining a better understanding of
mirror symmetry through a study of extended moduli spaces. It is a
remarkable fact (pointed out in \cite{sc}) that the theory of graded
D-branes allows us to give an {\em explicit} physical description of
such points.

To understand this, let us consider the case of graded D-branes $a$
and $b=a[n]$ on $L$, whose underlying flat bundles coincide
($E_a=E_b:=E$) and whose grades differ by $n$.  In this case, the
string field $\phi$ belongs to the space ${\cal
  H}^1=\Omega^1(End(E))^{\oplus 2}\oplus \Omega^{1-n}(End(E))\oplus
\Omega^{1+n}(End(E))$, and the classical equation of motion
(\ref{eom}) is replaced by the Maurer-Cartan equation for the
commutator algebra of the boundary algebra ${\cal H}$, which gives the
equation of motion (\ref{eom}) for the string field theory
(\ref{action_pair}): \be
\label{mc}
d\phi+\frac{1}{2}[\phi,\phi]_\bullet=0~~, \ee with the gauge group
generated by transformations of the form (\ref{gauge}), which in our
case become: \bea
\label{gauge_pair}
\delta\phi^{(1)}_{aa}~~~&=&-d\alpha^{(0)}_{aa}-
[\phi^{(1)}_{aa},\alpha^{(0)}_{aa}]_{\bullet}-
\phi^{(1+n)}_{ba}\bullet\alpha^{(-n)}_{ab}-
\alpha^{(n)}_{ba}\bullet\phi^{(1-n)}_{ab}~~\nn\\
\delta\phi^{(1)}_{bb}~~~&=&-d\alpha^{(0)}_{bb}-
[\phi^{(1)}_{bb},\alpha^{(0)}_{bb}]_{\bullet}-
\phi^{(1-n)}_{ab}\bullet
\alpha^{(n)}_{ba}-\alpha^{(-n)}_{ab}\bullet\phi^{(1+n)}_{ba}~~\nn\\
\delta\phi^{(1+n)}_{ba}&=&-d\alpha^{(n)}_{ba}-
\phi^{(1+n)}_{ba}\bullet\alpha^{(0)}_{bb}-\alpha^{(0)}_{aa}\bullet
\phi^{(1+n)}_{ba}-
\phi^{(1)}_{aa}\bullet\alpha^{(n)}_{ba}-\alpha^{(n)}_{ba}\bullet\phi^{(1)}_{bb}~~\\
\delta\phi^{(1-n)}_{ab}&=&-d\alpha^{(-n)}_{ab}-
\phi^{(1-n)}_{ab}\bullet\alpha^{(0)}_{aa}-\alpha^{(0)}_{bb}
\bullet\phi^{(1-n)}_{ab}-
\phi^{(1)}_{bb}\bullet\alpha^{(-n)}_{ab}-\alpha^{(-n)}_{ab}
\bullet\phi^{(1)}_{aa}~~,\nn \eea with a parameter
$\alpha=\left[\begin{array}{cc}\alpha^{(0)}_{aa}& \alpha^{(n)}_{ba}\\
    \alpha^{(-n)}_{ab}&\alpha^{(0)}_{bb}\end{array}\right]$, where
forms of rank outside of the range $0..3$ are of course vanishing.

As in (\ref{d'}), the graded commutator is now taken with respect to
the boundary product $\bullet$ and worldsheet charge $|~.~|$.  For
$n=0$, this equation describes deformations of the block-diagonal flat
connection $A=A_a\oplus A_a$ to a flat connection on the bundle
$E\oplus E$ (which need not have block-diagonal form).  However, the
moduli space of solutions of (\ref{mc}) for $0<|n|<3$ contains slices
of the {\em extended} moduli space of flat connections on $E$.  For
example a graded D-brane system with $n=1$ contains deformations
generated by $H^0(End(E))$ and $H^2(End(E))$, while for $n=2$ we
obtain deformations generated by $H^3(End(E))$.  Hence {\em one
  represents extended deformations of $E$ through condensation of
  boundary condition changing operators in a pair of graded
  topological D-branes based on $E$} (figure 2).

\hskip 1.0 in
\begin{center} 
  \scalebox{0.5}{\input{moduli.pstex_t}}
\end{center}
\begin{center} 
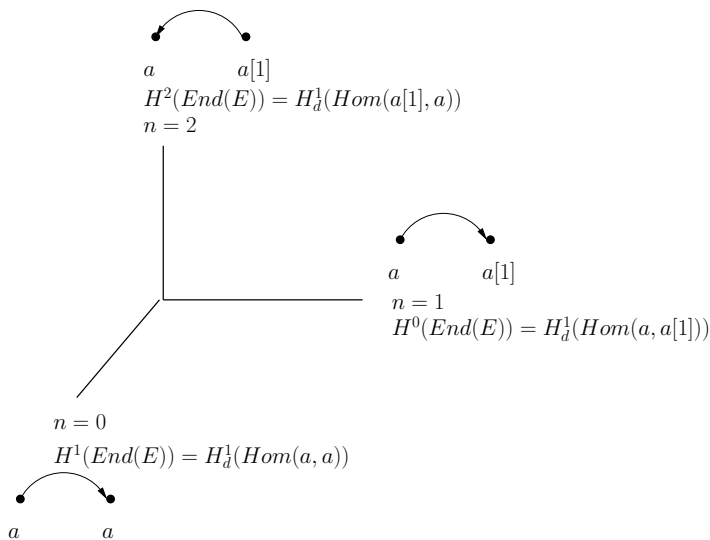
  Figure 2. {\footnotesize Points in the extended moduli space of an
    ungraded D-brane $a$ can be represented through condensation of
    boundary condition changing operators between $a$ and one of its
    higher shifts $b=a[n]$. For example, deformations along
    $H^0(End(E))$, $H^1(End(E))$ and $H^2(End(E))$ can be represented
    by condensing operators in the sectors $Hom(a,a[1])$, $Hom(a,a)$
    and $Hom(a[1],a)$ respectively.}
\end{center}

\

A complete description of the extended moduli space requires
consideration of all graded branes $a[n]$ with $n\in \Z$, and is
discussed at that level of generality in \cite{sc}; the condensates
resulting from a given graded D-brane pair only describe various
slices through ${\cal M}_e$. One approach to the physical significance
of points in ${\cal M}_e-{\cal M}$ is to study the quantum dynamics of
the associated string field theories. A first step in this direction
is the BV analysis carried out in this paper.  This interpretation of
graded D-brane condensates is the main reason for our interest in
graded Chern-Simons theory.

\subsection{Acyclic composites}

The composite resulting from a condensation process can happen to be
{\em acyclic}, i.e. the cohomology of the shifted BRST operator $d'$
may vanish in all degrees.  In this case, the theory expanded around
the composite contains only BRST trivial states, and the resulting
point in the (extended) moduli space can be interpreted as a closed
string vacuum.

\paragraph{Observation} The axioms of string field theory imply that the 
bilinear form $\langle .,. \rangle$ induces a perfect pairing between
the BRST cohomology groups $H^k_{d'}({\cal H})$ and
$H^{3-k}_{d'}({\cal H})$ for any worldsheet $U(1)$ charge $k$.  In
particular, acyclicity of a composite can be established by checking
vanishing of only half of the cohomology groups $H^k_{d'}({\cal H})$.

\subsection{Acyclic composites for $n=1$} 

\subsubsection{Assumptions}

In this subsection we discuss a particular class of condensation
processes which is somewhat similar to tachyon condensation in bosonic
string theory in that it produces acyclic composites. Condensation
processes of the type discussed below can take place for a graded
D-brane pair satisfying the conditions:

\ 

(1) the relative grade of the pair is $n=1$

\ 
 
(2) $E_a$ and $E_b$ are isomorphic {\em as flat bundles}.

\ 

(3)
$H^1(End(E_a))=H^1(End(E_b))=H^1(Hom(E_a,E_b))=H^2(Hom(E_a,E_b))=0$.

\ 

The argument presented below is closely related to ideas of
\cite{Vafa_cs}, though our physical realization seems to be different.

It is not hard to see that condition (2) amounts to the existence of a
{\em covariantly constant} isomorphism between $E_a$ and $E_b$. This
means a bundle isomorphism $f:E_a\rightarrow E_b$ which satisfies
$df=0$ \footnote{It is not hard to see that such a map gives an
  isomorphism of the underlying flat structures. In the language of
  flat connections, the condition $df=0$ implies that $f$ induces a
  gauge transformation which takes $A_a$ to $A_b$ (one obtains
  $A_b=fA_af^{-1}-(df)f^{-1}$ upon choosing local frames for $E_a$ and
  $E_b$). Alternately, $f$ gives an isomorphism between flat
  trivializations of $E_a$ and $E_b$.}  (remember that the
differential $d$ on $Hom(E_a, E_b)$ is coupled to the background flat
connections $A_a$ and $A_b$).  The constraints on the first cohomology
of $E_a$ and $E_b$ mean that these flat connections possess no
deformations.

\subsubsection{Construction of acyclic composites}

With the assumptions discussed above, one can construct a solution
$\phi$ of the string field equations of motion which leads to an
acyclic D-brane composite. Indeed, let us look for a solution of the
form $\phi=\left[\begin{array}{cc}0&0\\\phi_{ab}^{(0)} &0
\end{array}\right]$. In this case, the equations of motion 
$d\phi+\phi\bullet\phi=0$ reduce to $d\phi_{ab}^{(0)}=0$, which says
that $\phi_{ab}^{(0)}$ is covariantly constant section of
$Hom(E_a,E_b)$.  We shall further require that $\phi^{(0)}_{ab}$ is a
bundle isomorphism, which is possible in view of our second
assumption. Condensation of the boundary condition changing state
$\phi^{(0)}_{ab}$ leads to a new D-brane background which can be
viewed as composite of the D-branes $a$ and $b$.  The theory expanded
around this background is endowed with the shifted BRST operator given
by $d'_\phi~u=d\phi+[\phi,u]_\bullet$.  We wish to show that $d'_\phi$
has trivial cohomology. Since $n=1$, the total boundary space of our
D-brane system has non-vanishing components of charge $k=-1,0,1,2,3$
and $4$.  Moreover, Poincar\'e duality identifies the worldsheet BRST
cohomologies in degrees $k$ and $3-k$, so that it suffices to show
vanishing of $H^k_{d'}({\cal H})$ for $k=-1, 0$ and $1$.

\subsubsection{Proof of acyclicity}

\paragraph{Vanishing of $H^{-1}_{d'_\phi}({\cal H})$}

A state of charge $-1$ has the form $u=\left[\begin{array}{cc} 0 &
    u_{ba}^{(0)}\\0 & 0 \end{array}\right]$. In this case, the BRST
closure condition $d'_\phi u=0\Leftrightarrow du+\phi\bullet
u+u\bullet \phi=0$ is equivalent with: \be
u^{(0)}_{ba}\bullet\phi^{(0)}_{ab}=0~~,~~ du^{(0)}_{ba}=0~~,~~
\phi_{ab}^{(0)}\bullet u_{ba}^{(0)}=0~~.  \ee Since $\phi^{(0)}_{ab}$
is a bundle isomorphism, this is equivalent with $u_{ba}^{(0)}=0$ i.e.
$u=0$, which shows that $H^{-1}_{d'_\phi}({\cal H})=0$.

\paragraph{Vanishing of $H^0_{d'_\phi}({\cal H})$}

Given a charge zero state $u=\left[\begin{array}{cc} u_{aa}^{(0)}&
    u_{ba}^{(1)}\\0 & u_{bb}^{(0)} \end{array}\right]$, the BRST
closure condition $d'_\phi u=0\Leftrightarrow du+\phi\bullet
u-u\bullet \phi =0$ reads: \be
\label{closed0}
du_{aa}^{(0)}=u^{(1)}_{ba}\bullet\phi^{(0)}_{ab}~~,~~
du^{(1)}_{ba}=0~~,~~ \phi_{ab}^{(0)}\bullet
u^{(0)}_{aa}=u^{(0)}_{bb}\bullet \phi^{(0)}_{ab}~~,~~
du_{bb}^{(0)}=-\phi_{ab}^{(0)}\bullet u_{ba}^{(1)}~~.  \ee On the
other hand, the exactness condition
$u=d'_\phi\alpha=d\alpha+\phi\bullet \alpha+\alpha\bullet\phi$ (for an
element $\alpha=\left[\begin{array}{cc} 0 & \alpha_{ba}^{(0)}\\0 & 0
  \end{array}\right]$ of charge $-1$) is equivalent with: \be
\label{exact0}
u^{(0)}_{aa}=\alpha^{(0)}_{ba}\bullet \phi^{(0)}_{ab}~~,~~
u^{(1)}_{ba}=d\alpha^{(0)}_{ba}~~,~~
u^{(0)}_{bb}=\phi^{(0)}_{ab}\bullet \alpha^{(0)}_{ba}~~.  \ee

The assumption $H^2(Hom(E_a,E_b))=0\Leftrightarrow
H^1(Hom(E_b,E_a))=0$ allows us to solve the second equation of
(\ref{closed0}) in the form: \be
\label{ex1}
u^{(1)}_{ba}=d\alpha^{(0)}_{ba}~~, \ee with $\alpha^{(0)}_{ba}$ a
section of $Hom(E_b,E_a)$, determined up to \be
\label{amb}
\alpha^{(0)}_{ba}\rightarrow \alpha^{(0)}_{ba}+f^{(0)}_{ba}~~, \ee
where $f^{(0)}_{ba}$ is a {\em covariantly constant} section of
$Hom(E_b, E_a)$.  Upon substituting (\ref{ex1}) into the first and
fourth equations of (\ref{closed0}), these can be solved as: \be
\label{ex2}
u^{(0)}_{aa}=\alpha^{(0)}_{ba}\bullet \phi^{(0)}_{ab}+f^{(0)}_{aa}~~
\ee and \be
\label{ex3}
u^{(0)}_{bb}=\phi^{(0)}_{ab}\bullet \alpha^{(0)}_{ba}+f^{(0)}_{bb}~~,
\ee where $f^{(0)}_{aa}$ and $f^{(0)}_{bb}$ are covariantly constant
sections of $End(E_a)$ and $End(E_b)$ and we used
$d\phi^{(0)}_{ab}=0$.  Using these expressions in the third equation
of (\ref{closed0}) gives: \be \label{ex20} \phi^{(0)}_{ab}\bullet
f^{(0)}_{aa}=f^{(0)}_{bb}\bullet \phi^{(0)}_{ab}~~, \ee a condition
which can be satisfied upon choosing
$f^{(0)}_{bb}=\phi^{(0)}_{ab}f^{(0)}_{aa}(\phi^{(0)}_{ab})^{-1}$ (this
is covariantly constant since each of the factors is).  With this
choice, both $f^{(0)}_{aa}$ and $f^{(0)}_{bb}$ can be eliminated from
(\ref{ex2}) and (\ref{ex3}) upon using the freedom (\ref{amb}) to
redefine: \be\label{shift0} \alpha^{(0)}_{ba}\rightarrow \alpha^{'(0)}_{ba}:=
\alpha^{(0)}_{ba}+ f^{(0)}_{aa}\bullet (\phi^{(0)}_{ab})^{-1}~~.  \ee
This shows that equations (\ref{ex1}), (\ref{ex2}) and (\ref{ex3}) can
be brought to the form (\ref{exact0}). We conclude that
$H^0_{d'_\phi}({\cal H})$ vanishes as well.

\paragraph{Vanishing of $H^1_{d'_\phi}({\cal H})$}

If $u=\left[\begin{array}{cc} u_{aa}^{(1)}& u_{ba}^{(2)}\\u_{ab}^{(0)}
    & u_{bb}^{(1)} \end{array}\right]$ is a state with $|u|=1$, then
the BRST closure condition $d'_\phi u=0\Leftrightarrow
du+[\phi,u]_\bullet =0$ is equivalent with the system: \bea
\label{closed}
du_{aa}^{(1)}+u^{(2)}_{ba}\bullet\phi^{(0)}_{ab}&=&0~~\nn\\
du^{(2)}_{ba}&=&0~~\nn\\
du_{ab}^{(0)}+\phi_{ab}^{(0)}\bullet u^{(1)}_{aa}+u^{(1)}_{bb}\bullet
\phi^{(0)}_{ab}&=&0~~\\
du_{bb}^{(1)}+\phi_{ab}^{(0)}\bullet u_{ba}^{(2)}&=&0~~.\nn \eea

Using the assumption $H^2(Hom(E_b,E_a))=0$, the second equation can be
solved as: \be
\label{exact1}
u_{ba}^{(2)}=d\alpha_{ba}^{(1)}~~, \ee for some one form
$\alpha_{ba}^{(1)}$ with coefficients in $Hom(E_b,E_a)$. Upon
substituting this into the first equation and using
$d\phi^{(0)}_{ab}=0$, one obtains: \be
\label{exact2}
u_{aa}^{(1)}=d\alpha_{aa}^{(0)}-\alpha^{(1)}_{ba}\bullet\phi^{(0)}_{ab}~~,
\ee for some section $\alpha_{aa}^{(0)}$ of $End(E_a)$.  To arrive at
this equation, we made use of the assumption $H^1(End(E_a))=0$.  The
section $\alpha^{(0)}_{aa}$ is determined up to transformations of the
form: \be
\label{ambiguity}
\alpha_{aa}^{(0)}\rightarrow \alpha_{aa}^{(0)}+f^{(0)}_{aa}~~, \ee
with $f_{aa}^{(0)}$ a {\em covariantly constant} (flat) section of
$End(E_a)$.  Using the solution for $u^{(2)}_{ba}$ and the condition
$d\phi^{(0)}_{ab}=0$, we can similarly solve the last equation of
(\ref{closed}): \be
\label{exact3}
u_{bb}^{(1)}=d\alpha_{bb}^{(0)}+\phi_{ab}^{(0)}\bullet
\alpha^{(1)}_{ba}~~, \ee where we used the assumption
$H^1(End(E_b))=0$. Finally, combining (\ref{exact1}) and
(\ref{exact2}) allows us to solve the third equation in
(\ref{closed}): \be
\label{exact4_intmd}
u_{ab}^{(0)}=\phi_{ab}^{(0)}\bullet
\alpha_{aa}^{(0)}-\alpha^{(0)}_{bb} \bullet
\phi^{(0)}_{ab}+f^{(0)}_{ab}~~, \ee with $f_{ab}^{(0)}$ a covariantly
constant section of $Hom(E_a,E_b)$.

Since both sections $f^{(0)}_{ab}$ and $\phi_{ab}^{(0)}$ of
$Hom(E_a,E_b)$ are covariantly constant, the section
$g^{(0)}_{aa}:=(\phi_{ab}^{(0)})^{-1}\circ f_{ab}^{(0)}$ of $End(E_a)$
satisfies $dg^{(0)}_{aa}=0$. Upon using the ambiguity
(\ref{ambiguity}), we can therefore absorb $f_{aa}^{(0)}$ in
$\alpha_{aa}^{(0)}$ by defining: \be
\alpha_{aa}^{'(0)}=\alpha_{aa}^{(0)}+g_{aa}^{(0)}~~.  \ee This allows
us to re-write (\ref{exact4_intmd}) in the
form: \be u_{ab}^{(0)}=\phi_{ab}^{(0)}\bullet
\alpha_{aa}^{'(0)}-\alpha_{bb}^{(0)} \bullet \phi^{(0)}_{ab}~~.  \ee

Hence we can assume without loss of generality that
$\alpha_{aa}^{(0)}$ has been chosen such that: \be
\label{exact4}
u_{ab}^{(0)}=\phi_{ab}^{(0)}\bullet
\alpha_{aa}^{(0)}-\alpha^{(0)}_{bb} \bullet \phi^{(0)}_{ab}~~.  \ee
The last step is to define the state $\alpha=\left[\begin{array}{cc}
    \alpha_{aa}^{(0)}&\alpha_{ba}^{(1)}\\0 & \alpha_{bb}^{(0)}
  \end{array}\right]$, which has charge zero.  It is then easy to
check that equations (\ref{exact1}), (\ref{exact2}) and (\ref{exact3})
and (\ref{exact4}) are equivalent with the condition: \be
u=d'_\phi\alpha=d\alpha+[\phi,\alpha]_\bullet~~.  \ee Thus every
$d'_\phi$-closed degree one state is $d'_\phi$-exact, which means that
the BRST complex of $d'_\phi$ is also trivial in degree one. This
shows that the composite obtained by condensation of $\phi_{ab}^{(0)}$
is acyclic.

Note that the boundary condition changing state $\phi^{(0)}_{ab}$ is
physical, since it is BRST-closed and has charge $1$.  In fact, the
(`unshifted') BRST cohomology $H^1_d({\cal H}_{ab})$ in degree one
coincides with the space $H^0_d(Hom(E_a,E_b))$ of covariantly constant
sections of $Hom(E_a,E_b)$.

\paragraph{Observations} 

(1) We wish to stress that one does {\em not} need to consider the
extended action (\ref{extended_action}), nor the extended boundary
space ${\cal H}_e$, in order to understand condensation processes.

(2) We mention that the assumption $E_a\approx E_b$ is necessary only 
for establishing acyclicity in the sector with $U(1)$ charge $-1$.
In the sector with vanishing $U(1)$ charge,
this condition can be avoided by choosing
$f_{aa}^{(0)}=g_{ba}^{(0)}\bullet \phi^{(0)}_{ab}$ and 
$f_{bb}^{(0)}=\phi^{(0)}_{ab}\bullet g_{ba}^{(0)}$ 
in equations (\ref{ex2}) and (\ref{ex3}), where
$g_{ba}$ is a covariantly constant section of
$Hom(E_b,~E_a)$. Then equation (\ref{ex20}) is automatically satisfied
and one shifts $\alpha_{ba}^{(0)}$ to $\alpha_{ba}^{'(0)}=\alpha_{ba}^{(0)}+
g_{ba}^{(0)}$ instead of equation (\ref{shift0}). 
Hence vanishing of $H_{d'}^0({\cal H})$ 
can be established without requiring 
invertibility of $\phi^{(0)}_{ab}$. A similar 
modification of the argument is possible when showing 
vanishing of $H^1_{d'}({\cal H})$; namely one chooses $f_{ab}^{(0)}$ 
to be of the form $f_{ab}^{(0)}=\phi_{ab}^{(0)}\bullet g_{aa}^{(0)}$,
with  $g_{aa}^{(0)}$ a covariantly constant section of $End(E_a)$.
However, the condition that $\phi_{ab}$ is invertible is crucial 
for assuring vanishing of $H^{-1}_{d'}({\cal H})$. 

(3) The fact that a D-brane pair satisfying our conditions can
condense to an acyclic composite gives some justification for viewing
these systems as `topological brane-antibrane pairs'.  However, we
were able to establish the existence of such processes only under
restrictive topological assumptions on the original background. To
understand just how restrictive our hypotheses are, consider the case
of two singly-wrapped graded D-branes, for which $E_a=E_b={\cal O}_L$,
the trivial line bundle over $L$ (endowed with the trivial flat
connection $A_a=A_b=0$).  In this situation, our argument requires
vanishing of the cohomology group $H^1(L)\approx H^2(L)$, which in
particular means that the cycle $L$ must be a rational homology sphere. 
It is well-known that Calabi-Yau threefolds contain a wealth of 
special Lagrangian cycles which do not satisfy this 
condition (for example, special Lagrangian
3-tori, which according to the conjecture of \cite{SYZ} should give a
fibration of $X$ if $X$ admits a geometric mirror). Our argument does
not apply to such cycles, even in the singly-wrapped case. It is an
interesting problem to study composite formation in this more general
situation.

\section{Direct construction of the master action for D-brane pairs} 

In this section we give a direct construction of the classical BV
actions for the string field theories of graded D-brane pairs. This
will allow us to recover the extended theory (\ref{extended_action})
by applying the more familiar cohomological formalism.  Moreover, we
shall give a BV level proof that there are precisely six inequivalent
classical BV actions, namely those associated to the relative gradings
$n=0,\,\pm1,\,\pm2$ and $(n\ge 3,\,\,n\le-3)$, thus leading to a
$\Z_6$ multiplicity of such systems. This gives the string field
theory realization of an observation made in
\cite{Douglas_Kontsevich}.

As already mentioned at the end of Section 4, the BV algorithm
prolongs a basis for the space of classical fields to a basis ${\cal
  B}= (e^{*\alpha}, e_{\alpha})$ for the space of extended fields,
which is Darboux for the underlying odd symplectic form. Since the
coefficients of the latter are particularly simple in this basis, the
BV bracket takes the form (\ref{Darboux_bracket}).

In this section we apply the BV algorithm in a component approach, and
carry out the computations in terms of the usual wedge product of
forms (which we shall write as juxtaposition in order to simplify
notation), rather than in terms of the extended boundary product $*$
used in the previous section.  This makes for an explicit
presentation, at the cost of introducing somewhat complicated
formulae. A more synthetic description (which uses the results of the
geometric formalism in order to simplify certain steps) is given in
the next section.  As discussed in Section 5 \ref{sec:gradedpairs}, it
is enough to consider the cases with relative grading $n\equiv
n_b-n_a\ge 0$, since the rest can be obtained by reversing the roles
of $a$ and $b$ \footnote{Two D-brane pairs with relative gradings $n$
  and $-n$ are related by the conjugation operation discussed in
  \cite{sc}. While this is a `symmetry' at the classical level, two
  systems related in this manner should not be identified. The
  conjugation symmetry of \cite{sc} is akin to the charge conjugation
  of particle physics, and conjugated configurations must be
  considered physically distinct.}.

\subsection{The BV action for $n=1$ }

$${}$$

For the case $n=1$, the classical fields are two one-forms, a two-form
and a zero-form: \be {\phi}=\pmatrix{ \phi^{(1)} & \phi^{(2)}\cr
  \phi^{(0)} & \phi^{'(1)} \cr}~~.
\label{eq:n=1}
\ee In terms of the usual wedge product (\ref{wedge}), the classical
action (\ref{action_pair}) reads: \bea S(\phi)=\int_L &&tr_a\left[
  {1\ov 2}\left(\p{1}d\p{1}-\p{2}d\p{0}\right)+{1\ov
    3}\left(\p{1}\p{1}\p{1}+
    \p{1}\p{2}\p{0}\right)\right]~~  \\
-\int_{L}&&tr_b\left[{1\ov
    2}\left(\hp{1}d\hp{1}-\p{0}d\p{2}\right)+{1\ov 3}\left(
    \hp{1}\hp{1}\hp{1}+\hp{1}\p{0}\p{2}\right)\right]~~.\nn \eea

Upon substituting $\alpha=-C_1\lambda$ into the gauge transformations
(\ref{gauge_pair0}), where $\lambda$ is a Grassmann-odd constant
parameter and: \be C_1=\pmatrix{ c_1^{(0)} & c_1^{(1)}\cr 0&
  {c'}_1^{(0)} \cr}~~,
\label{eq:n1gen1}
\ee is the corresponding matrix of ghosts, we find the BRST
transformations of the physical fields: \bea &&\delta^{(1)}\p{1}=
\left[dc_1^{(0)}+[\p{1},\,c_1^{(0)}]- c_1^{(1)}\p{0}\right]\lambda
\nonumber\\
&&\delta^{(1)}\p{2}=
\left[dc_1^{(1)}+(\p{1}c_1^{(1)}+\p{2}{c'}_1^{(0)} -c_1^{(0)}\p{2}+
  c_1^{(1)}\hp{1})\right]\lambda
\nonumber\\
&&\delta^{(1)}\p{0}= (\p{0}c_1^{(0)}-{c'}_1^{(0)}\p{0})\lambda\nonumber\\
&&\delta^{(1)} \hp{1}=\left[d{c'}_1^{(0)}
  +[\hp{1},\,{c'}_1^{(0)}]-\p{0}c_1^{(1)}\right] \lambda
\label{brst}~~.
\eea Requiring nilpotence of $\delta^{(1)}$ leads to the following
ghost transformations: \be \delta^{(1)} c_1^{(0)}= c_1^{(0)}
c_1^{(0)}\lambda~~, ~~\delta^{(1)} {c'}_1^{(0)}= {c'}_1^{(0)}
{c'}_1^{(0)}\lambda~~, ~~\delta^{(1)}
c_1^{(1)}=(c_1^{(0)}c_1^{(1)}+c_1^{(1)}{c'}_1^{(0)})\lambda~~.
\label{ghn1brst0}
\ee

It is easy to see that the gauge transformations (\ref{gauge_pair})
vanish on-shell for a particular choice of parameters.
Oversimplifying, consider the transformation\footnote{We remind the
  reader that the cohomological (BRST) approach extends the gauge
  algebra ${\bf g}$ of Subsection 2.2.2 to the jet bundle associated
  with our fields. Thus, one views the various partial derivatives of
  a generator as being independent quantities.  This is why we can
  consider `differential zero modes', i.e. expressions involving the
  exterior derivative of some generator. The jet bundle extension is a
  standard device for describing the on-shell algebraic structure of
  gauge transformations and observables, and forms the heart of the
  modern approach to BRST quantization. We refer the reader to
  \cite{Stasheff_bv} for an elegant review of the jet bundle formalism
  and further references.}  $\delta\p2 = -d\alpha^{(1)} -\p1
\alpha^{(1)}+ \alpha^{(1)}\hp1$ and take $\alpha^{(1)}$ to be of the
form $\alpha^{(1)}
=-(d\beta^{(0)}+\p{1}\beta^{(0)}+\beta^{(0)}\hp{1})$, with
$\beta^{(0)}$ a section of $Hom(E_b,E_a)$.  Substituting this into the
gauge transformation, one finds that $\delta\p2$ vanishes when $\p1$
and $\hp1$ satisfy their equations of motion (in this exemplification
we ignored the role of $\alpha^{(0)}$ and $\alpha^{'(0)}$ in the BRST
transformation).  It follows that our gauge algebra is {\em
  reducible}.  In such a situation, one must introduce ghosts for
ghosts.  Accordingly, we consider a second generation ghost: \be
C_2=\pmatrix{ 0 & c_2^{(0)}\cr 0 & 0\cr}~~,
\label{eq:n1gen2}
\ee which allows us to extend the BRST transformation of $c_1^{(1)}$
by adding: \bea
&&{\delta}_2c_1^{(1)}=\left(dc_2^{(0)}+\p{1}c_2^{(0)}-c_2^{(0)}\hp{1}
\right) \lambda~~.  \eea Allowing $\delta_2$ to act trivially on the
physical fields $\phi$ and requiring nilpotence of $\delta^{(2)}\equiv
\delta^{(1)}+\delta_2$ leads to the following corrections to the
transformations of $c_1^{(0)}$ and ${c'}_1^{(0)}$: \be
{\delta}_2c_1^{(0)}=c_2^{(0)}\p{0}\lambda~~,~~
{\delta}_2{c'}_1^{(0)}=\p{0}c_2^{(0)}\lambda~~ \ee and specifies the
BRST transformation of the second generation ghost: \bea \delta_2
c_2^{(0)}=(-c_1^{(0)}c_2^{(0)}+c_2^{(0)}{c'}_1^{(0)})\lambda~~.
\label{ghn1brst1}
\eea Since there are no further zero modes, the gauge algebra is first
order reducible and the full BRST transformations are given by: \bea
\delta\equiv \delta^{(2)}=\delta^{(1)}+\delta_2
\label{eq:fullBRSTn1}
\eea where we let $\delta^{(1)}$ act trivially on second generation
ghosts: $\delta^{(1)} c_2^{(0)}=0$.

The BV construction now introduces antifields $\Phi^*, C_1^*$ and
$C_2^*$ for each of the fields $\phi, C_1$ and $C_2$: \be
\Phi^*=\pmatrix{ \phi^{*(2)}&\phi^{*(3)}\cr
  \phi^{*(1)}&\phi'^{*(2)}\cr}~~,~~ C_1{}^*=\pmatrix{ c_1^{*(3)} & 0
  \cr c_1^{*(2)} & {c'}_1^{*(3)}\cr}~~,~~ C_2{}^*=\pmatrix{ 0 & 0\cr
  c_2^{*(3)} & 0\cr}~~.  \ee
\noindent 
Note that the antifield of each matrix block sits in the transposed
position in the full matrix, for example ($\phi^{(1)}$,
$\phi^{*(2)}$), ($\phi^{'(1)}$, $\phi^{'*(2)}$), ($\phi^{(2)}$,
$\phi^{*(1)}$) and ($\phi^{(0)}$, $\phi^{*(3)}$) form field-antifield
pairs.  The classical action $S$ is extended by adding a term of the
form $S_1=tr(\Phi^* \delta\phi/\lambda+ C_1^* \delta
C_1/\lambda+\dots)$~.  The Grassmann parities of antifields are chosen
such that $S_1$ is even: $g({\rm field})+g({\rm antifield})={\hat
  1}~(mod~2)$.

Following this procedure we write the first order action: \bea S_1&&
=\int_L\,tr_a\Big[~
\phi^{*(2)}\left(dc_1^{(0)}+[\p{1},\,c_1^{(0)}]-c_1^{(1)}\p{0}\right)
\nonumber\\
&&~~~~~~~~ +c_1^{*(3)}\left(c_1^{(0)}c_1^{(0)}+c_2^{(0)}\p{0}\right)
+\phi^{*(3)}\left( \p{0}c_1^{(0)}-{c'}_1^{(0)}\p{0} \right)\Big]\nn\\
&&+\int_L\,tr_b\Big[~ \phi^{*(1)}\left(
  dc_1^{(1)}+(\p{1}c_1^{(1)}+\p{2}{c'}_1^{(0)}-c_1^{(0)}\p{2}+
  c_1^{(1)}\hp{1}) \right)\nn\\
&&~~~~~~~~+
\phi'^{*(2)}\left(d{c'}_1^{(0)}+[\hp{1},\,{c'}_1^{(0)}]-\p{0}c_1^{(1)}\right)
\nonumber\\
&&~~~~~~~~+
c_1^{*(2)}\left(c_1^{(0)}c_1^{(1)}+c_1^{(1)}{c'}_1^{(0)}+dc_2^{(0)}+\p{1}c_2^{(0)}-
  c_2^{(0)}\hp{1}\right)
\nonumber\\
&& {}~~~~~~~+
{c'}^{*(3)}_1\left({c'}_1^{(0)}{c'}_1^{(0)}+\p{0}c_2^{(0)}\right) +
c_2^{*(3)}\left( -c_1^{(0)}c_2^{(0)}+c_2^{(0)}{c'}_1^{(0)} \right)
\Big]~~.  \eea

Higher order terms in the antifield expansion $S_{BV}=S_0+S_1+S_2+...$
are constructed from the requirement that $S_{BV}$ satisfies the
master equation.  Since vanishing of $\left\{S,\,S_1\right\}$ is
guaranteed by gauge invariance of the classical action, the next step
is to compute the BV bracket $\left\{S_1,\,S_1\right\}$ and find a
non-vanishing result, linear in antifields\footnote{In fact, the BRST
  transformation of $c_2^{(0)}$ can also be inferred from the
  requirement $\left\{S_1,\,S_1\right\}$ contains no terms which are
  independent of antifields. }.  Thus, we have to supplement $S_1$ by
a further term $S_2$ quadratic in antifields, such that $\{ S_1,
S_1\}+2\{S ,S_2\}=0$.  It is not very hard to check that the choice:
\be S_2=\int_L\,tr_a\left(c_2^{(0)}\phi^{*(1)} \phi^{*(2)}-
  c_2^{(0)}\phi'^{*(2)}\phi^{*(1)}\right) ~~, \ee assures that the BV
action: \bea
\label{bv1}
S_{BV}=S+S_1+S_2 \eea obeys the master equation. Some details of the
relevant computation are given in Appendix C.

To show that the BV action coincides with the coordinate-free
expression (\ref{extended_action}), we set: \bea
\label{coords1}
{\hat \phi}&=&\pmatrix{ c_1^{(0)}+\phi^{(1)}-\phi^{*(2)} +c_1^{*(3)}&
  c_2^{(0)}+c_1^{(1)}+\p{2} -\phi^{*(3)}\cr
  \phi^{(0)}+\phi^{*(1)}+c_1^{*(2)}+c_2^{*(3)}& {c'}_1^{(0)}
  +\hp{1}+{\phi'}^{*(2)}-{c'}_1^{*(3)}
  \cr}\nn\\
&=&c_2+c_1+\phi+\phi^*+c_1^*+c_2^*~~,
\label{eq:superfieldn1}
\eea where we defined: \be \phi^*=\pmatrix{
  -\phi^{*(2)}&-\phi^{*(3)}\cr \phi^{*(1)}&\phi'^{*(2)}\cr}~~,~~
c_1{}^*=\pmatrix{ c_1^{*(3)} & 0 \cr c_1^{*(2)} &
  -{c'}_1^{*(3)}\cr}~~,~~ \ee and: \be
c_1=C_1~~,~~c_2=C_2~~,~~c_2{}^*=C_2^*~~,
\label{notghn1}
\ee notations which will be useful in the next section.  Counting
worldsheet $U(1)$ charges and Grassmann parities shows that ${\hat
  \phi}$ is an element of $M={\cal H}_e^1$.  Relation (\ref{coords1})
can be viewed as the expression of the vector ${\hat \phi}$ in the
particular linear coordinates on $M$ built by the BV algorithm.

One can check by direct computation that substitution of
(\ref{coords1}) into the extended action (\ref{extended_action})
recovers the expanded form (\ref{bv1}) upon expressing everything in
terms of the usual wedge product.  This shows that the BV action
(\ref{bv1}) is simply the form of the extended action in the
particular linear coordinates (\ref{coords1}).  To show equivalence
with the geometric formalism of Section 4, we must also check that
(\ref{coords1}) gives Darboux coordinates for the odd symplectic form
(\ref{omega}).  As mentioned in Subsection 4.8.1, the odd symplectic
form is completely determined by its restriction to {\em even} vector
fields, which can be identified with the form $\omega_0$ of equation
(\ref{omega0_wedge}). It thus suffices to check that
(\ref{omega0_wedge}) reduces to Darboux form in the coordinates
(\ref{coords1}).  This follows by direct computation upon substituting
(\ref{coords1}) into (\ref{omega0_wedge}) \footnote{To obtain Darboux
  coordinates per se, one must choose bases for the spaces of
  bundle-valued forms corresponding to each block in the decomposition
  of our matrices of morphisms. We leave the details to the reader.}:
\bea \omega_0=tr_a\Big[ &&(\delta c_1^{*(3)}\we \delta
c_1^{(0)}-\delta c_1^{(0)}\we\delta c_1^{*(3)})+
(\delta\phi^{*(2)}\we \delta\p1 - \delta\p1\we\delta\phi^{*(2)})\nn\\
+&&(\delta c_2^{*(3)}\we\delta c_2^{(0)}-\delta c_2^{(0)}\we \delta
c_2^{*(3)})
+(\delta c_1^{*(2)}\we\delta c_1^{(1)}-\delta c_1^{(1)}\we \delta c_1^{*(2)})\nn\\
+&&(\delta\phi^{*(1)}\we \delta\p2-\delta\p2\we\delta\phi^{*(1)})\Big]\nn\\
+tr_b\Big[&& (\delta {c'}_1^{*(3)}\we\delta {c'}_1^{(0)}-\delta
{c'}_1^{(0)}\we\delta {c'}_1^{*(3)})
+(\delta\phi'^{*(2)}\we \delta\hp1-\delta\hp1\we\phi'^{*(2)})\nn\\
+&&(\delta\phi^{*(3)}\we
\delta\p0-\delta\p0\we\delta\phi^{*(3)})\Big]~~.  \eea We conclude
that the homological construction for a D-brane pair with relative
grading $n=1$ agrees with the BV system discussed in Section 4.

\subsection{The BV action for $n=2$}

We next consider D-brane pairs with relative grading $n=2$. For such
systems, the classical fields are two one-forms and a three form: \be
\phi =\pmatrix{ \phi^{(1)} & \phi^{(3)}\cr 0 & {\phi'}^{(1)} \cr}~~.
\ee Since $n$ is even, the relative grading plays no role in the
boundary product $\bullet$, which reduces to the usual wedge product
(\ref{wedge}).  The classical action is: \bea
\label{action_2}
\!\!\!\!\!\!\!\!\!\!\!\!\!\!\!\!  S(\phi)&&=\int_L {tr_a\left[\frac 12
    \p1 d\p1+\frac 13 \p1\p1\p1\right]} +\int_{L}{tr_b\left[\frac 12
    \hp1 d\hp1+\frac 13 \hp1\hp1\hp1 \right]}~~.  \eea Note that the
field $\phi^{(3)}$ does not appear in the action, and thus its
components define classical flat directions in the moduli space of the
theory.  We stress that, even though the action (\ref{action_2}) is
invariant with respect to the shift symmetry $\phi^{(3)}\rightarrow
\phi^{(3)}+b^{(3)}$ (with $b^{(3)}$ an arbitrary 3-form valued in the
bundle $Hom(E_b,E_a))$, we do {\em not} include such symmetries in our
gauge algebra. The reason is that a nonzero background value of
$\phi^{(3)}$ corresponds to condensation of boundary condition
changing states between the graded branes $a$ and $b$, and thus leads
to a {\em physically relevant} shift of the string vacuum.  This
forbids us to identify $\phi^{(3)}$ and $\phi^{(3)}+b^{(3)}$.  This
means that we cannot treat shifts of $\phi^{(3)}$ as {\em gauge}
symmetries.

Since one of the classical fields is a massless 3-form, we expect
first, second and third generation ghosts: \be C_1=\pmatrix{
  c_1^{(0)}&c_1^{(2)}\cr 0&{c'}_1^{(0)}\cr}~~,~~ C_2=\pmatrix{
  0&c_2^{(1)}\cr 0&0\cr}~~,~~ C_3=\pmatrix {0&c_3^{(0)}\cr 0&0\cr}~~.
\ee The BRST transformations of the classical fields read: \bea
\delta^{(1)}\p1=\left(dc_1^{(0)}+[\p{1},\,c_1^{(0)}]\right)\lambda~~,~~
\delta^{(1)}\hp1=\left(d{c'}_1^{(0)}+[\hp{1},\,{c'}_1^{(0)}]\right)\lambda\\
\delta^{(1)}\p3=\left(dc_1^{(2)}-c_1^{(2)}\hp{1}+\p{1}c_1^{(2)}-c_1^{(0)}\p{3}+
  \p{3}{c'}_1^{(0)}\right)\lambda~~.
\label{brst2o}~~\nn
\eea The transformations of first generation ghosts are derived by
requiring nilpotence of $\delta^{(1)}$ on $\phi$: \be \delta^{(1)}
c_1^{(2)}=\left(c_1^{(0)}c_1^{(2)}+c_1^{(2)}{c'}_1^{(0)}\right)\lambda~~,~~
\delta^{(1)} c_1^{(0)}=c_1^{(0)}c_1^{(0)}\lambda~~,~~ \delta^{(1)}
c_1^{'(0)}=c_1^{'(0)}{c'}_1^{(0)}\lambda~~.  \ee Once again,
(\ref{brst2o}) has on-shell zero modes, which requires us to extend
the BRST transformation of $c_1^{(2)}$ by adding the variation: \be
\delta_2 c_1^{(2)}
=(dc_2^{(1)}+\p{1}c_2^{(1)}+c_2^{(1)}\hp{1})\lambda~~.
\label{eq:2ndgengh}
\ee Allowing $\delta_2$ to act trivially on the physical fields
($\delta_2\phi=0$) and requiring nilpotence of
$\delta^{(2)}=\delta^{(1)}+\delta_2$ implies that the second
generation ghosts must transform as \be {\delta_2
  c_2^{(1)}}=(-c_1^{(0)}c_2^{(1)}+ c_2^{(1)}{c'}_1^{(0)})\lambda~~.
\ee Equation (\ref{eq:2ndgengh}) has further zero modes which lead to
the third generation ghosts and to the transformations: \be {\delta_3
  c_2^{(1)}}=\left(dc_3^{(0)}-c_3^{(0)}\hp{1}+\p{1}c_3^{(0)}\right)\lambda~~,~~
\delta_3
c_3^{(0)}=\left(c_1^{(0)}c_3^{(0)}+c_3^{(0)}{c'}_1^{(0)}\right)\lambda~~.
\label{eq:3rdgengh}
\ee Since there are no further zero modes, the theory is second order
reducible and we conclude that the full BRST transformations are given
by: \be
\delta\equiv\delta^{(3)}=\delta^{(2)}+\delta_3=\delta^{(1)}+\delta_2+\delta_3
\label{eq:fullBRSTn2}
\ee where $\delta^{(1)}$ acts trivially on the second and third
generation ghosts, $\delta_2$ acts trivially on the physical fields
and the third generation ghosts and $\delta_3$ acts trivially on the
physical fields and the first generation ghosts.

We next add the antifields: \be \Phi^*=\pmatrix{ \phi^{*(2)}& 0\cr
  \phi^{*(0)}&{\phi'}^{*(2)}\cr}~~,~~ C_1^*=\pmatrix{ c_1^{*(3)}&0\cr
  c_1^{*(1)} &{c'}_1^{*(3)}\cr}~~,~~ C_2^*=\pmatrix{ 0&0\cr c_2^{*(2)}
  &0\cr}~~,~~ C_3^*=\pmatrix{ 0&0\cr c_3^{*(3)}&0\cr}~~, \ee and build
the first order action $S_1$ as in the previous section.

Since $\{S_1, S_1\}$ does not vanish, the last step is to add a term
$S_2$ quadratic in antifields, such that the classical master equation
is satisfied.  This leads to the full BV action: \bea
\label{bv2}
S_{BV}=\int_L\, tr_a&&\Big[{1\ov 2}\p{1}d\p{1}+{1\ov 3}\p{1}\p{1}\p{1}
+\phi^{*(2)}\left(dc_1^{(0)}+[\p{1},\,c_1^{(0)}]\right)\nn\\
&&+c_1^{*(3)}\,c_1^{(0)}c_1^{(0)}
-c_3^{(0)}\left({\phi'}^{*(2)}c_1^{*(1)}-c_1^{*(1)}\phi^{*(2)}
  +\phi^{*(0)}
  c_1^{*(3)}-{c'}^{*(3)}_1\phi^{*(0)} \right)\nn\\
&&+c_2^{(1)}\left({\phi'}^{*(2)}\phi^{*(0)}+\phi^{*(0)}\phi^{*(2)}\right)\Big]
+\nn\\
+\int_L\,tr_b&&\Big[ {1\ov 2}\hp{1}d\hp{1}+{1\ov 3}\hp{1}\hp{1}\hp{1}
+{\phi'}^{*(2)}\left(d{c'}_1^{(0)}+[\hp{1},{c'}_1^{(0)}]\right)\nn\\
&&+\phi^{*(0)}\left(dc_1^{(2)}-c_1^{(2)}\hp{1}+\p{1}c_1^{(2)}-c_1^{(0)}
  \p{3}+\p{3}c_1^{'(0)}\right)\nn\\
&&
+c_1^{*(1)}\left(c_1^{(0)}c_1^{(2)}+c_1^{(2)}{c'}_1^{(0)}+dc_2^{(1)}+\p{1}c_2^{(1)}
  +c_2^{(1)}\hp{1}\right)
\nonumber\\
&&
+c_2^{*(2)}\left(dc_3^{(0)}-c_3^{(0)}\hp{1}+\p{1}c_3^{(0)}-c_1^{(0)}c_2^{(1)}
  +c_2^{(1)}{c'}_1^{(0)}\right)
\nonumber\\
&&+{c'}^{*(3)}_1{c'}_1^{(0)}{c'}_1^{(0)}-
c_3^{*(3)}\,\left(c_1^{(0)}c_3^{(0)}+c_3^{(0)}{c'}_1^{(0)}\right)
\Big]~~.  \eea A brief discussion on the master equation can be found
in Appendix C.

This action can be cast into the form (\ref{extended_action}) provided
that we identify the extended string field with: \bea
\label{coords2}
{\hat \phi}&=&\pmatrix{ c_1^{(0)}+\phi^{(1)}-\phi^{*(2)} +c_1^{*(3)}&
  c_3^{(0)}+c_2^{(1)}+c_1^{(2)}+ \phi^{(3)}\cr -\phi^{*(0)}+c_1^{*(1)}
  -c_2^{*(2)}+c_3^{*(3)}&
  {c'}_1^{(0)} +\hp{1}-{\phi'}^{*(2)}+ {c'}^{*(3)}_1\cr}~~\nn\\
\nn\\
&=&c_3+c_2+c_1+\phi+\phi^*+c_1^*+c_2^* +c_3^*~~,
\label{eq:superfieldn2}
\eea where we defined: \be \phi^*=\pmatrix{-\phi^{*(2)}& 0\cr
  -\phi^{*(0)}&-{\phi'}^{*(2)}\cr}~~,~~ c_2^*=\pmatrix{ 0&0\cr
  -c_2^{*(2)} &0\cr}~~ \ee and \be
c_1=C_1~~,~~c_2=C_2~~,~~c_3=C_3~~,~~c_1^*=C_1^*~~,~~c_3^*=C_3^*~~.
\label{ghn2}
\ee We have $deg{\hat \phi}={\hat 1}$, as required.  Note that for
$n=2$, the graded trace $str_e$ in (\ref{extended_action}) coincides
with the ordinary trace.  One checks again by direct computation that
(\ref{coords2}) gives Darboux coordinates for the odd symplectic form
and that the extended action (\ref{extended_action}) reduces to
(\ref{bv2}) in this basis.

\subsection{The BV action for $n=0$ }

We next consider D-brane pairs with relative grading $n=0$.  In this
case, the gauge algebra is irreducible and one needs only first
generation ghosts.  The physical fields and their ghosts are given by:
\be \phi=\pmatrix{ \phi^{(1)} & {\check\phi}^{(1)}\cr
  {\tilde\phi}^{(1)} & {\phi'}^{(1)} \cr} ~~~~~~~~~,~~~~~~~~~
c_1=\pmatrix{ c_1^{(0)} & {\check c}_1^{(0)}\cr {\tilde c}_1^{(0)}&
  {c'}_1^{(0)} \cr}
\label{fields0}~~.
\ee Because the relative shift vanishes, the product $\bullet$ is the
usual wedge product of forms and the classical action is the CS action
for $E_a\oplus E_b$.  Axelrod and Singer \cite{AS1} quantized the same
classical theory, in the Faddeev-Popov approach. They expressed the
gauged fixed action as an `extended CS action', where the top form in
the extended string field $\hat\phi$ is constrained to vanish. Here we
show that the BV construction yields the same extended action, but for
us the top form is unconstrained \footnote{This BV action is of course
  well-known (see, for example, \cite{Ikemori}).}.

The BV action for this system is: \be
\label{bv0}
S_{BV}=\int_L~tr_{E_a\oplus E_b}\left[{1\ov 2}\phi d\phi+{1\ov
    3}\phi\phi\phi +\phi{}^*\left(dc_1+\phi
    c_1-c_1\phi\right)+c_1^*c_1c_1 \right]~~, \ee where \be
\Phi^*=\pmatrix{ \phi^{*(2)} & \phi^{*(2)}\cr \phi^{*(2)}&
  {\phi'}^{*(2)} \cr} ~~{\rm~and~}~~ c_1^*=\pmatrix{ c_1^{*(3)} &
  c_1^{*(3)}\cr c_1^{*(3)} & c^{'*(3)}_1 \cr}
\label{eq:antifn=0}
\ee are the antifields associated with (\ref{fields0}).

To present $S_{BV}$ in the form (\ref{extended_action}), one defines
the extended string field: \be
\label{coords0}
\hat \phi=c_1+ \phi + \phi^* + c_1^*~~, \ee where $\phi^*:=-\Phi^*$.
Upon substituting this into (\ref{extended_action}), we recover the BV
action (\ref{bv0}).  It is also easy to check that (\ref{coords0})
gives Darboux coordinates for the odd symplectic form.

\subsection{The BV action for $n\geq 3$}

For relative grading $n\ge 3$, the physical field contains two
one-forms: \be {\phi}=\pmatrix{ \phi^{(1)} & 0 \cr 0 & {\phi'}^{(1)}
  \cr} \ee and the action appears as the direct sum or difference
(according to the supertrace prescription) of two Chern-Simons terms:
\bea S=\int_{L}{tr_a\left[
    \frac{1}{2}{\phi}{}^{(1)}d{\phi}{}^{(1)}+\frac{1}{3}
    ({\phi}{}^{(1)})^3\right]}+(-1)^n \int_{L}{tr_b\left[
    \frac{1}{2}{\phi'}{}^{(1)}d{\phi'}{}^{(1)}+\frac{1}{3}
    ({\phi'}{}^{(1)})^3\right]}~~.\label{truncation} \eea

Based on this form, it would naively seem that we can obtain the BV
action simply as a sum or difference of the classical master actions
associated with the two Chern-Simons theories,
$S_{BVnaive}=S_{BV}^a+S_{BV}^b~$, where, according to the discussion
of the previous subsection, the BV actions for the two sectors have
the extended Chern-Simons form.  However, this conclusion does not
take into account the physically correct form of the gauge algebra.
Indeed, a direct sum construction of the BV action is only justified
if {\em both} the action and the gauge algebra are of direct sum form.

A systematic approach to the analysis of pairs with $n\geq 3$ is
afforded by the device of `gauging zero' \cite{warren}, which in our
case amounts to formally applying the BRST procedure to forms of ranks
higher than $3$. For this, we shall pretend that an $(n+1)$-form is
not identically vanishing in three dimensions\footnote{For the
  rigorous reader, we note that this procedure can be justified in the
  jet bundle approach \cite{Stasheff_bv, Barnich}.  Consider the
  bundle ${\cal W}=End({\bf E})= End(E_a\oplus E_b)$, whose typical
  fiber we denote by $W$.  One defines local coordinates $x^i$ on the
  base manifold and $u^a$ on the fiber. The infinite jet bundle
  $J^\infty {\cal W}$ is a prolongation of ${\cal W}$ to an
  infinite-dimensional vector bundle with coordinates $u_I^a=(u^a,
  u_{i_1}^a, u_{i_1 i_2}^a, \dots)$ where $i_k\in \{1, 2, 3\}$ and the
  indices are symmetrized; the coordinates $u_I=u_{i_1..i_k}$, for a
  symmetric multi-index $I=(i_1\dots i_k)$, are viewed as formal
  partial derivatives of $u_a$ and regarded as functionally
  independent. A section $\sigma$ of the jet bundle decomposes as
  $\sigma =\sum_{k=0}^\infty{\sigma_k}$, according to the number $k$
  of formal derivatives; the component $\sigma_0$ defines a section of
  the unextended bundle ${\cal W}$.  A section $s$ of ${\cal W}$ (with
  components $u^a\circ s =s^a$) induces a section of $J^\infty {\cal
    W}$, the {\em associated jet} $j^\infty s$, which has the property
  $u_I^a\circ j^\infty s = \partial_{i_1}\partial_{i_2}\dots
  \partial_{i_r}s^a$. This decomposes in a sum $j^\infty s=(j^\infty
  s)_0+(j^\infty s)_1+(j^\infty s)_2+\dots~$.  The zeroth component of
  the jet is the original section, $(j^\infty s)_0=s$. One can also
  define forms on the infinite jet bundle: an element $\omega^{(r,s)}$
  of the {\em variational bicomplex} $\Omega^{*,*}(J^\infty {\cal W})$
  can be locally expressed in the basis $\theta_{I_1}^{a_1} \wedge
  \dots\theta_{I_r}^{a_r}\wedge dx^{i_1}\wedge\dots \wedge dx^{i_s}$
  where the {\em contact forms} $\theta_I^a=du_I^a-\sum_{i=1}^3
  u_{Ii}^a dx^i$ span the vertical directions, and have the property
  $(j^\infty s)^*(\theta^a_I)=d(\partial_I\varphi^a)-
  (\partial_{Ii}s^a)dx^i =0$.  The space ${\cal H}$ of classical
  fields is the space of sections of ${\cal W}\otimes \Lambda^*T^*L$,
  which prolongs to the space $\Omega^{0,*}(J^\infty{\cal W})$ of {\em
    horizontal forms}.  Hence given a classical field
  $\phi=\sum_{s=1}^3{\phi_{i_1\dots i_s} dx^{i_1}\wedge \dots \wedge
    dx^{i_s}}$, we can define its jet $j^\infty(\phi)= \sum_{s=1}^3{
    j^\infty(\phi_{i_1\dots i_s})dx^{i_1}\wedge \dots \wedge
    dx^{i_s}}\in \Omega^{0,*}(J^\infty{\cal W})$, whose zeroth
  component $j(\phi)_0$ coincides with $\phi$.  The trick of
  introducing forms of rank higher than three can be understood as
  working with elements of $\Omega^{*,*}(J^\infty {\cal W})$. For our
  purpose, a `formal $k$-form' $\omega$ is an element of the
  variational bicomplex having total degree $p+q=k$. This can be
  decomposed as $\omega=\omega^{(k,0)}\oplus \omega^{(k-1,1)} \oplus
  \omega^{(k-2,2)} \oplus \omega^{(k-3,3)}$ and has a horizontal
  component $\omega^{(0,k)}$ if and only if the formal rank $k$ lies
  between $0$ and $3$.  A similar construction can be given for the
  various ghost generations, upon considering Grassmann-valued forms.
  At the end, we shall `evaluate' all such forms, i.e. take the zeroth
  jet component of their horizontal projection, which is a
  (Grassmann-valued) section of the physical bundle ${\cal W}\otimes
  \Lambda^*T^*L$. This defines a BV field, i.e. an element of the
  extended boundary space ${\cal H}_e$.  It is clear that `evaluation'
  in this sense applied on $\omega$ will produce zero unless the
  formal rank $k$ belongs to $0..3$, in which case the result of
  evaluation will be the zeroth jet component of $\omega^{(0,k)}$,
  i.e. a `true' $k$-form field. This recovers the procedure of
  \cite{warren}.}, and declare that the physical fields are \be
{\phi}=\pmatrix{ \phi^{(1)} & \p{n+1} \cr 0 & {\phi'}^{(1)} \cr}~~.
\ee

We next perform the BV construction taking this as a starting point.
Forms of ranks higher than 3 will be set to zero only in the final
result.

Correspondingly, we have the BRST variations: \bea
\delta^{(1)}\p1=\left(dc^{(0)}_1+[\p{1},\,c_1^{(0)}]\right)\lambda~~,~~
\delta^{(1)}\hp1=\left(d c'{}^{(0)}_1+[\hp{1},\, c'{}_1^{(0)}]\right)
\lambda~~~~~~~\nn\\
\delta^{(1)}\p{n+1}=\left(d c^{(n)}_1 +(-1)^{1+n}
  c^{(n)}_1\hp{1}+\p{1} c^{(n)}_1+\p{n+1} c'{}^{(0)}_1-
  c^{(0)}_1\p{n+1}\right)\lambda~~.~~~~~~~ \eea

These extend to nilpotent transformations provided that we use the
following variations of the ghosts: \bea \delta^{(1)}
c^{(0)}_1&=&c^{(0)}_1c^{(0)}_1\lambda ~~;~~~ \delta^{(1)}
c'{}^{(0)}_1=c'{}^{(0)}_1c'{}^{(0)}_1\lambda ~~;~~~ \delta^{(1)}
c'{}^{(n)}_1=(c{}^{(0)}_1c{}^{(n)}_1+c{}^{(n)}_1c'{}^{(0)}_1)\lambda
\label{eq:lev1}~~.~~~~~~~ 
\eea One can write the following more synthetic form of the first
level BRST transformations: \be \delta^{(1)}
\phi=\left(dC_1+[\phi,\,C_1]_\bullet\right)\lambda~~,~{\rm with}~~~~
C_1=\pmatrix{ c^{(0)}_1& c^{(n)}_1\cr 0& c'{}_1^{(0)}\cr}~~.
\label{eq:lev1mat}
\ee

On the classical shell the transformation $\delta^{(1)}\p{n+1}$ has a
residual local invariance given by \be \delta_2 c{}^{(n)}_1=\left(d
  c{}^{(n-1)}_2 +\p{1} c^{(n-1)}_2+ (-1)^n
  c^{(n-1)}_2\hp{1}\right)\lambda~~, \ee where $c^{(n-1)}_2$ is the
second generation ghost.  With the notations introduced in Sections
7.1 and 7.2, the full BRST transformations
$\delta^{(2)}\equiv\delta^{(1)}+\delta_2$ at the second level coincide
with $\delta^{(1)}$ when acting on $\phi^{(1)}$, ${\phi'}{}^{(1)}$,
$c_1^{(0)}$, ${c'}{}_1^{(0)}$, and act on $c_1^{(n)}$ as: \bea
\label{unu}
\delta^{(2)}c^{(n)}_1 =(d c{}^{(n-1)}_2 +\p{1} c^{(n-1)}_2+
(-1)^nc^{(n-1)}_2\hp{1} +
c{}^{(0)}_1c{}^{(n)}_1+c{}^{(n)}_1c'{}^{(0)}_1)\lambda~~.  \eea The
transformations of second generation ghosts are determined by
requiring nilpotence of (\ref{unu}).  On the other hand, on-shell zero
modes of (\ref{unu}) lead to ghosts of the next generation.

Since we want to discuss all cases with $n\geq 3$, we give an
inductive proof that the BRST transformations of the $k^{th}$
generation ghosts have the form: \bea
\label{k1}
\delta^{(k)}c{}^{(n-k+1)}_{k}&=&\left(d c{}^{(n-k)}_{k+1} +\p{1}
  c^{(n-k)}_{k+1}+ (-1)^{n-k+2}
  c^{(n-k)}_{k+1}\hp{1}\right.\nonumber\\
&{}&~~~~+(-1)^{k+1}\left.c{}^{(0)}_1 c{}^{(n-k+1)}_k+c{}^{(n-k+1)}_k
  c'{}^{(0)}_1\right)\lambda~~.  \eea Assume that (\ref{k1}) holds for
the $k-1$ ghosts: \bea \delta^{(k-1)}c{}^{(n-k+2)}_{k-1}&=&\left(d
  c{}^{(n-k+1)}_{k} +\p{1} c^{(n-k+1)}_{k}+ (-1)^{n-k+1}
  c^{(n-k+1)}_{k}\hp{1}\right.\nonumber\\
&+&(-1)^{k}\left.c{}^{(0)}_1 c{}^{(n-k+2)}_{k-1}+c{}^{(n-k+2)}_{k-1}
  c'{}^{(0)}_1\label{k2}\right)\lambda\label{doi1}~~.  \eea Then
nilpotence of the BRST transformation of the $(k-1)^{th}$ generation
ghost implies: \be \delta_{k-1} c_k{}^{(n-k+1)}=\left((-1)^{k+1}
  c_1^{(0)}c_k{}^{(n-k+1)} +c_k{}^{(n-k+1)} c'{}_1^{(0)}
\right)\lambda
\label{doi2}~~.
\ee We finally note that the BRST transformations (\ref{doi2}) have
the following on-shell zero modes: \be \delta_k
c_{k}^{(n-k+1)}=\left(d c{}^{(n-k)}_{k+1} +\p{1} c^{(n-k)}_{k+1}+
  (-1)^{n-k+2} c^{(n-k)}_{k+1}\hp{1} \right)\lambda\label{trei}~~.
\ee We recover the full BRST transformation (\ref{k1}) as the sum of
(\ref{doi2}) and (\ref{trei}),
$\delta^{(k)}c_{k}^{(n-k+1)}=\sum_{j=1}^k \delta_j c_{k}^{(n-k+1)}
=(\delta_{k-1} +\delta_k) c_{k}^{(n-k+1)}$, where we used the fact
that all transformations with the exception those of levels $k-1$ and
$k$ act trivially on $c_{k}^{(n-k+1)}$.

Note that the Grassmann parity of the ghosts depends on the generation
number \be
\label{gparity}
g(c_k^{(n-k+1)})=(-)^k~~.  \ee

We now set to zero all forms of ranks higher than $3$, which
eliminates all ghosts except for those of ranks between zero and
three.  The ansatz: \bea
\label{coords_general}
\!\!\!\!\!\!\!\!\!  \!\!\!\!\!\!\!\!\!  &&{\hat\phi}=
\nonu\\
\!\!\!\!\!\!\!\!\!\!\!  \!\!\!\!\!\!\!\!\! &&\!\!\!\!\!  \pmatrix{
  c_1^{(0)}+\phi^{(1)} - \phi^{*(2)}+c_1^{*(3)}&
  c_{n+1}^{(0)}+c_{n}^{(1)}+c_{n-1}^{(2)}+c_{n-2}^{(3)} \cr
  (-)^{n+1}c_{n-2}^{*(0)}+c_{n-1}^{*(1)}-(-)^{n}c_{n}^{*(2)}+
  c_{n+1}^{*(3)} & c_1^{'(0)}+\phi^{'(1)}-(-)^{n}\phi^{'*(2)} +(-)^n
  c_1^{'*(3)}
  \cr}\nonumber\\
\eea shows that the resulting BV action coincides with the extended
action (\ref{extended_action}). It is also easy to check that
(\ref{coords_general}) define Darboux coordinates for the odd
symplectic form (\ref{omega}). Moreover, because of equation
(\ref{gparity}), the degree of the extended string field is~$1$.

\subsection{On the equivalence of the extended actions with $n\ge 3$}

Given the fact that the classical actions and physical fields of the
theories with relative grading $n\ge 3$ are identical, we could
attempt to relate the extended actions by a canonical transformation.
As shown in the previous subsection, in these cases the extended
action can be obtained by adding a vanishing $(n+1)$-form in the
sector $Hom(E_a,\,E_b)$.  Without this trick, one would conclude that
the BV action is the graded sum of the two master actions
corresponding to $E_a$ and $E_b$.  Therefore, it seems natural to
inquire whether the extended action can be expressed in this form.  We
will first isolate from the extended action the graded sum part and
then search for a canonical transformation which annihilates the
remnant.

Let the extended field ${\hat\phi}$ for some grade difference $n\ge 3$
be \be {\hat \phi}={\hat\phi_0}+{\hat c}+{\hat c}{}^*~~, \ee where:
\bea
{\hat\phi}_0&=&\pmatrix{c_1^{(0)}+\phi^{(1)}+\phi^{*(2)}+c_1^{*(3)}&0\cr
  0& c'{}_1^{(0)}+\phi'{}^{(1)}+\phi'{}^{*(2)}+c'{}_1^{*(3)} \cr}
\equiv
\pmatrix{{\phi}&0\cr 0&{\phi}'\cr}   \nonumber \\
\nn\\
\nn\\
{\hat
  c}&=&\pmatrix{0&c_{k+3}^{(0)}+c_{k+2}^{(1)}+c_{k+1}^{(2)}+c_{k}^{(3)}
  \cr 0&0\cr}\equiv \pmatrix{0&c\cr 0&0\cr} \nn\\
\nn\\
{\hat c}{}^*&=&\pmatrix{0&0\cr
  c_{k+3}^{*(0)}+c_{k+2}^{*(1)}+c_{k+1}^{*(2)}+c_{k}^{*(3)}
  &0\cr}\equiv \pmatrix{0& 0 \cr c^*&0\cr}~~.  \eea Then the extended
action is: \be S_e =S_e ({\hat \phi}_0)+\int_L\,str_e\Big[ {\hat c}^*
*\big(d{\hat c}+{\hat \phi}_0*{\hat c}+{\hat c}*{\hat
  \phi}_0\big)\Big]~~, \ee where $S_e({\hat \phi}_0)$ is the graded
sum of the two BV actions corresponding to $E_a$ and $E_b$. We now
look for a canonical transformation which annihilates the last term of
the previous equation: \bea
\label{Delta_S}
\Delta S_e \equiv S_e({\hat\phi})-S_e({\hat\phi}_0)&=&(-1)^n
\int_L\,tr_b\Big[ c{}^**\left(dc+\phi * c+c *\phi'\right)
\Big]\nn\\
&=&\int_L\,tr_a\Big[ c*\left(-dc{}^*+\phi' * c{}^*+c{}^* * \phi\right)
\Big]~~.  \eea

Let $\Psi$ be the generator of this canonical transformation, a
Grassmann-odd function of fields and antifields whose ghost number
equals $-1$.  If $ad_\Psi(F):=\{\Psi,F\}$ is the left adjoint action
of $\Psi$ on Grassmann-valued functions $F$ of fields and antifields,
then the associated one-parameter group of canonical transformations
acts as $e^{t ad_\Psi}F$.  Notice that the BV bracket of two monomials
of degrees $p$ and $q$ in fields and antifields gives either zero or a
homogeneous polynomial of degree $p+q-2$.  Since (\ref{Delta_S})
contains terms of degrees $2$ and $3$ in fields and antifields, it
follows that the adjoint action of a term of order $p$ in the Taylor
expansion of $\Psi$ will produce zero or a sum of polynomials of
degrees $p$ and $p+1$. This implies that the only source of quadratic
terms in $e^{t ad_\Psi}\Delta S_e$ is the quadratic term in the
expansion of $\Psi$ acting an arbitrary number of times on the
quadratic term in the expansion of $\Delta S_e$.  Therefore, a crucial
property of the desired generator $\Psi$ is that the action of $e^{t
  ad_\Psi}$ on the quadratic part of $\Delta S_e$ vanishes.

If we also also require that $S_e({\hat \phi}_0)$ be invariant under
the canonical transformation we are looking for, it follows that the
only choice (up to a factor) for the quadratic term in $\Psi$ is \be
\Psi_2=\int_L str_e[{\hat c}*{\hat c}{}^*]=
\int_Ltr_a[c*c{}^*]=-(-)^n\int_Ltr_b[c{}^* * c]~~, \ee since
$S_e({\hat \phi}_0)$ is independent of the pair $({\hat c}, {\hat
  c}^*)$.  The fact that $\Delta S_e$ is linear in fields and
antifields, \be tr_a[c{\partial \ov \partial c}\Delta S_e]=\Delta
S_e~~~~~~~ tr_b[c{}^*{\partial \ov \partial c{}^*}\Delta S_E]=\Delta
S_e \ee implies that: \be ad_{\Psi_2}\Delta S_e\equiv \{\Psi_2,\,
\Delta S_e\}=-(1+(-1)^n)\Delta S_e~~.  \ee It follows that a finite
transformation of $\Delta S_e$ has the form: \be e^{t
  ~ad_{\Psi_2}}\Delta S_e = {1\ov 2} \left(1-(-1)^n\right)\Delta S_e
+{1\ov 2} \left(1+(-1)^n\right)e^{-2t}\Delta S_e~~.
\label{eq:cannytransf}
\ee We therefore find that for odd $n$ the canonical transformation
generated by $\Psi_2$ maps $\Delta S_e$ to itself. In particular,
since $\Psi_2$ was unique up to a factor, this implies that $\Delta
S_e$ cannot be removed. Hence $S_e$ cannot be canonically transformed
into a difference of CS actions for $n=2k+1\geq 3$.

For even $n$, the transformation (\ref{eq:cannytransf}) multiplies
$\Delta S_e$ by the factor $e^{-2t}$ which is non-vanishing for any
finite $t$. In the limit $t\rightarrow +\infty$, $\Delta S_e$ is
mapped to zero. Thus, by choosing $\Psi=\Psi_2$ it seems that we can
remove the full $\Delta S_e$. Unfortunately, the limit $t\rightarrow
+\infty$ describes a point in the closure of the space of canonical
transformations and it is not clear whether such a transformation is
indeed allowed.

Thus, the extended actions with relative grading $n\ge 3$ (with $n$ of
a given parity) do not seem to be canonically equivalent, even though
their classical counterparts are. However, this does not necessarily
mean that the physical phenomena described by them are different,
since the differences between the actions arise, as shown in the
previous section, by gauging the symmetries associated to a vanishing
field. In \cite{warren} a rather similar situation was analyzed (the
classical action was taken to vanish and the BV action was constructed
using the gauge symmetries of a 5-form field strength coupled to
gravity, in four dimensions). The conclusion of the analysis was that
the vanishing of physical observables survives quantization and the
fields introduced by the BV construction contribute only to gauge
dependent correlation functions. The analogy with the situation in
that paper suggests that once quantization is performed, the
off-diagonal sector $Hom(E_a,\,E_b)$ does not contribute to any
observable, and the equivalence of classical actions with $n\geq 3$
extends to an equivalence between the algebras of observables at the
quantum level.

\section{Relation between the constructive and geometric approach}

As discussed in Sections 4 and 7, the role of $S_e$ as a tree-level BV
action for graded D-brane pairs can be understood from two quite
different perspectives, a `top-down' approach which makes use of the
geometric formalism and a `bottom-up' approach which uses the standard
homological approach.  While the geometric formalism has the advantage
of extreme generality (for example, it applies to an arbitrary number
of graded D-branes, for which direct computation in components is
impractical) and conceptual elegance, the discussion of Section 7 is
more explicit and constructive (providing, in particular, a deeper
understanding of the origin of various ghosts). In this section we
explain the relation between the two descriptions, and give a more
synthetic formulation of the recursive computations of the previous
section.

It is clear from the direct analysis of the previous section that the
gauge algebra of our models is generally reducible.  Therefore, the
BRST procedure extends $S$ by adding the first order action $S_1$,
which is built recursively from a tower of ghosts of various
generations, accompanied by the corresponding antifields.  This leads
to successive extension of the string field $\phi$ to enlarged
collections of fields $\boldphi^{(k)}$ and antifields
$\boldphi^{*(k)}$.  At each step $k$, one builds the $k^{th}$
approximation $S_1^{(k)}$ (defined on ${\cal L}^{(k)}\oplus {\cal
  N}^{(k)}$) to the first order action: \be
\label{Sk}
S^{(k)}_1(\boldphi^{(k)}, \boldphi^{*(k)})= -\omega(\boldphi^{*(k)},
{\bf q}^{(k)}(\boldphi^{(k)}))~~, \ee where ${\bf q}^{(k)}$ is the
BRST generator at order $k$ (identified with a nonlinear operator
acting on the linear space of truncated fields).  The procedure stops
at the step $k_m=\sigma+1$ dictated by order of reducibility $\sigma$
of the gauge algebra. Then ${\bf q}^{(k_m)}={\bf q}$ and
$S_1=S^{(k_m)}_1=-\omega(\boldphi^*, {\bf q}(\boldphi))$. In view of
(\ref{Sexp}), the full master action $S_{BV}=S_e$ is obtained from
$S+S_1$ by adding the following term quadratic in in antifields: \be
S_2(\boldphi^*,\boldphi)=\omega(\boldphi, \boldphi^**\boldphi^*)~~.
\ee To describe this in the geometric language of Section 4, we
consider the expansion $\boldphi=\sum_{s\geq 0}{\phi_s}$, with
$\phi_s\in M_s$.  The BRST operator (\ref{BRST}) has the form: \be
{\bf q}(\boldphi)=\oplus_{s\geq 0}{{\bf q}_s(\boldphi)}~~, \ee with
${\bf q}_s(\boldphi)\in \Pi M_s=P_s$.  Expanding (\ref{BRST}) gives:
\be {\bf q}_s(\boldphi)=-d\phi_{s+1}-\sum_{i+j=s+1}{\phi_i*\phi_j}~~.
\ee Consider the subspaces: \be {\cal L}^{(k)}:=\oplus_{0\leq s\leq
  k}{M_{s}}~~,~~ {\cal R}^{(k)}:=\oplus_{s>k}{M_{s}}~~, \ee which give
complementary ascending and descending sequences of approximations for
${\cal L}$: \bea
\label{chains}
& &{\cal L}^{(0)}~\subset {\cal L}^{(1)}~\subset {\cal
  L}^{(2)}~\subset ...
~\subset{\cal L}^{(k_m)}={\cal L}~~\nn\\
& &0={\cal R}^{(k_m)}\subset {\cal R}^{(k_{m-1})}\subset
...\subset {\cal R}^{(1)}\subset{\cal R}^{(0)}~~,\\
& &{\cal L}^{(k)}\oplus {\cal R}^{(k)}={\cal L}~~.\nn \eea We also
define ${\cal N}^{(k)}= \oplus_{-k\leq s<0}{M_{s}}$, which give an
ascending sequence of approximations for ${\cal N}$.  Then the $k$-th
approximation to ${\bf q}$ is obtained as: \be {\bf
  q}^{(k)}=P^{(k)}{\bf q}|_{{\cal L}^{(k)}}~~, \ee where $P^{(k)}$ is
the projector of $\Pi{\cal L}$ onto $\Pi {\cal L}^{(k)}$, parallel
with the subspace $\Pi{\cal R}^{(k)}$.  Defining: \be
\boldphi^{(k)}:=\sum_{0\leq s\leq k}{\phi_s}\in {\cal L}^{(k)}~~,~~
\boldphi^{*(k)}:=\sum_{0\leq s\leq k}{\phi^*_s}\in {\cal N}^{(k)}~~,
\ee we can write: \be {\bf q}^{(k)}(\boldphi^{(k)})= \oplus_{0\leq
  s\leq k}{{\bf q}^{(k)}_s(\boldphi^{(k)})}~~, \ee with: \be
\label{qk}
{\bf q}^{(k)}_s(\boldphi^{(k)})= -d\phi_{s+1}-\sum_{\tiny
  \begin{array}{c}i+j=s+1\\0\leq i,j \leq k
\end{array}}{\phi_i*\phi_j}~~.
\ee This explicit expression for ${\bf q}^{(k)}$ allows us to recover
the truncation $S_1^{(k)}$ of the first order action upon applying the
prescription (\ref{Sk}).  For comparison with the case of D-brane
pairs, we list the first three approximations $k=1,2,3$.

\paragraph{First approximation}
        
The first order extended field and antifield are: \bea
{\boldphi}^{(1)}=\phi+c_1~~,~~{\boldphi}^{*(1)}=\phi^*+c_1^*~~.  \eea
Expanding (\ref{qk}) gives: \bea
\label{brst1}
-{\bf q}_0^{(1)}(\boldphi^{(1)})&=&dc_1+\phi*c_1+c_1*\phi~~\nn\\
-{\bf q}_1^{(1)}(\boldphi^{(1)})&=&c_1*c_1~~.  \eea \footnote{In the
  previous section we used the infinitesimal BRST transformation
  denoted by $\delta$. As usual, the infinitesimal transformations are
  given by the action of the generator multiplied by the parameter of
  the transformation.  In our case, at the first step in the iterative
  procedure, we have: \bea \delta\boldphi^{(1)}\equiv -{\bf
    q}^{(1)}({\boldphi}^{(1)})\lambda~~, \nn \eea where $\lambda$ is a
  Grassmann-odd constant which plays the role of parameter for the
  BRST transformations. Projecting this relation onto $M_0({\hat 0})$
  and $M_1({\hat 1})$ we find the infinitesimal transformations used
  previously: \bea
\label{brst1lambda}
\delta^{(1)}\phi=(dc_1+\phi*c_1+c_1*\phi)\lambda~~,
~~\delta^{(1)}c_1=(c_1*c_1)\lambda~~.\nn \eea }.  For D-brane pairs
with relative grading $n=0$ the gauge algebra is irreducible ($o=0$)
and $S_1=S_1^{(1)}$.

\paragraph{Second approximation}

For this, one introduces second generation ghosts $c_2$ and antighost
$c^*_2$.  The BRST transformations of the extended field
${\boldphi}^{(2)}=\phi+c_1+c_2$ have the expanded form: \bea
\label{brst2}
-{\bf q}^{(2)}_0(\boldphi^{(2)})&=&dc_1+\phi*c_1+c_1*\phi~~\nn\\
-{\bf q}^{(2)}_1(\boldphi^{(2)})&=&dc_2+\phi*c_2+c_2*\phi+c_1*c_1~~\\
-{\bf q}^{(2)}_2(\boldphi^{(2)})&=&c_2*c_1+c_1*c_2~~.\nn \eea The
second approximation $S_1^{(2)}$ to $S_1$ is given by (\ref{Sk}) with
the antifields ${\boldphi}^{*(2)}=\phi^*+c_1^*+c_2^*$.  For D-brane
pairs of relative grading $n=1$, the gauge algebra is first stage
reducible ($o=1$) and $S_1=S_1^{(2)}$.

\paragraph{Third approximation}

We continue by adding a third generation ghost $c_3$ and antifield
$c_3^*$.  This gives the extended field
${\boldphi}^{(3)}=\phi+c_1+c_2+c_3$, whose components has the BRST
transformations (\ref{qk}): \bea
\label{brst3}
-{\bf q}^{(3)}_0(\boldphi^{(3)})&=&dc_1+\phi*c_1+c_1*\phi~~\nn\\
-{\bf q}^{(3)}_1(\boldphi^{(3)})&=&dc_2+\phi*c_2+c_2*\phi+c_1*c_1~~\nn\\
-{\bf q}^{(3)}_2(\boldphi^{(3)})&=&dc_3+\phi*c_3+c_3*\phi+c_2*c_1+c_1*c_2\\
-{\bf q}^{(3)}_3(\boldphi^{(3)})&=&c_2*c_2+c_1*c_3+c_3*c_1~~.\nn \eea
The third approximation $S_1^{(3)}$ is constructed from (\ref{Sk})
with the antifields ${\boldphi}^{*(3)}=\phi^*+c_1^*+c_2^*+c_3^*$. This
coincides with the full first order action for the case of relative
grading $n=3$.  For D-brane pairs with relative grading $n=2$, the
gauge algebra is second order reducible ($o=2$) and $S_1=S_1^{(3)}$.

To check agreement with previous computations, let us first consider
the case $n=1$.  It is not hard to check that upon inserting
(\ref{coords1}) into (\ref{brst2}) one recovers equation
(\ref{eq:fullBRSTn1}) of the previous section.  For $n=2$ one recovers
the BRST transformations (\ref{eq:fullBRSTn2}) upon inserting
(\ref{coords2}) into (\ref{brst3}).  It is also easy to check that the
difference between $\delta^{(k)}=-{\bf q}^{(k)}\lambda$ and
$\delta^{(k-1)}=-{\bf q}^{(k-1)}\lambda$ of this section produces
$\delta_{k}$ of the previous section.  It is clear that the covariant
formulation given above can be generalized to systems containing more
than two graded D-branes.

\section{Conclusions and directions for further research}

We studied graded D-brane systems along the lines proposed in
\cite{sc}, showing that the extended action written down in that paper
plays the role of classical master action for the string field theory
of such backgrounds. We gave a completely general proof of the master
equation by making use of a certain $\Z$-graded version of the
geometric BV formalism, which is based on the concept of graded
supermanifolds recently introduced by T. Voronov. We argued that
graded supermanifolds are the correct framework for a covariant
description of BV systems, and discussed the basics of the geometric
approach within this context. These results are of independent
interest for foundational studies of BV quantization.

We also performed a direct construction of the master action for the
case of graded D-brane pairs. This allowed us to identify the various
components of the extended string field as ghosts and antifields
produced by the BV procedure, and explain their origin in the
reducibility of the gauge algebra.  Upon using the formalism of
\cite{com1} and \cite{com3}, we analyzed formation of D-brane
composites in systems of two graded D-branes, and in particular gave a
rigorous construction of acyclic condensates for the case of unit
relative grading. This gives a detailed implementation and
generalization of ideas proposed in \cite{Vafa_cs}, though in a
somewhat different context.

For the case of graded D-brane pairs, we showed that the six classical
theories corresponding to various relative grades arise through
different choices of ghost grading and classical gauge for two
underlying master actions, the extended Chern-Simons and extended
super-Chern-Simons functionals.  We also showed that these theories
are inequivalent in the case when the underlying special Lagrangian
three-cycle is topologically nontrivial.  This sheds new light on the
`mod~6~periodicity' of the D-brane grade discussed from a worldsheet
perspective in \cite{Douglas_Kontsevich}.

Our results can be viewed as a starting point for the quantization of
such systems.  As already pointed out in \cite{sc}, the string field
theory of graded D-branes gives a concrete representation of points of
the {\em extended} boundary moduli space, thereby holding the promise
for a better understanding of its physical significance. This should
be of direct relevance for the homological mirror symmetry programme.
A quantum analysis of our theories around such backgrounds should lead
to new physical information, as well as to certain extensions of
various geometric invariants. This and related matters are the subject
of ongoing research. Here we only note that a thorough study away from
the large radius limit must take into account destabilization and
other instanton effects \cite{Fukaya, Fukaya2, boundary}.

Let us finally mention that a similar analysis can be carried out for
the open B-model, in which case one deals with (graded) D6-branes
wrapping the entire Calabi-Yau manifold. In that case, the relevant
string field theory is a graded version of holomorphic Chern-Simons
theory \cite{com3, Diaconescu}. It is clear that all of our
constructions have a parallel in such situations, provided that one
makes the appropriate modifications.  Since the B model does not
receive instanton corrections, the BV systems associated to graded
B-type branes should be viewed as a description of the associated
deformation theory; for example, they allow for a classification of
the local string field observables, which can be used to deform such
models. Some of these issues are currently under investigation.

\acknowledgments{ We thank W.~Siegel, M.~Ro\v{c}ek , S.~Vandoren, and
  D.~E.~Diaconescu for interesting conversations and interest in our
  work.  The authors are supported by the Research Foundation under
  NSF grant PHY-9722101.

\appendix

\section{Graded supermodules and super-bimodules}

Consider an associative superalgebra $G$ ($\Z_2$-graded associative
algebra with a unit).  We remind the reader that a {\em right/left
  supermodule} $U$ over $G$ is a $\Z_2$-graded right/left G-module
such that the scalar multiplication $U\times G\rightarrow U$
(respectively $G\times U\rightarrow U$) is homogeneous of degree zero:
\be deg(u\alpha)=degu+deg\alpha~~{\rm~respectively~}deg(\alpha
u)=degu+deg\alpha \ee where $deg$ denotes the grading on $U$ or $G$. A
{\em super-bimodule} $U$ is simultaneously a left and right $G$-module
such that the left and right scalar multiplications are compatible:
\be \alpha u=(-1)^{deg\alpha~deg u}u\alpha~~.  \ee It is clear that
this relation determines one scalar multiplication given the other, so
a left or right supermodule can be made into a super-bimodule in a
unique manner.

\section{Left and right vector fields}

Consider a DeWitt-Rogers supermanifold $M$ modeled over the algebra of
constants $G$. The space ${\cal F}(M,G)$ of G-valued (smooth)
functions is then a super-bimodule over $G$ with respect to pointwise
scalar multiplication: \be (F\alpha)(p)=F(p)\alpha~~,~~(\alpha
F)(p)=\alpha F(p)~~.  \ee It is also a ring with respect to pointwise
multiplication of functions: \be (FG)(p)=F(p)G(p)~~.  \ee Combining
the two structures, we obtain a superalgebra over $G$; this
statement means that one has relations such as $\alpha FG=
(-1)^{\epsilon_\alpha\epsilon_F}F\alpha G=
(-1)^{\epsilon_\alpha(\epsilon_F+\epsilon_G)}FG\alpha$.

A left (right) vector field on $M$ is a left (right) graded derivation
of this algebra. For left vector fields, this means: \be
X_l(F\alpha)=X_l(F)\alpha~~,~~X_l(FG)=X_l(F)G+
(-1)^{\epsilon_F\epsilon_{X_l}}FX_l(G)~~, \ee while for right vector
fields one requires: \be X_r(\alpha F)=\alpha
X_r(F)~~,~~X_r(FG)=FX_r(G)+ (-1)^{\epsilon_G \epsilon_{X_r}}X_r(F)G~~.
\ee In these relations, $\epsilon_{X_l}$ and $\epsilon_{X_r}$ are the
vector field parities. One also defines scalar multiplications: \be
(\alpha X_l)(F):=\alpha X_l(F)~~,~~(Y_l\alpha)(F)=Y_l(F)\alpha~~.  \ee
With these operations and grading (and the obvious definition of
addition), the space ${\cal X}_l(M)$ of left derivations of ${\cal
  F}(M)$ is a left $G$-supermodule, while the space ${\cal X}_r(M)$ of
right derivations is a right $G$-supermodule.

It is common procedure to identify left and right vector fields on $M$
in the following manner. If $X_l$ is a left vector field, one defines
$\phi(X_l)$ by: \be \phi(X_l)(F):=(-1)^{\epsilon_X\epsilon_F}X_l(F)~~.
\ee It is the easy to check that $\phi$ is a degree zero isomorphism
between the graded Abelian groups $(X_l(M),+)$ and $(X_r(M),+)$.
Moreover, one has the property: \be
\label{phi_mult}
\phi(\alpha~X_l)=(-1)^{\epsilon_\alpha\epsilon_X}\phi(X_l)\alpha~~.
\ee This allows one to identify ${\cal X}_l(M)$ and ${\cal X}_r(M)$ to
a single abstract space ${\cal X}(M)$, the space of vector fields on
$M$.  This space inherits left and right module structures from
$X_l(M)$ and $X_r(M)$, and relation (\ref{phi_mult}) shows that the
two structures are compatible. Thus ${\cal X}(M)$ is a super-bimodule
over $G$. An element $X$ of ${\cal X}(M)$ can be viewed as a pair
$(X_l, X_r)$ with $X_l\in {\cal X}_l(G)$ and $X_r\in {\cal X}_r(G)$,
where $X_l$ and $X_r$ are $\phi$-related, i.e.  $X_r=\phi(X_l)$. In
this situation, one uses the notation: \be
X_l:=\stackrel{\rightarrow}{X}~~,~~X_r:=\stackrel{\leftarrow}{X}~~,
\ee and writes $\stackrel{\rightarrow}{X}F$ for $X_l(F)$ and
$F\stackrel{\leftarrow}{X}$ for $X_r(F)$.  With these notations, the
$\phi$-relatedness condition becomes: \be
\stackrel{\rightarrow}{X}F=(-1)^{\epsilon_X\epsilon_F}
F\stackrel{\leftarrow}{X}~~.  \ee This is the rigorous construction
behind the conventions used in Section 4.1. We warn the reader that
the vector fields $\partial^l_a$ and $\partial^r_a$ defined in that
section do {\em not} form a $\phi$-related pair, in spite of the
similarity with the notation used in this appendix.  In the body of
this paper, we use exclusively the convention that left and right
vector fields are identified to abstract objects $X$, and the left and
right components of $X$ are denoted by superscript arrows as explained
above.

\section{Direct check of the master equation for D-brane pairs}

In this appendix we exemplify the steps needed for the proof that the
BV action satisfies the master equation.  We choose to take here an
easier (but equivalent) path, namely we prove that the BRST
transformations are nilpotent. More precisely, we show that: \be
\delta({\lambda_2})\delta({\lambda_1}) \hat\phi=\{\{\hat\phi,
S_{BV}\lambda_1\},S_{BV}\lambda_2\}=0~~, \ee where $\lambda_1,
\lambda_2$ are anti-commuting constants \footnote{Each block component
  in the matrix $\hat\phi$ can be viewed locally as a matrix of forms.
  For example, $(\phi^{(1)})^{ji}$ and its antifield
  $(\phi^{*(2)})^{ij}$ have indices $i,j=1,\dots rk E_a $ while
  $(\phi^{(0)})^{ik}$ and its antifield $(\phi^{(3)})^{ki}$ have
  indices $i=1,\dots rk E_a$, $k=1,\dots rk E_b$. To avoid complicated
  notation we suppress all such indices.}  .  For the case with
relative grading $n=1$, consider the BRST transformations of $\p1$:
\bea
\label{sss}
\delta({\lambda_2})\delta({\lambda_1}) \p1&&=\{\left(dc_1^{(0)}
  -c_1^{(0)}\p1+\p1c_1^{(0)}-c_1^{(1)}\p0-c_2^{(0)}\phi^{*(1)}\right)\lambda_1,
S_{BV}\lambda_2\}\nonu\\
&&=\left[ -d((c_1^{(0)})^2+c_2^{(0)}\p0) +
  ((c_1^{(0)})^2+c_2^{(0)}\p0)\p1
  -\p1((c_1^{(0)})^2+c_2^{(0)}\p0)\right.
\nonu\\
&& +(dc_1^{(0)} -c_1^{(0)}\p1+\p1c_1^{(0)}-c_1^{(1)}\p0-c_2^{(0)}\phi^{*(1)})c_1^{(0)}\nonu\\
&&+ c_1^{(0)}(dc_1^{(0)} -c_1^{(0)}\p1+\p1c_1^{(0)}-c_1^{(1)}\p0-c_2^{(0)}\phi^{*(1)})\nonu\\
&& +(c_1^{(0)}c_1^{(1)}+c_1^{(1)}c_1^{'(0)}+dc_2^{(0)}+\p1c_2^{(0)}-c_2^{(0)}\hp1)\p0\nonu\\
&& +(c_1^{(0)}c_2^{(0)}-c_2^{(0)}c_1^{'(0)})\phi^{*(1)} +c_1^{(1)}(\p0c_1^{(0)}-c_1^{'(0)}\p0)\nonu\\
&&
\left.-c_2^{(0)}(-d\p0+\p0\p1-\hp1\p0-\phi^{*(1)}c_1^{(0)}-c_1^{'(0)}\phi^{*(1)})
\right]\lambda_1\lambda_2=0~~.  \eea We similarly have: \bea
\delta({\lambda_2})\delta({\lambda_1})
\p0&&=\{\left( \p0c_1^{(0)}+c_1^{'(0)}\p0\right)\lambda_1, S_{BV} \lambda_2\}\nonu\\
&&=\left[-(\p0c_1^{(0)}-c_1^{'(0)}\p0)c_1^{(0)}+\p0(c_2^{(0)}\p0+(c_1^{(0)})^2)\right.\nonu\\
&&\left.-((c_1^{'(0)})^2+\p0c_2^{(0)})\p0-
  c_1^{'(0)}(\p0c_1^{(0)}-c_1^{'(0)}\p0)\right]\lambda_1\lambda_2=0~~,
\eea as well as: \bea \delta({\lambda_2})\delta({\lambda_1})
\p2&&=\{\left(dc_1^{(1)}+\p2c_1^{'(0)}-c_1^{(0)}\p2+\p1c_1^{(1)}+c_1^{(1)}\hp1
  +\phi^{*(2)}c_2^{(0)}+c_2^{(0)}\phi^{'*(2)}\right)\lambda_1, S_{BV} \lambda_2\}\nonu\\
&&=\left[-(d(c_1^{(0)}c_1^{(1)}+c_1^{(1)}c_1^{'(0)}+dc_2^{(0)}+\p1c_2^{(0)}-c_2^{(0)}\hp1)\right.\nonu\\
&&-(\p1(c_1^{(0)}c_1^{(1)}+c_1^{(1)}c_1^{'(0)}+dc_2^{(0)}+\p1c_2^{(0)}-c_2^{(0)}\hp1)\nonu\\
&&-(c_1^{(0)}c_1^{(1)}+c_1^{(1)}c_1^{'(0)}+dc_2^{(0)}+\p1c_2^{(0)}-c_2^{(0)}\hp1)\hp1\nonu\\
&&+(dc_1^{(1)}+\p2c_1^{'(0)}-c_1^{(0)}+c_1^{(1)}\hp1+\p1c_1^{(1)}+\phi^{*(2)}c_2^{(0)}+c_2^{(0)}\phi^{'*(2)})c_1^{'(0)}\nonu\\
&&+c_1^{(0)}(dc_1^{(1)}+\p2c_1^{'(0)}-c_1^{(0)}+c_1^{(1)}\hp1+\p1c_1^{(1)}+\phi^{*(2)}c_2^{(0)}+c_2^{(0)}\phi^{'*(2)})\nonu\\
&&+((c_1^{(0)})^2+c_2^{(0)}\p0)\p2-\p2((c_1^{'(0)})^2+\p0c_2^{(0)})\nonu\\
&&-c_1^{(1)}(dc_1^{'(0)}+\hp1c_1^{'(0)}-c_1^{'(0)}\hp1-\p0c_1^{(1)}-\phi^{*(1)}c_2^{(0)})\nonu\\
&&(dc_1^{(0)}+\p1c_1^{(0)}-c_1^{(0)}\p1+c_1^{(1)}\p0-c_2^{(0)}\phi^{*(1)})c_1^{(1)}\nonu\\
&&+\phi^{*(2)}(c_1^{(0)}c_2^{(0)}-c_2^{(0)}c_1^{'(0)})-(c_1^{(0)}c_2^{(0)}-c_2^{(0)}c_1^{'(0)})\phi^{'*(2)}\nonu\\
&&+(d\p1+\p1{}^2+\p2\p0-\phi^{*(2)}c_1^{(0)}-c_1^{(0)}\phi^{*(2)}-c_1^{(1)}
\phi^{*(1)}+c_2^{(0)}
c_1^{*(2)})c_2^{(0)}\nonu\\
&&+c_2^{(0)}(-d\hp1-\hp1{}^2-\p0\p2-c_1^{'(0)}\phi^{'*(2)}
-\phi^{'*(2)}c_1^{'(0)}+\phi^{*(1)}c_1^{(1)}\nonu\\
&&\left.-c_1^{*(2)}c_2^{(0)})\right]\lambda_1\lambda_2=0~~.  \eea

Nilpotence of the BRST transformations in the ghost sector follows
from similar calculations. For instance: \bea
\delta_{\lambda_2}\delta_{\lambda_1}c_2^{(0)}&=&\{(c_1^{(0)}c_2^{(0)}-c_2^{(0)}c_1^{'(0)})\lambda_1, S_{BV} \lambda_2\}\nonu\\
&=&\left[c_1^{(0)}(c_1^{(0)}c_2^{(0)}-c_2^{(0)}c_1^{'(0)})-((c_1^{(0)})^2+c_2^{(0)}\p0)c_2^{(0)}\right.\nonu\\
&+&\left.c_2^{(0)}((c_1^{'(0)})^2+\p0c_2^{(0)})+((c_1^{(0)}c_2^{(0)}-c_2^{(0)}c_1^{'(0)})c_1^{'(0)})\right]
\lambda_1\lambda_2=0~~.  \eea To prove the master equation, one must
also analyze the transformation of antifields.

Let us now exemplify nilpotence of the bracket for the case of
relative grading $n=2$. For $\phi^{(1)}$, one has: \bea
\delta_{\lambda_2}\delta_{\lambda_1}\p1&&=\{\left(dc_1^{(0)}
  -c_1^{(0)}\p1+\p1c_1^{(0)}-c_2^{(0)}c_1^{*(1)}-c_2^{(1)}
  \phi^{*(0)}\right)\lambda_1,
S_{BV}\lambda_2\}\nonu\\
&&=\left[-d((c_1^{(0)})^2-c_2^{(0)}\phi^{*(0)})-\p1 ((c_1^{(0)})^2-c_2^{(0)}\phi^{*(0)})\right.\nonu\\
&&+((c_1^{(0)})^2-c_2^{(0)}\phi^{*(0)})\p1+c_1^{(0)}( dc_1^{(0)}
-c_1^{(0)}\p1+\p1c_1^{(0)}-c_2^{(0)}c_1^{*(1)}-c_2^{(1)}\phi^{*(0)})
\nonu\\
&&+(dc_1^{(0)}
-c_1^{(0)}\p1+\p1c_1^{(0)}-c_2^{(0)}c_1^{*(1)}-c_2^{(1)}\phi^{*(0)})c_1^{(0)}+(c_1^{(0)}c_2^{(0)}+c_2^{(0)}c_1^{'(0)})c_1^{*(1)}
\nonu\\
&&-c_2^{(0)}(d\phi^{*(0)}+\hp1 \phi^{*(0)}-\phi^{*(0)}\p1-c_1^{*(1)}c_1^{(0)}+c_1^{'(0)}c_1^{*(1)})\nonu\\
&&-(dc_2^{(0)}-c_2^{(0)}\hp1+\p1c_2^{(0)}-c_1^{(0)}c_2^{(1)}+c_2^{(1)}c_1^{'(0)})\phi^{*(0)}\nonu\\
&&+\left.c_2^{(1)}(\phi^{*(0)}c_1^{(0)}+
  c_1^{'(0)}\phi^{*(0)})\right]\lambda_1\lambda_2=0~~.  \eea

\end{document}

%% file: pair.pstex_t
\begin{picture}(0,0)%
\includegraphics{pair.pstex}%
\end{picture}%
\setlength{\unitlength}{4144sp}%
\begingroup\makeatletter\ifx\SetFigFont\undefined%
\gdef\SetFigFont#1#2#3#4#5{%
  \reset@font\fontsize{#1}{#2pt}%
  \fontfamily{#3}\fontseries{#4}\fontshape{#5}%
  \selectfont}%
\fi\endgroup%
\begin{picture}(4393,4123)(783,-4769)
\put(2656,-1366){\makebox(0,0)[lb]{\smash{\SetFigFont{17}{20.4}{\familydefault}{\mddefault}{\updefault}
$a$
}}}
\put(3106,-2581){\makebox(0,0)[lb]{\smash{\SetFigFont{17}{20.4}{\familydefault}{\mddefault}{\updefault}
$Hom(a,a)$
}}}
\put(2701,-961){\makebox(0,0)[lb]{\smash{\SetFigFont{17}{20.4}{\familydefault}{\mddefault}{\updefault}
$b$
}}}
\put(4096,-871){\makebox(0,0)[lb]{\smash{\SetFigFont{17}{20.4}{\familydefault}{\mddefault}{\updefault}
$Hom(a,b)$
}}}
\put(5176,-1681){\makebox(0,0)[lb]{\smash{\SetFigFont{17}{20.4}{\familydefault}{\mddefault}{\updefault}
$Hom(b,b)$
}}}
\put(5131,-3166){\makebox(0,0)[lb]{\smash{\SetFigFont{17}{20.4}{\familydefault}{\mddefault}{\updefault}
$Hom(b,a)$
}}}
\end{picture}

%% file: moduli.pstex_t
\begin{picture}(0,0)%
\includegraphics{moduli.pstex}%
\end{picture}%
\setlength{\unitlength}{4144sp}%
\begingroup\makeatletter\ifx\SetFigFont\undefined%
\gdef\SetFigFont#1#2#3#4#5{%
  \reset@font\fontsize{#1}{#2pt}%
  \fontfamily{#3}\fontseries{#4}\fontshape{#5}%
  \selectfont}%
\fi\endgroup%
\begin{picture}(5820,6310)(91,-5956)
\put(5761,-2851){\makebox(0,0)[lb]{\smash{\SetFigFont{17}{20.4}{\familydefault}{\mddefault}{\updefault}
$a[1]$
}}}
\put(4636,-2851){\makebox(0,0)[lb]{\smash{\SetFigFont{17}{20.4}{\familydefault}{\mddefault}{\updefault}
$a$
}}}
\put(4681,-3211){\makebox(0,0)[lb]{\smash{\SetFigFont{17}{20.4}{\familydefault}{\mddefault}{\updefault}
$n=1$
}}}
\put(2836,-421){\makebox(0,0)[lb]{\smash{\SetFigFont{17}{20.4}{\familydefault}{\mddefault}{\updefault}
$a[1]$
}}}
\put(1711,-421){\makebox(0,0)[lb]{\smash{\SetFigFont{17}{20.4}{\familydefault}{\mddefault}{\updefault}
$a$
}}}
\put(1711,-1096){\makebox(0,0)[lb]{\smash{\SetFigFont{17}{20.4}{\familydefault}{\mddefault}{\updefault}
$n=2$
}}}
\put(631,-4651){\makebox(0,0)[lb]{\smash{\SetFigFont{17}{20.4}{\familydefault}{\mddefault}{\updefault}
$n=0$
}}}
\put(1216,-5956){\makebox(0,0)[lb]{\smash{\SetFigFont{17}{20.4}{\familydefault}{\mddefault}{\updefault}
$a$
}}}
\put( 91,-5956){\makebox(0,0)[lb]{\smash{\SetFigFont{17}{20.4}{\familydefault}{\mddefault}{\updefault}
$a$
}}}
\put(4681,-3526){\makebox(0,0)[lb]{\smash{\SetFigFont{17}{20.4}{\familydefault}{\mddefault}{\updefault}
$H^0(End(E))=H^1_d(Hom(a,a[1]))$
}}}
\put(631,-5056){\makebox(0,0)[lb]{\smash{\SetFigFont{17}{20.4}{\familydefault}{\mddefault}{\updefault}
$H^1(End(E))=H^1_d(Hom(a,a))$
}}}
\put(1711,-781){\makebox(0,0)[lb]{\smash{\SetFigFont{17}{20.4}{\familydefault}{\mddefault}{\updefault}
$H^2(End(E))=H^1_d(Hom(a[1],a))$
}}}
\end{picture}